\documentclass[aps,prx,twocolumn,superscriptaddress,nofootinbib]{revtex4-1}
\usepackage{array}
\usepackage{booktabs}
\usepackage{multirow}
\usepackage{amsmath}
\usepackage{amssymb}
\usepackage{graphicx}
\usepackage{esint}
\usepackage[unicode=true,pdfusetitle,
 bookmarks=true,bookmarksnumbered=false,bookmarksopen=false,
 breaklinks=false,pdfborder={0 0 1},backref=false,colorlinks=true,citecolor={blue}]
 {hyperref}

\makeatletter

\providecommand{\tabularnewline}{\\}

\usepackage{graphicx}
\usepackage[section]{placeins}%%put table in its section
\usepackage{tabularx}
\usepackage{amsmath}
\usepackage{amstext}
\usepackage{amssymb}
\usepackage{xfrac}
\usepackage{graphicx}
\usepackage{amsmath}
\usepackage{amstext}
\usepackage{amssymb}
\usepackage{amsfonts}
\usepackage{longtable,booktabs}
\usepackage{url}
\usepackage{subfigure}
\usepackage{dsfont}
\usepackage{booktabs}
\usepackage{amsbsy}
\usepackage{dcolumn}
\usepackage{amsthm}
\usepackage{bm}
\usepackage{esint}
\usepackage{multirow}
\usepackage{cleveref}
\usepackage{mathrsfs}
\usepackage{amsfonts}
\usepackage{amsbsy}
\usepackage{dcolumn}
\usepackage{bm}
\usepackage{multirow}
\usepackage{color}
\usepackage{extarrows}
\usepackage{datetime}
\usepackage{comment}
\usepackage[super]{nth}

\newcommand{\di}{\mathrm{d}}
\newcommand{\ii}{\mathrm{i}}

\newcommand{\comments}[1]{}

\def\Z{\mathbb{Z}}

\makeatother

\begin{document}
\title{Non-Abelian Fusion, Shrinking and Quantum Dimensions  of  Abelian Gauge Fluxes}
\author{Zhi-Feng Zhang}
\affiliation{School of Physics, State Key Laboratory of Optoelectronic Materials and Technologies, and Guangdong Provincial Key Laboratory of Magnetoelectric Physics and Devices, Sun Yat-sen University, Guangzhou, 510275, China}
\author{Qing-Rui Wang}
\email{wangqr@mail.tsinghua.edu.cn}
\affiliation{Yau Mathematical Sciences Center, Tsinghua University, Haidian, Beijing, China}
\author{Peng Ye}
\email{yepeng5@mail.sysu.edu.cn}
\affiliation{School of Physics, State Key Laboratory of Optoelectronic Materials and Technologies, and Guangdong Provincial Key Laboratory of Magnetoelectric Physics and Devices, Sun Yat-sen University, Guangzhou, 510275, China}
\date{{\color{blue}\textbf{\today}}\\\textbf{{\small \color{blue}   version/20221031~~~~RevTex}}}
\begin{abstract}
Braiding and fusion rules of topological excitations are indispensable  topological invariants  in    topological quantum computation  and   topological orders. While   excitations in 2D   are always particle-like anyons, those in 3D incorporate not only particles but also loops---spatially nonlocal objects---making it novel and challenging to study topological invariants in higher dimensions.   While 2D fusion rules  have been well understood from bulk Chern-Simons  field theory and edge conformal field theory, it is yet to be thoroughly  explored for 3D fusion rules  from higher dimensional bulk topological  field theory. Here, we perform a field-theoretical study on  {(i)} how loops that carry Abelian gauge fluxes fuse and  {(ii)} how loops are shrunk into particles in the path integral, which generates fusion rules, loop-shrinking rules, and   descendent invariants, e.g., quantum dimensions.  We first assign  a  gauge-invariant Wilson operator to each excitation and determine the   number of distinct   excitations  through equivalence classes of Wilson operators. Then, we adiabatically shift two  Wilson operators together   to observe how they fuse and are split in the path integral; despite the Abelian nature of the gauge fluxes   carried by loops, their fusions may be   of non-Abelian nature.  Meanwhile,   we  adiabatically deform  world-sheets of unknotted  loops   into world-lines  and examine the shrinking outcomes;  we find that the resulting loop-shrinking rules   are algebraically consistent to fusion rules.  Interestingly,  fusing a pair of loop and   anti-loop may generate multiple vacua, but    fusing a pair of anyon and anti-anyon   in 2D has one vacuum only.   By   establishing a    field-theoretical ground for fusion and shrinking  in 3D, this work leaves intriguing  directions for future exploration, e.g., symmetry enrichment,     quantum gates, and      physics of  braided monoidal 2-category of 2-group.
\end{abstract}

\maketitle
\tableofcontents{}

\clearpage
\section{Introduction\label{sec:Introduction}}

{\textbf{Topological order and topological excitations}.---}Topologically ordered phases which are beyond the paradigm of symmetry-breaking
theory have attracted lots of attentions for years~\citep{Wen2019,wen_stacking,RevModPhys.77.871,wenZootopoRMP,hartnoll2021quantum,2015arXiv150802595Z} from not only condensed matter physics,
but also high-energy physics, quantum information science, and mathematical
physics.
Experimentally confirmed by the fractional quantum Hall effect (FQHE),
topological order cannot be characterized by any local order parameters.
 At   low energies,
  topological quantum field theory (TQFT) is utilized as the effective
field theory to describe topologically ordered phases.
In addition, inspired by quantum information science, topological
order of a gapped many-body system is tightly connected to 
the pattern of long-range entanglement~\cite{2015arXiv150802595Z}. Recently, the algebraic theory for 2D topological
orders has also been explored deeply, making it intriguing to make joint efforts in condensed matter physics and mathematical category theory, see, e.g., concise introductory materials in Refs.~\citep{Barkeshli2014symmetry}. 
%To understand
%topological order or more generally quantum matter is a pursuit of not only condensed matter physics,
%but also high-energy physics, quantum information science, and mathematical
%physics~\citep{hartnoll2021quantum,2015arXiv150802595Z}.
%~\citep{hartnoll2021quantum,KitaevTopoEE,Levin_Wen_TEE,kong2020mathematical,kong2021mathematical,lantian3dto1,
%delcamp2018gauge,delcamp20192}.

  Topological excitations are essential ingredients of topologically
ordered phases. In absence of any symmetry-breaking order parameters, the topological properties, such as fusion and braiding statistics of topological excitations constitute the key observables of topological orders and also important processes in   Topological Quantum Computation (TQC)~\cite{sarma_08_TQC}. Analogous to quasi-particles in
solid-state physics, topological excitations are collective phenomena and can be created as   localized energy   lump above the ground state. 
In $2$D space, topological excitations are point-like particle excitations,
e.g., the anyon excitations in FQHE. In $3$D space,   topological excitations incorporate not only particle excitations but also   loop
excitations that  are spatially nonlocal. Moreover, a loop excitation can also be decorated
by a particle excitation, i.e., a particle excitation is attached to
a loop excitation, named \emph{decorated  loop} (see Fig.~\ref{fig:world-line_world_sheet}). For those loop excitations  {not} decorated by particle excitations, we call them \emph{pure loops}.\footnote{For simplicity, we use particle and loop to refer corresponding topological
excitations in the following main text when there is no ambiguity.}
 If we move forward to $4$D space, we would find that   topological
excitations there include $2$-dimensional closed-surface-like membrane
excitations~\citep{zhang2022topological,chen2021loops}.

{\textbf{Braiding statistics}.---}Let us first review braiding statistics of topological excitations. During a braiding process of particles and loops, an adiabatic
quantum phase is accumulated which is proportional to the linking
invariant of the link formed by world-lines of particles and world-sheets
of loops. For the braiding processes in $4$D space, the emergence of world-volumes
of membrane excitations generates a large variety of    exotic linking invariants~\citep{zhang2022topological}. These
adiabatic quantum phases are called braiding phases, serving as an
important data set to characterize topological order. TQFT, as the
low-energy effective theory of topological order, provides us a quantitative
approach to braiding phase~\citep{Witten1989}. For example, the braiding
phases of anyons in $2$D space are captured by the $\left(2+1\right)$D
Chern-Simons theory ($\sim tr (A\wedge dA+\frac{2}{3}A\wedge A\wedge A)$)~\citep{Witten1989,wen2004quantum,wen_stacking,blok1990effective}.
In $3$D space, braiding processes involve particles and loops. If
we consider a discrete gauge group $G=\prod_{i}\mathbb{Z}_{N_{i}}$,
all particles and loops can be labeled by periodic gauge charges and gauge 
fluxes respectively. The braiding processes can be divided into three
classes: particle-loop braiding~\citep{hansson2004superconductors,abeffect,PRESKILL199050,PhysRevLett.62.1071,PhysRevLett.62.1221,ALFORD1992251},
multi-loop braiding~\citep{wang_levin1,PhysRevLett.114.031601,2016arXiv161209298P,YeGu2015,ye16_set,string4,2016arXiv161008645Y,ning2022fractionalizing,PhysRevB.99.235137,jian_qi_14,string5,PhysRevX.6.021015,string6,corbodism3,3loop_ryu,string10,Tiwari:2016aa,peng2020gauge,PhysRevB.99.205120},
and particle-loop-loop braiding {[}i.e., Borromean rings (BR) braiding{]}
\citep{yp18prl}. In each class, there are different braiding phases
depending on different assignments of gauge group. The TQFTs describing
these braiding processes are expressed as the combination of a multi-component
$BF$ term~\citep{Horowitz1990,Baez:2006un,bullivant2019representations,ye16a,Ye:2017aa,Cho20111515} with twisted terms. The $BF$ term in $\left(3+1\right)$D
is written as $B\wedge dA$ where $B$ and $A$ are $2$- and $1$-form
$\mathbb{U}\left(1\right)$ gauge fields respectively. For multi-loop
braiding, the twisted terms $AAdA$ and $AAAA$ ($\wedge$ is omitted) ~\citep{2016arXiv161209298P}
correspond to $3$-loop and $4$-loop braidings, respectively. For particle-loop-loop
braiding (BR braiding, see Fig.~\ref{fig:Borromean_rings}), the twisted term is $AAB$~\citep{yp18prl}.
If we consider a topologically ordered system that supports particle-loop
and/or multi-loop braiding, the TQFT is consistent with the Dijkgraaf-Witten
(DW) cohomological classification $\mathcal{H}^{4}\left(G,U\left(1\right)\right)$ for gauge group $G$.
Nevertheless, once we demand the system to support BR braiding as
well, some multi-loop braidings would be excluded in the sense that
no legitimate DW TQFT describing all these braidings can be constructed.
Such incompatibility between BR braiding and multi-loop braiding can
be traced back to the requirement of gauge invariance for TQFT~\citep{zhang2021compatible}.
For the purpose of this paper, we denote a system is equipped with \emph{Borromean rings topological
order} (BR topological order) if it supports BR braiding.

{\textbf{Fusion rules and loop-shrinking rules}.---}Fusion rules of topological excitations form another important
set of topological invariants for 3D topological order. Pictorially, the fusion of two
topological excitations is to adiabatically bring them together in space and the combined object behaves like another topological
excitation. To be more precise, each topological excitation $\mathsf{e}_{i}$
corresponds to a fusion space $\mathcal{V}\left(M_{d};\mathsf{e}_{i}\right)$
where $M_{d}$ is the spatial manifold supporting all topological
excitations~\citep{wen_stacking}. The bases of $\mathcal{V}\left(M^{d};\mathsf{e}_{i}\right)$
are degenerate ground states of $H+\delta H_{i}$ with $H_{i}$ is
non-zero only near the location of $\mathsf{e}_{i}$. If the dimension
of $\mathcal{V}\left(M_{d};\mathsf{e}_{i}\right)$ cannot be altered
by any local perturbation near the location of $\mathsf{e}_{i}$,
the type of $\mathsf{e}_{i}$ is \emph{simple}. Otherwise, the type
of $\mathsf{e}_{i}$ is \emph{composite}. The fusion space of a composite
topological excitation can be decomposed as a direct sum of that of
other simple topological excitations. The fusion of two simple topological
excitations, e.g., $\mathsf{a}$ and $\mathsf{b}$, corresponds to the direct
product of their fusion space: $\mathcal{V}\left(M_{d};\mathsf{a}\right)\otimes\mathcal{V}\left(M_{d};\mathsf{b}\right)$.
The resulting fusion space may correspond to another simple excitation,
e.g., $\mathsf{c}$, and this fusion is called \emph{Abelian fusion}: $\mathcal{V}\left(M_{d};\mathsf{a}\right)\otimes\mathcal{V}\left(M_{d};\mathsf{b}\right)=\mathcal{V}\left(M_{d};\mathsf{c}\right)$.
It may also correspond to a direct sum of fusion spaces of \emph{multiple} simple
excitations, e.g., $\mathcal{V}\left(M_{d};\mathsf{a}\right)\otimes\mathcal{V}\left(M_{d};\mathsf{b}\right)=\mathcal{V}\left(M_{d};\mathsf{d}\right)\oplus\mathcal{V}\left(M_{d};\mathsf{f}\right)$,
and such fusion is called \emph{non-Abelian fusion}. The fusion rules are
just simplified notations for the previous algebraic relations: $\mathsf{a}\otimes\mathsf{b}=\mathsf{c}$
or $\mathsf{a}\otimes\mathsf{b}=\mathsf{d}\oplus\mathsf{f}$. In a
more general setting, the fusion rule of two simple topological excitations
can be given by $\mathsf{a}\otimes\mathsf{b}=\oplus_{i}N_{\mathsf{e}_{i}}^{\mathsf{a}\mathsf{b}}\mathsf{e}_{i}$
where $N_{\mathsf{e}_{i}}^{\mathsf{a}\mathsf{b}}$ is a non-negative
integer and the type of $\mathsf{e}_{i}$ is simple. In this paper, unless otherwise specified, ``excitations'' are always of simple type.

In the literature, fusion in $2$D topological orders   has been studied extensively from exactly solvable models, field theory, to mathematical foundation. For example, fusion rules of anyons are encoded in the mathematical concept of unitary fusion tensor categories \citep{turaev2016quantum,bakalov2001lectures}. On the other hand, just like the role of loops in exotic   braiding statistics reviewed above,   loop excitations that are entirely absent in $2$D topological orders, are also expected to contribute nontrivial fusion rules in 3D topological orders.  The nature of non-locality of loop excitations may significantly complicate but meanwhile significantly enrich the analysis of fusion rules (see, e.g., the cartoons in Figs.~\ref{fig:fusion_illustrate}, \ref{fig:4_fusion_examples}). \emph{Firstly}, combinatorially, we   need to analyze  fusions of (i) two particles, (ii) two loops, and (iii) one particle plus one loop. \emph{Secondly}, as loops can be either pure loops or decorated loops as reviewed above, the resulting fusion data are expected to be further enriched. \emph{Thirdly}, one can also consider self-knotted or mutually linked loops (see, e.g., Fig.~1 of Ref.~\cite{ypdw1}) and study their fusion rules. \emph{Fourthly}, while there have been  intensive discussions in realization and manipulation of  Majorana zero modes (see, e.g., incomplete reference list: Refs.~\cite{PhysRevLett.86.268,fusiontest,PhysRevB.73.014505}), it will be of great interests to explore how to implement fusion rules of loop excitations (``loop-like errors/defects'' by following nomenclature in quantum information science) in TQC gates of stabilizer codes.   All in all,  fusion rules for loops  in 3D topological orders deserve  a thorough  study from various aspects.

 Besides, the presence of loops provides us with another indispensable topological invariants---\emph{loop-shrinking rules}. A loop excitation, if it is unknotted, can be smoothly shrunk to a point (see Fig.~\ref{fig:loop_shrinking_pic}). Notice that this process  is apparently meaningless for   particle excitations that are already point-like, so the exploration of such topological invariants should be at least starting from 3D topological orders. Interestingly, such a shrinking process can be alternatively  understood as the consequence of observing  a loop when an observer stands far away from the loop such that   the loop   ``looks like''  a point. It is curious to ask what is the consequence of such a loop-shrinking operation? How can we analytically describe this process, e.g., by means of field theory? Can we obtain another set of meaningful topological invariants from such an operation?  To answer questions of such kinds, it is highly worthwhile to conduct an in-depth study into the outcomes of such a  {loop-shrinking operation}, which are encoded in the  {loop-shrinking rules} that are lacking in $2$D topological orders. 
     All in all, we expect that the presence of non-local loops in 3D topological orders will  lead to not only nontrivial braiding statistics as studied before but also fruitful quantum phenomena encoded in   fusion rules and   loop-shrinking rules. This line of efforts  will be of great help for deeply understanding topological orders of all dimensions, theoretically developing TQC for all dimensions, and  proposing experimental manipulation of braiding, fusion, and shrinking of non-local topological excitations in the future.

     From the tradition of condensed matter physics and also by following the spirit of renormalization group and universality, it is always vital to explore  the long-distance low-energy effective field theory of   underlying   phases of matter, and further ask how to systematically define and compute observables  from such effective field theories. It has been  known that Ginzburg-Landau perturbative field theories are used to describe symmetry-breaking phases, but for topological orders, topological field theories are the correct field-theoretical language.   While it has been well established  that  the topological invariants (e.g., fusion rules and braiding) of 2D topological orders can be systematically extracted from $(2+1)$D bulk Chern-Simons field theory as well as   edge CFT (conformal field theory), it is still yet to be   thoroughly explored for fusion rules and loop-shrinking rules of 3D topological orders from $(3+1)$D field theories that are generally TQFT of certain types. Thus, it is important  to perform such a topological-field-theoretical study.

Motivated by, but not limited to, above discussions,  in this paper, we aim to perform a topological-field-theoretical study on      fusion rules and loop-shrinking rules of
$3$D topological orders when all loops carry Abelian gauge fluxes.  Specially,
we start with the BR topological order with Abelian gauge group $G=\left(\mathbb{Z}_{N}\right)^{3}$. In details, we first construct and classify topologically distinct  Wilson operators for all types of particles
and loops by means of path-integral quantization. Then, the number of topological excitations is just the number of Wilson operators, collected in Tables \ref{tab:AAB_particle_nonequivalent}, \ref{tab:AAB_pure_loop}, and \ref{tab:AAB_decorated_loop}.    
Next, we study   fusion rules in terms of path integral.  In practice, we spatially fuse two Wilson operators together, which leads to nontrivial splitting in the path integral formalism. We collect all fusion coefficients in Table~\ref{tab:fusion_AAB_Z2Z2Z2_full}, where there exist   non-Abelian fusion processes despite the Abelian nature of the gauge fluxes carried by loops. From the fusion coefficients, we can also extract \emph{quantum dimensions} for all excitations, as collected in Table~\ref{tab:qu_dim_AAB_Z2Z2Z2}.  Then, we   compute shrinking rules for loop excitations (see Table~\ref{tab:shrinking rules_AAB_z2z2z2}), i.e., the process of shrinking an unknotted loop excitation into   particles,  which are found to be algebraically consistent with the fusion rules, and are critical in establishing an anomaly-free topological order.      We also find an interesting phenomenon that  fusing a loop and   anti-loop may generate more than one vacuum, which is different from 2D topological orders where fusing a pair of particle and antiparticle   has one vacuum only. At last,  we  generalize the above analysis to    topological  orders with gauge group $G=\prod_{i}^{n}\Z_{N_i}$ ($n=1,2,3$) where various  interesting braiding statistics  can be realized.     This work   establishes a  continuum field-theoretical ground for fusion, shrinking and quantum dimensions in 3D TO, and also  future explorations.

{\textbf{Outline}.---}This paper is organized as follows. In Sec. \ref{sec:operator}, we
review the TQFT action of BR topological order and
construct Wilson operators for topological excitations. The number
of topological excitations is consistent with that computed from a
lattice cocycle model. In Sec.~\ref{sec:fusion rule}, fusion rules
of excitations are derived via Wilson operators and path integral.
Besides, the shrinking rules for loops are also studied, which shows consistency with the fusion data. In Sec. \ref{sec:compatible_fusion},
the relation between fusion rules and combinations of compatible braiding
processes is studied, which generalizes the above analysis to  topological  orders with gauge group $G=\prod_{i}^{n}\Z_{N_i}$ ($n=1,2,3$). Discussion and outlook  are given in Sec. \ref{sec:discussion_outlook}. Technical
details are collected in the Appendices.

\section{Wilson operators for topological excitations \label{sec:operator}}

In order to study fusion rules in TQFT, we first need to express
topological excitations in the field-theoretical formalism. For each
topological excitation carrying a specific amount of gauge charges and
gauge fluxes, it is uniquely represented by a Wilson operator that
is invariant under gauge transformations. In this section, we begin
with reviewing the TQFT action for BR topological order
with gauge group $G=\prod_{i=1}^{3}\mathbb{Z}_{N_{i}}$. Then, by
considering $N_{1}=N_{2}=N_{3}=2$ as  a typical  example,
we construct Wilson operators for topological
excitations, i.e., particles, pure loops, and decorated loops. In
this case, there are $2^{3}\times2^{3}=64$ different combinations
of gauge charges and gauge fluxes, which seems to indicate that there are $64$ different
topological excitations, i.e., Wilson operators. Nevertheless, we find
that some   Wilson operators have the same correlation function
with an arbitrary operator. In this sense,  such Wilson operators
belong to the same \emph{equivalence class}. Finally, among $64$
possible Wilson operators we find only $19$ nonequivalent ones, i.e.,
$19$ essentially different topological excitations, for BR topological order with $G=\left(\mathbb{Z}_{2}\right)^{3}$. These
$19$ nonequivalent Wilson operators corresponding to $19$ topological
excitations are listed in Table.~\ref{tab:AAB_particle_nonequivalent} (particles), Table.~\ref{tab:AAB_pure_loop} (pure loops) , and Table.~\ref{tab:AAB_decorated_loop} (decorated loops) which are the cornerstone  of fusion rules in Sec.~\ref{sec:fusion rule}.

\subsection{TQFT action for BR topological order}

\begin{figure}
\includegraphics[scale=0.2]{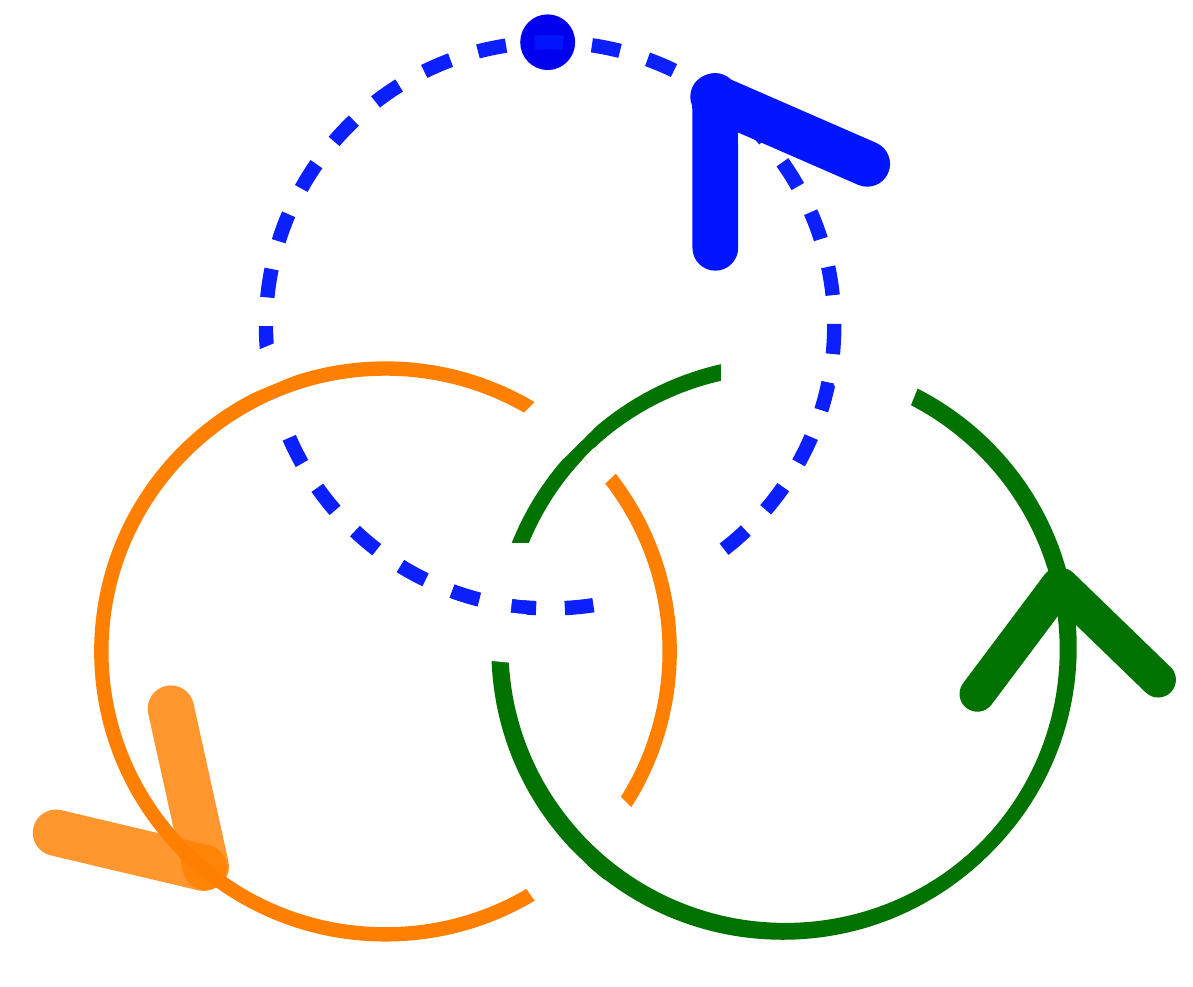}

\caption{Illustration of Borromean rings braiding that is realized in the topological order described by Eq.~(\ref{eq:action_AAB}). In this braiding process, Borromean rings is formed by the spatial trajectory of the particle (blue)
and two loops (orange and green). \label{fig:Borromean_rings}}
\end{figure}

BR topological order~\citep{yp18prl} is featured with
a special braiding process of one particle and two loops which carry
gauge charge or fluxes from three different gauge subgroups. In this braiding
process, the spatial trajectory of particle and two loops form Borromean
rings (or general Brunnian link) in $3$D space, as shown in Fig.~\ref{fig:Borromean_rings}. The braiding phase of this process is proportional to the Milnor's triple linking number \citep{yp18prl,milnor1954link,mellor2003geometric}.
This Borromean rings braiding cannot be classified by   cohomology group
$\mathcal{H}^{4}\left(G,U\left(1\right)\right)$ for gauge group $G$. The latter is   applicable
only for particle-loop braidings and multi-loop braidings. In addition, a Borromean rings braiding is compatible with specific multi-loop braidings only for a given gauge group $G$. In other words, a legitimate DW TQFT can only describe Borromean rings braiding and some, but not all, of multi-loop braidings simultaneously. By legitimacy we mean that the DW TQFT is a theory with well-defined gauge transformations~\citep{zhang2021compatible}.

In the following,
we consider gauge group $G=\prod_{i=1}^{3}\mathbb{Z}_{N_{i}}$.   The action for BR topological
order is
\begin{equation}
S=\int\sum_{i=1}^{3}\frac{N_{i}}{2\pi}B^{i}dA^{i}+qA^{1}A^{2}B^{3},\label{eq:action_AAB}
\end{equation}
where $A^{i}$ and $B^{i}$are $1$- and $2$-form $\mathbb{U}\left(1\right)$
gauge fields respectively. The coefficient $q=\frac{pN_{1}N_{2}N_{3}}{\left(2\pi\right)^{2}N_{123}}$
with $p\in\mathbb{Z}_{N_{123}}$, where $N_{123}$ is the greatest
common divisor (GCD) of $N_{1}$, $N_{2}$ and $N_{3}$. The quantization
of $q$ is the result of the large gauge invariance. In action (\ref{eq:action_AAB}), $B^{1}$, $B^{2}$, and $A^{3}$ serve as the Lagrange multipliers which locally enforce the flat-connection conditions: $dA^{1}=0$, $dA^{2}=0$, and $dB^{3}=0$. The gauge transformations
for the action (\ref{eq:action_AAB}) are given by
\begin{align}
A^{1}\rightarrow & A^{1}+d\chi^{1},\\
A^{2}\rightarrow & A^{2}+d\chi^{2}, \\
A^{3}\rightarrow & A^{3}+d\chi^{3}+X^{3},\\
B^{1}\rightarrow & B^{1}+dV^{1}+Y^{1}, \\
B^{2}\rightarrow & B^{2}+dV^{2}+Y^{2}, \\
B^{3}\rightarrow & B^{3}+dV^{3},\label{eq:GT_AAB}
\end{align}
with nontrivial shifts
\begin{align}
X^{3}= & -\frac{2\pi q}{N_{3}}\left(\chi^{1}A^{2}+\frac{1}{2}\chi^{1}d\chi^{2}\right)\nonumber \\
 & +\frac{2\pi q}{N_{3}}\left(\chi^{2}A^{1}+\frac{1}{2}\chi^{2}d\chi^{1}\right),\\
Y^{1}= & -\frac{2\pi q}{N_{1}}\left(\chi^{2}B^{3}-A^{2}V^{3}+\chi^{2}dV^{3}\right),\\
Y^{2}= & \frac{2\pi q}{N_{2}}\left(\chi^{1}B^{3}-A^{1}V^{3}+\chi^{1}dV^{3}\right),
\end{align}
where $\chi^{i}$ and $V^{i}$ are respectively $0$-form and $1$-form gauge parameters with $\int d\chi^{i}\in2\pi\mathbb{Z}$
and $\int dV^{i}\in2\pi\mathbb{Z}$.

\subsection{$G=\left(\mathbb{Z}_{2}\right)^{3}$: Operators for topological excitations
and their equivalence classes\label{subsec:Operators_equivalence class}}

\begin{figure}
\includegraphics[scale=0.4]{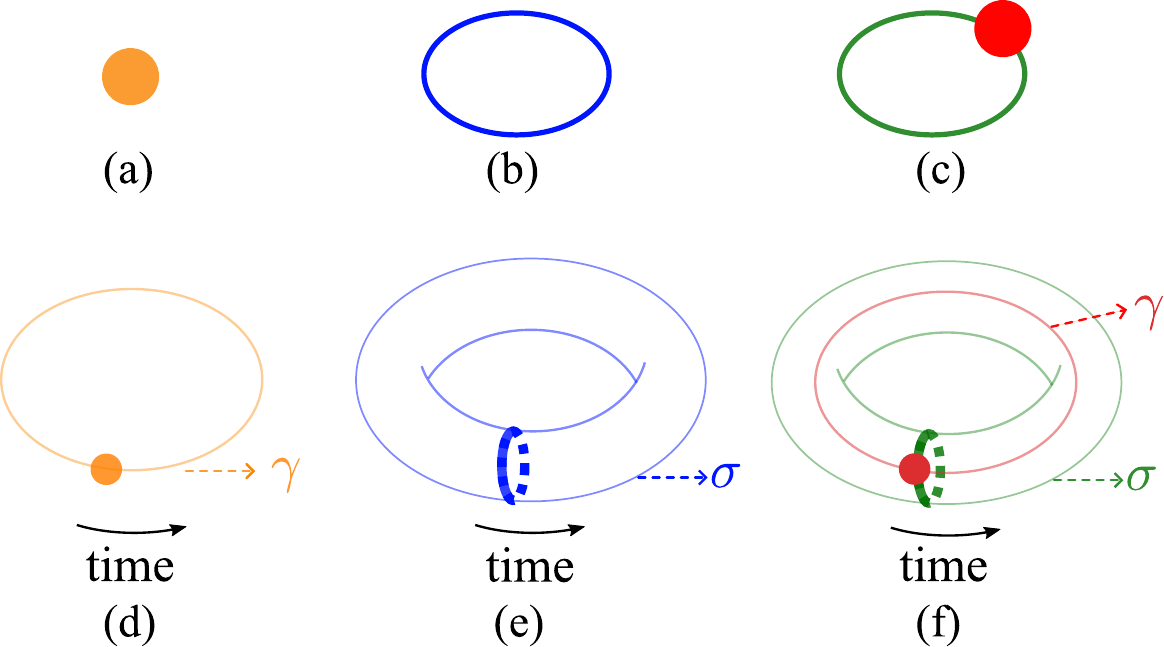}

\caption{(a) Particle excitation carrying gauge charge in $3$D space. (b)
Pure loop excitation carrying gauge flux in $3$D space. In this paper,
we only consider loop excitations which are unknotted and thus can be  deformed to $S^{1}$
smoothly. (c) Decorated loop carrying gauge flux and charge simultaneously. A decorated loop can be viewed as a pure loop with a particle attached to it. This particle (red solid circle) can be located at any position on the loop. (d) The closed world-line of particle denoted by $\gamma$. (e) The closed world-sheet of pure loop denoted by $\sigma$. (f) The space-time trajectories of decorated loop.
In $\left(3+1\right)$D spacetime, $\sigma$ is a torus and $\gamma$ is a closed curve that is not self-knotted. For the particle attached to a loop excitation, its world-line $\gamma$ would be the
non-contractible path on $\sigma$ circling along the time direction.
\label{fig:world-line_world_sheet}}

\end{figure}

Since our TQFT action (\ref{eq:action_AAB}) is a gauge theory, it is expected
that the operators for topological excitations are gauge-invariant.
Notice that the gauge group is $G=\prod_{i=1}^{3}\mathbb{Z}_{N_{i}}$,
the topological excitations include particles carrying $\mathbb{Z}_{N_{i}}$ gauge
charges, loops carrying $\mathbb{Z}_{N_{i}}$ gauge flux only (\emph{pure}
loop), and loops simultaneously carrying $\mathbb{Z}_{N_{i}}$ gauge flux and $\mathbb{Z}_{N_{j}}$ gauge charge
(\emph{decorated} loops; $i$ and $j$ can be same or different),
as illustrated in Fig.~\ref{fig:world-line_world_sheet}. The $\Z_{N_i}$ gauge charges and $\Z_{N_i}$ gauge fluxes are group representations and conjugacy classes of $\Z_{N_i}$ gauge subgroup. Only \emph{simple} (see Introduction) topological excitations are considered in this paper. In this section, we explain how to label topological excitations by Wilson operators. Furthermore, we show that some topological excitations are equivalent   in the path integral formalism, which  leads to the notion of equivalence class among Wilson operators.

First, if we consider a particle with one unit of $\mathbb{Z}_{N_{1}}$
gauge charge (a $\Z_{N_1}$ particle), we can use the following operator to represent it:
\begin{equation}
\mathsf{P}_{100}=\exp\left({\rm i}\int_{\gamma}A^{1}\right),
\end{equation}
where the closed $1$-dimensional $\gamma$ can be understood as the
closed world-line of particle in $\left(3+1\right)$D spacetime. $\gamma$ is a closed curve in $\left(3+1\right)$D spacetime and can be deformed to $S^{1}$ smoothly, as shown in Fig.~\ref{fig:world-line_world_sheet}. The
capital letter $\mathsf{P}$ stands for particle excitation and the
subscript $n_{1}n_{2}n_{3}$ of $\mathsf{P}_{n_{1}n_{2}n_{3}}$ denotes the number of $\mathbb{Z}_{N_{1}}$,
$\mathbb{Z}_{N_{2}}$, and $\mathbb{Z}_{N_{3}}$ gauge charges respectively.
For instance, the subscript of $\mathsf{P}_{100}$ denotes that this
particle excitation carries one unit of $\mathbb{Z}_{N_{1}}$ gauge
charge and vanishing $\mathbb{Z}_{N_{2}}$ or $\mathbb{Z}_{N_{3}}$ gauge
charge. The anti-particle of $\mathsf{P}_{100}$ is represented by
\begin{equation}
\bar{\mathsf{P}}_{100}=\mathsf{P}_{\left(-1\right)00}=\exp\left(-{\rm i}\int_{\gamma}A^{1}\right).
\end{equation}
For simplicity, we let the gauge group to be $G=\left(\mathbb{Z}_{2}\right)^{3}$,
i.e., $N_{1}=N_{2}=N_{3}=2$. We consider
\begin{align}
\left\langle \mathsf{P}_{\left(-1\right)00}\right\rangle = & \frac{1}{\mathcal{Z}}\int\mathcal{D}\left[A^{i}\right]\mathcal{D}\left[B^{i}\right]\exp\left({\rm i}S\right)\nonumber \\
 & \times\exp\left(-{\rm i}\int_{\gamma}A^{1}\right),
\end{align}
where $\mathcal{Z}=\int\mathcal{D}\left[A^{i}\right]\mathcal{D}\left[B^{i}\right]\exp\left({\rm i}S\right)$ is the partition function.
Integrating out $B^{1}$ leads to the constraint
\begin{equation}
\oint A^{1}=\frac{2\pi m_{1}}{N_{1}}=\frac{2\pi m_{1}}{2},m_{1}\in\mathbb{Z}.
\end{equation}
This constraint implies $\exp\left({\rm i}2\int_{\gamma}A^{1}\right)=1$.
With this fact, the expectation value of $\mathsf{P}_{\left(-1\right)00}$
can be written as
\begin{align}
\left\langle \mathsf{P}_{\left(-1\right)00}\right\rangle = & \frac{1}{\mathcal{Z}}\int\mathcal{D}\left[A^{i}\right]\mathcal{D}\left[B^{i}\right]\exp\left({\rm i}S\right)\nonumber \\
 & \times\exp\left(-{\rm i}\int_{\gamma}A^{1}\right)\times\exp\left({\rm i}2\int_{\gamma}A^{1}\right)\nonumber \\
= & \left\langle \mathsf{P}_{100}\right\rangle .
\end{align}
In the sense of path integral, we can see that the anti-particle
of $\mathsf{P}_{100}$ is itself when $G=\left(\mathbb{Z}_{2}\right)^{3}$. This result is easy to understand since the
particle carries gauge charge of cyclic $\mathbb{Z}_{2}$ group.

Next, we consider a \emph{pure} loop carrying one unit of $\mathbb{Z}_{N_{3}}$
flux, denoted as $\Z_{N_3}$-loop for simplicity. The corresponding operator is
\begin{equation}
\mathsf{L}_{001}=\exp\left({\rm i}\int_{\sigma}B^{3}\right),
\end{equation}
where $\sigma$ is a closed $2$-dimensional surface as the
closed world-sheet of a loop. In details, $\sigma$ is a $2$-torus formed by circling the loop along the time direction, as shown in Fig.~\ref{fig:world-line_world_sheet}.
The letter $\mathsf{L}$ stands for loop excitations. For pure loop
excitations, the subscript denotes the gauge fluxes carried by the
loop. Similarly, for a pure loop carrying one (mod $2$) unit of $\mathbb{Z}_{2}$
flux, its anti-loop is itself, e.g., $\bar{l}_{001}=l_{00\left(-1\right)}=\exp\left({\rm i}\int_{\sigma}B^{3}\right)$.

Last, we consider a \emph{decorated} loop [see Fig.~\ref{fig:world-line_world_sheet}(c)]. For instance, a $\mathbb{Z}_{N_{3}}$-loop decorated by a $\mathbb{Z}_{N_{1}}$-particle
is represented by
\begin{equation}
\mathsf{L}_{001}^{100}=\exp\left({\rm i}\int_{\sigma}B^{3}+{\rm i}\int_{\gamma}A^{1}\right).
\end{equation}
For decorated loop excitations, the superscript (e.g., ``$100$'' in $\mathsf{L}_{001}^{100}$) denotes the charge decoration, i.e., the gauge
charges carried by the particle attached to the loop. Such decoration of particle
on a loop requires that the particle's world-line $\gamma$ lies on
the loop's world-sheet $\sigma$. This requirement is reasonable:
imagine a loop moving in $\left(3+1\right)$D spacetime, the decorated
particle also moves together with the loop, thus its world-line becomes a non-contractible path on the world-sheet of loop, as illustrated in Fig.~\ref{fig:world-line_world_sheet}.

One may notice that in gauge transformations (\ref{eq:GT_AAB}), some
gauge fields transform by a shift term, i.e., $X^{3}$, $Y^{1}$,
or $Y^{2}$. These shift terms indicate that the gauge-invariant operators
of these gauge fields needs to be treated carefully. For example,
we consider the operator for $B^{1}$ gauge field which corresponds to a pure loop
carrying $\mathbb{Z}_{N_{1}}$flux:
\begin{align}
 &\mathsf{L}_{100} \nonumber\\
= & 2\exp\left[{\rm i}\int_{\sigma}B^{1}+\frac{1}{2}\frac{2\pi q}{N_{1}}\left(d^{-1}A^{2}B^{3}+d^{-1}B^{3}A^{2}\right)\right]\nonumber\\
 & \times\delta\left(\int_{c}A^{2}\right)\delta\left(\int_{\sigma}B^{3}\right)
\end{align}
with $d^{-1}A^{2}=\int_{\left[a,b\right]\in c}A^{2}$ and $d^{-1}B^{3}=\int_{\mathcal{A}\in\sigma}B^{3}$
where $\left[a,b\right]$ is an open interval on a closed curve $c$
and $\mathcal{A}$ is an open area on $\sigma$. The normalization factor $2$ in the front of $\mathsf{L}_{100}$ is explained in Appendix~\ref{appendix:derivation_normalization_factor}. These two Kronecker
delta functions are\begin{equation}
\delta\left(\int_{c}A^{2}\right)=\begin{cases}
1, & \int_{c}A^{2}=0\mod2\pi\\
0, & \text{else }
\end{cases}
\end{equation}
 and
\begin{equation}
\delta\left(\int_{\sigma}B^{3}\right)=\begin{cases}
1, & \int_{\sigma}B^{3}=0\mod2\pi\\
0, & \text{else }
\end{cases}.
\end{equation}
These constraints ensure that $d^{-1}A^{2}$ and $d^{-1}B^{3}$ are
well-defined: for this purpose, we need $\int_{\forall c\in\sigma}A^{2}=0\mod2\pi$
and $\int_{\sigma}B^{3}=0\mod2\pi$.\footnote{In order to properly define $d^{-1}A^2$, we required $A^2$ to be exact on $\sigma$. This is equivalent to that the integral of $A^2$ over any $1$-dimensional closed submanifold is zero. Therefore, $\int_{\forall c\in\sigma}A^{2}=0\mod2\pi$ is imposed. For $d^{-1}B^3$, the argument is similar.} Since we have $\gamma\in\sigma$ (see Fig.~\ref{fig:world-line_world_sheet}),
we can choose $c=\gamma$ such that the constraint becomes $\int_{\gamma}A^{2}=0\mod2\pi$
and the expression of $\mathsf{L}_{100}$ becomes
\begin{align}
 & \mathsf{L}_{100}\nonumber \\
= & 2\exp\left[{\rm i}\int_{\sigma}B^{1}+\frac{1}{2}\frac{2\pi q}{N_{1}}\left(d^{-1}A^{2}B^{3}+d^{-1}B^{3}A^{2}\right)\right]\nonumber \\
 & \times\delta\left(\int_{\gamma}A^{2}\right)\delta\left(\int_{\sigma}B^{3}\right).
\end{align}
In fact, $\delta\left(\int_{\gamma}A^{2}\right)$ behaves as a projector
in path integral:
\begin{align}
\delta\left(\int_{\gamma}\widetilde{A}^{2}\right)= & \delta\left(\frac{2\pi m_{2}}{N_{2}}\right)=\frac{1}{2}\left[1+\exp\left(\frac{{\rm i}2\pi m_{2}}{N_{2}}\right)\right]
\end{align}
where $\widetilde{A}^{2}$ is the configuration of $A^{2}$ after
integrating out $B^{2}$ in path integral and satisfies the constraint
$\int_{\gamma}\widetilde{A}^{2}=\frac{2\pi m_{2}}{N_{2}}$ with $m_{2}\in\mathbb{Z}$.
For $\delta\left(\int_{\sigma}B^{3}\right)$, the discussion is similar. In other words,
these two Kronecker delta functions require that $\exp\left({\rm i}\int_{\gamma}\widetilde{A}^{2}\right)=1$
and $\exp\left({\rm i}\int_{\sigma}\widetilde{B}^{3}\right)=1$ otherwise the operator $\mathsf{L}_{100}$ is trivial.

These Kronecker delta functions are important when discussing Wilson
operators for topological excitations. They introduce an equivalence
relation between seemingly different operators. As an example, we
consider the $\mathbb{Z}_{N_{1}}$-loop decorated by a $\mathbb{Z}_{N_{2}}$-particle
and write down the operator for this decorated loop excitation:
\begin{align}
\mathsf{L}_{100}^{010}= & 2\exp\left[{\rm i}\int_{\gamma}A^{2}+{\rm i}\int_{\sigma}B^{1}\right.\nonumber \\
 & \left.+{\rm i}\int_{\sigma}\frac{1}{2}\frac{2\pi q}{N_{1}}\left(d^{-1}A^{2}B^{3}+d^{-1}B^{3}A^{2}\right)\right]\nonumber \\
 & \times\delta\left(\int_{\gamma}A^{2}\right)\delta\left(\int_{\sigma}B^{3}\right)
\end{align}
The correlation function of $\mathsf{L}_{100}^{010}$ and an arbitrary
operator $\mathcal{O}$ is given by
\begin{align}
\left\langle \mathcal{O}\mathsf{L}_{100}^{010}\right\rangle= & \frac{1}{\mathcal{Z}}\int\mathcal{D}\left[A^{i}\right]\mathcal{D}\left[B^{i}\right]\exp\left({\rm i}S\right)\times\mathcal{O}\times\mathsf{L}_{100}^{010}\nonumber \\
= & \widetilde{\mathcal{O}}\times2\exp\left[{\rm i}\int_{\gamma}\widetilde{A}^{2}+{\rm i}\int_{\sigma}\widetilde{B}^{1}\right.\nonumber \\
 & \left.+{\rm i}\int_{\sigma}\frac{1}{2}\frac{2\pi q}{N_{1}}\left(d^{-1}\widetilde{A}^{2}\widetilde{B}^{3}+d^{-1}\widetilde{B}^{3}\widetilde{A}^{2}\right)\right]\nonumber \\
 & \times\delta\left(\int_{\gamma}\widetilde{A}^{2}\right)\delta\left(\int_{\sigma}\widetilde{B}^{3}\right)\nonumber \\
= & \left\langle \mathcal{O}\mathsf{L}_{100}\right\rangle
\end{align}
where $\widetilde{A}^{i}$, $\widetilde{B}^{i}$, and $\widetilde{O}$
are obtain by integrating out corresponding Lagrange multipliers.
$\delta\left(\int_{\gamma}\widetilde{A}^{2}\right)=1$ guarantees that
$\exp\left({\rm i}\int_{\gamma}\widetilde{A}^{2}\right)=1$. We see that $\mathsf{L}_{100}^{010}$
and $\mathsf{L}_{100}$ behave as a same operator in path integral
and we regard that they belong to the same \emph{equivalence class}.
In fact, $\delta\left(\int_{\gamma}A^{2}\right)$ enforces the $\mathbb{Z}_{N_{2}}$-particle on loop $\mathsf{L}_{100}$ to behave as a trivial particle. Similarly,
we can prove that this equivalence class also includes the following two topological excitations: the pure loop carrying
$\mathbb{Z}_{N_{1}}$ and $\mathbb{Z}_{N_{3}}$ fluxes,
\begin{align}
\mathsf{L}_{101}= & 2\exp\left[{\rm i}\int_{\sigma}B^{3}+{\rm i}\int_{\sigma}B^{1}\right.\nonumber \\
 & \left.+{\rm i}\int_{\sigma}\frac{1}{2}\frac{2\pi q}{N_{1}}\left(d^{-1}A^{2}B^{3}+d^{-1}B^{3}A^{2}\right)\right]\nonumber \\
 & \times\delta\left(\int_{\gamma}A^{2}\right)\delta\left(\int_{\sigma}B^{3}\right)
\end{align}
and the $\left(\mathbb{Z}_{N_{1}},\mathbb{Z}_{N_{3}}\right)$-loop decorated
by a $\mathbb{Z}_{N_{2}}$-particle,
\begin{align}
\mathsf{L}_{101}^{010}= & 2\exp\left[{\rm i}\int_{\gamma}A^{2}+{\rm i}\int_{\sigma}B^{3}+{\rm i}\int_{\sigma}B^{1}\right.\nonumber \\
 & \left.+{\rm i}\int_{\sigma}\frac{1}{2}\frac{2\pi q}{N_{1}}\left(d^{-1}A^{2}B^{3}+d^{-1}B^{3}A^{2}\right)\right]\nonumber \\
 & \times\delta\left(\int_{\gamma}A^{2}\right)\delta\left(\int_{\sigma}B^{3}\right)
\end{align}
In conclusion, we have
\begin{equation}
\mathsf{L}_{100}=\mathsf{L}_{100}^{010}=\mathsf{L}_{101}=\mathsf{L}_{101}^{010}.
\end{equation}
Let us consider a general topological excitation $\mathsf{a}$. If
its operator is equipped with Kronecker delta function, it is
free to attach specific excitations (determined by Kronecker delta
functions) to $\mathsf{a}$ without altering the result of path integral
involving $\mathsf{a}$. Once an excitation is attached to $\mathsf{a}$ (this in fact is a fusion), by definition $\mathsf{a}$ becomes another excitation, say, labeled by $\mathsf{b}$.
In this manner, an equivalence relation may be established between $\mathsf{a}$
and $\mathsf{b}$. One should keep in mind that such equivalence relation is discussed in the sense of path integral.
	Respecting the principle of gauge-invariance and treating the Kronecker
delta functions carefully, we obtain $19$ nonequivalent operators for topological excitations of BR topological order with $G=\left(\Z_{2}\right)^3$. These operators are listed in Table~\ref{tab:AAB_particle_nonequivalent} (particles), Table~\ref{tab:AAB_pure_loop} (pure loops), and Table~\ref{tab:AAB_decorated_loop} (decorated loops).

Among these $19$ nonequivalent operators (i.e., $19$ distinct topological
excitations), there are $4$ nontrivial particle excitations, $4$
nontrivial pure loop excitations, and $10$ nontrivial decorated loop
excitations. By the definition of topological excitation, the trivial particle and the trivial loop are regarded the same, i.e., they both correspond to the vacuum denoted by $\mathsf{1}$. The first row in Table~\ref{tab:AAB_particle_nonequivalent} (trivial particle) and that of Table~\ref{tab:AAB_pure_loop} (trivial loop) are both represented by the trivial Wilson operator $\exp\left(\mathrm{i}0\right)=1$. Therefore, the number of particle excitations (including trivial and nontrivial ones) is $5$. So is that of pure loop excitations.

The total number of excitations obtained from the above field-theoretical analysis
 agrees  with the lattice cocycle method~\cite{PhysRevB.95.205142}. The details can be found in Appendix~\ref{appendix:cocycle_model} and here we briefly sketch the main idea.
  After
integrating out the Lagrange multipliers in action (\ref{eq:action_AAB}),
the remaining gauge fields $A^{1}$, $A^{2}$, and $B^{3}$ are discretized
into $\mathbb{Z}_{N_{i}}$. We are motivated to define the following
lattice model with $1$-form and $2$-form cocycles on arbitrary $\left(3+1\right)$D
spacetime manifold triangulation $M_{4}$:
 $\mathcal{Z}_{k}\left(M_{4}\right)=\sum_{a_{1},a_{2}  }\exp\left({\rm i}2\pi\frac{k}{N}\int_{M_{4}}a_{1}a_{2}b\right)
 $
where $a_{1},a_{2}\in Z^{1}\left(M_{4},\mathbb{Z}_{N}\right)\,,b\in Z^{2}\left(M_{4},\mathbb{Z}_{N}\right)$, and we have assumed $N_{i}=N$ ($i=1,2,3$) for simplicity. $Z^{1}\left(M_{4},\mathbb{Z}_{N}\right)$
and $Z^{2}\left(M_{4},\mathbb{Z}_{N}\right)$ are the sets of $1$-
and $2$-cocycles on $M_{4}$ respectively. The 1-cocycles $a_{1}$ and $a_{2}$
map each link $\left\langle ij\right\rangle \in M_{4}$ to $a_{ij}\in\mathbb{Z}_{N}$;
the 2-cocycle $b$ maps each triangle $\left\langle ijk\right\rangle \in M_{4}$
to $b_{ijk}\in\mathbb{Z}_{N}$. If we choose the spacetime manifold
to be $M_{4}=S^{1}\times M_{3}$ where $S^{1}$ is the time circle,
the topological partition function $\mathcal Z_k^\mathrm{top}[M_4]$, obtained by appropriate normalization of $\mathcal Z_k[M_4]$, is a trace of identity operator in the ground
state subspace. Therefore, it equals to the ground state degeneracy
(GSD) on the space manifold $M_{3}$:
 $\text{GSD}_{k}\left(M_{3}\right)=\mathcal{Z}_{k}^\mathrm{top}\left(S^{1}\times M_{3}\right).
 $
Furthermore, the GSD on space manifold $M_{3}=S^{1}\times S^{2}$
equals to the number of particle excitations, and the number of pure
loop excitations. For the example of $N=2$ and $k=1$ theory, we have $\text{GSD}_{k}\left(S^{1}\times S^{2}\right)=\mathcal{Z}_{k}\left(T^{2}\times S^{2}\right)=5$.
This is the exactly the number of nonequivalent particles and pure loop
excitations discussed above and summarized in Table~\ref{tab:AAB_particle_nonequivalent}
(particles) and Table~\ref{tab:AAB_pure_loop} (pure loops).  

\onecolumngrid
\begin{table*}
\caption{Operators for nonequivalent particle excitations in BR topological
order with $G=\left(\mathbb{Z}_{2}\right)^{3}$. The $\Z_{N_i}$ gauge charges are representations of elements in $\Z_{N_i}$ gauge subgroup. The subscript in $\mathsf{P}_{n_1 n_2 n_3}$ indicates that the particle carries $n_i$ units of $\Z_{N_i}$ gauge charges where $i=1,2,3$. In the first row, the particle excitation carrying vanishing gauge charge (trivial particle) is the vacuum, denoted as $\mathsf{1}$. The trivial particle is the same topological excitation as the trivial pure loop (see Table.~\ref{tab:AAB_pure_loop} and the main text). There are in total $5$ nonequivalent
operators, corresponding to $5$ nonequivalent particle excitations. The result would not be changed if one replace an operator in path integral by its equivalent operator, as explained in Sec.~\ref{subsec:Operators_equivalence class}. \label{tab:AAB_particle_nonequivalent}}
\begin{tabular*}{\textwidth}{@{\extracolsep{\fill}}ccc}
\hline
\textbf{Charges} & \textbf{Operators for particle excitations} & \textbf{Equivalent operators}\tabularnewline
\hline
$0$ & $\mathsf{P}_{000}=\mathsf{1}=\exp\left({\rm i}0\right)=1$ & -\tabularnewline
\hline
$\mathbb{Z}_{N_{1}}$ & $\mathsf{P}_{100}=\exp\left({\rm i}\int_{\gamma}A^{1}\right)$ & -\tabularnewline
\hline
$\mathbb{Z}_{N_{2}}$ & $\mathsf{P}_{010}=\exp\left({\rm i}\int_{\gamma}A^{2}\right)$ & -\tabularnewline
\hline
$\mathbb{Z}_{N_{3}}$ & $\mathsf{P}_{001}=2\exp\left[{\rm i}\int_{\gamma}A^{3}+\frac{1}{2}\frac{2\pi q}{N_{3}}\left(d^{-1}A^{1}A^{2}-d^{-1}A^{2}A^{1}\right)\right]\delta\left(\int_{\gamma}A^{1}\right)\delta\left(\int_{\gamma}A^{2}\right)$ & $\mathsf{P}_{001}=\mathsf{P}_{101}=\mathsf{P}_{011}=\mathsf{P}_{111}$\tabularnewline
\hline
$\mathbb{Z}_{N_{1}},\mathbb{Z}_{N_{2}}$ & $\mathsf{P}_{110}=\exp\left({\rm i}\int_{\gamma}A^{1}+{\rm i}\int_{\gamma}A^{2}\right)$ & -\tabularnewline
\hline
\end{tabular*}
\end{table*}

\begin{table*}
\caption{Operators for nonequivalent \emph{pure} loop excitations in BR
topological order with $G=\left(\mathbb{Z}_{2}\right)^{3}$. The $\Z_{N_i}$ gauge fluxes correspond to conjugacy classes of $\Z_{N_i}$ gauge subgroup. The subscript
in $\mathsf{L}_{n_{1}n_{2}n_{3}}$ indicates that the pure loop carries
$n_{i}$ units of $\mathbb{Z}_{N_{i}}$ fluxes where $i=1,2,3$. In
the first row, the pure loop carrying vanishing gauge flux is actually the vacuum,
denoted as $\mathsf{1}$.  Trivial pure loop and trivial particle
 are in fact the same and represented by  the vacuum operator $\mathsf{1}$ (see Table~\ref{tab:AAB_particle_nonequivalent} and the main text). The number of
nonequivalent operators for pure loops is $5$, corresponding to $5$
different pure loop excitations.   In the path
integral, the operators for pure loops may behave like those for some
decorated loops, so they are equivalent as shown in Sec.~\ref{subsec:Operators_equivalence class}. The
superscript in $\mathsf{L}_{n_{1}n_{2}n_{3}}^{c_{1}c_{2}c_{3}}$ denotes
the charge decoration: $c_{i}$ means $c_{i}$ units of $\mathbb{Z}_{N_{i}}$
gauge charge; while the subscript $n_{1}n_{2}n_{3}$ indicates the flux
carried by the loop. Besides the pure loops and their equivalent decorated
loops, there are other nonequivalent decorated loops, as shown in
Table~\ref{tab:AAB_decorated_loop}. \label{tab:AAB_pure_loop}}
\begin{tabular*}{\textwidth}{@{\extracolsep{\fill}}cccc}
\hline
\textbf{Fluxes} & \textbf{Charge decoration} & \textbf{Operators for pure loop excitations} & \textbf{Equivalent operators}\tabularnewline
\hline
$0$ & $0$ & $\mathsf{L}_{000}=\mathsf{1}=\exp\left({\rm i}0\right)=1$ & -\tabularnewline
\hline
$\mathbb{Z}_{N_{1}}$ & $0$ & $\begin{alignedat}{1}\mathsf{L}_{100}= & 2\exp\left[{\rm i}\int_{\sigma}B^{1}+\frac{1}{2}\frac{2\pi q}{N_{1}}\left(d^{-1}A^{2}B^{3}+d^{-1}B^{3}A^{2}\right)\right]\\
 & \times\delta\left(\int_{\gamma}A^{2}\right)\delta\left(\int_{\sigma}B^{3}\right)
\end{alignedat}
$ & $\mathsf{L}_{100}=\mathsf{L}_{10n_{3}}^{0c_{2}0};c_{2},n_{3}=0,1$\tabularnewline
\hline
$\mathbb{Z}_{N_{2}}$ & $0$ & $\begin{alignedat}{1}\mathsf{L}_{010}= & 2\exp\left[{\rm i}\int_{\sigma}B^{2}-\frac{1}{2}\frac{2\pi q}{N_{2}}\left(d^{-1}B^{3}A^{1}+d^{-1}A^{1}B^{3}\right)\right]\\
 & \times\delta\left(\int_{\sigma}B^{3}\right)\delta\left(\int_{\gamma}A^{1}\right)
\end{alignedat}
$ & $\mathsf{L}_{010}=\mathsf{L}_{01n_{3}}^{c_{1}00};c_{1},n_{3}=0,1$\tabularnewline
\hline
$\mathbb{Z}_{N_{3}}$ & $0$ & $\mathsf{L}_{001}=\exp\left({\rm i}\int_{\sigma}B^{3}\right)$ & -\tabularnewline
\hline
$\mathbb{Z}_{N_{1}},\mathbb{Z}_{N_{2}}$ & $0$ & $\begin{alignedat}{1}\mathsf{L}_{110}= & 2\exp\left[{\rm i}\int_{\sigma}B^{1}+\frac{1}{2}\frac{2\pi q}{N_{1}}\left(d^{-1}A^{2}B^{3}+d^{-1}B^{3}A^{2}\right)\right.\\
 & +\left.{\rm i}\int_{\sigma}B^{2}-\frac{1}{2}\frac{2\pi q}{N_{2}}\left(d^{-1}B^{3}A^{1}+d^{-1}A^{1}B^{3}\right)\right]\\
 & \times\delta\left(\int_{\gamma}A^{2}-A^{1}\right)\delta\left(\int_{\sigma}B^{3}\right)
\end{alignedat}
$ & $\mathsf{L}_{110}=\mathsf{L}_{110}^{110}=\mathsf{L}_{111}=\mathsf{L}_{111}^{110}$\tabularnewline
\hline
\end{tabular*}
\end{table*}

\twocolumngrid

\begin{table*}
\caption{Operators for nonequivalent decorated loop excitations in BR
topological order with $G=\left(\mathbb{Z}_{2}\right)^{3}$. The $\Z_{N_i}$ gauge charges and fluxes correspond to group representations and conjugacy classes of $\Z_{N_i}$ gauge subgroup respectively. The superscript
in $\mathsf{L}_{n_{1}n_{2}n_{3}}^{c_{1}c_{2}c_{3}}$ denotes the charge decoration: $c_{i}$ means $c_{i}$ units of $\mathbb{Z}_{N_{i}}$ gauge charge;
while the subscript $n_{1}n_{2}n_{3}$ indicates the gauge fluxes carried
by the loop. There are in total $10$ nonequivalent decorated loops. Some decorated loops are in fact
equivalent to specific pure loops, as explained in Sec.~\ref{subsec:Operators_equivalence class}.
\label{tab:AAB_decorated_loop}}
\resizebox{\textwidth}{!}{
\begin{tabular}{cccc}
\hline

\hline

\hline

\textbf{Fluxes} & \textbf{Charge decoration} & \textbf{Operators for decorated loop excitations} & \textbf{Equivalent operators}\tabularnewline
\hline
$\mathbb{Z}_{N_{1}}$ & $\mathbb{Z}_{N_{1}}$ & $\begin{alignedat}{1}\mathsf{L}_{100}^{100}= & 2\exp\left[{\rm i}\int_{\gamma}A^{1}+{\rm i}\int_{\sigma}B^{1}+\frac{1}{2}\frac{2\pi q}{N_{1}}\left(d^{-1}A^{2}B^{3}+d^{-1}B^{3}A^{2}\right)\right]\\
 & \times\delta\left(\int_{\gamma}A^{2}\right)\delta\left(\int_{\sigma}B^{3}\right)
\end{alignedat}
$ & $\mathsf{L}_{100}^{100}=\mathsf{L}_{10n_{3}}^{1c_{2}0};c_{2},n_{3}=0,1$\tabularnewline
\hline
$\mathbb{Z}_{N_{1}}$ & $\mathbb{Z}_{N_{2}}$ & equivalent to $\mathsf{L}_{100}$ & $\mathsf{L}_{100}=\mathsf{L}_{10n_{3}}^{0c_{2}0};c_{2},n_{3}=0,1$\tabularnewline
\hline
$\mathbb{Z}_{N_{1}}$ & $\mathbb{Z}_{N_{3}}$ & $\begin{alignedat}{1}\mathsf{L}_{100}^{001}= & 4\exp\left[{\rm i}\int_{\sigma}B^{1}+\frac{1}{2}\frac{2\pi q}{N_{1}}\left(d^{-1}A^{2}B^{3}+d^{-1}B^{3}A^{2}\right)\right.\\
 & \left.+{\rm i}\int_{\gamma}A^{3}+\frac{1}{2}\frac{2\pi q}{N_{3}}\left(d^{-1}A^{1}A^{2}-d^{-1}A^{2}A^{1}\right)\right]\\
 & \times\delta\left(\int_{\gamma}A^{2}\right)\delta\left(\int_{\sigma}B^{3}\right)\delta\left(\int_{\gamma}A^{1}\right)
\end{alignedat}
$ & $\mathsf{L}_{100}^{001}=\mathsf{L}_{10n_{3}}^{c_{1}c_{2}1};c_{1},c_{2},n_{3}=0,1$\tabularnewline
\hline
$\mathbb{Z}_{N_{2}}$ & $\mathbb{Z}_{N_{1}}$ & equivalent to $\mathsf{L}_{010}$ & $\mathsf{L}_{010}=\mathsf{L}_{01n_{3}}^{c_{1}00};c_{1},n_{3}=0,1$\tabularnewline
\hline
$\mathbb{Z}_{N_{2}}$ & $\mathbb{Z}_{N_{2}}$ & $\begin{alignedat}{1}\mathsf{L}_{010}^{010}= & 2\exp\left[{\rm i}\int_{\gamma}A^{2}+{\rm i}\int_{\sigma}B^{2}-\frac{1}{2}\frac{2\pi q}{N_{2}}\left(d^{-1}B^{3}A^{1}+d^{-1}A^{1}B^{3}\right)\right]\\
 & \times\delta\left(\int_{\sigma}B^{3}\right)\delta\left(\int_{\gamma}A^{1}\right)
\end{alignedat}
$ & $\mathsf{L}_{010}^{010}=\mathsf{L}_{01n_{3}}^{c_{1}10};c_{1},n_{3}=0,1$\tabularnewline
\hline
$\mathbb{Z}_{N_{2}}$ & $\mathbb{Z}_{N_{3}}$ & $\begin{alignedat}{1}\mathsf{L}_{010}^{001}= & 4\exp\left[{\rm i}\int_{\sigma}B^{2}-\frac{1}{2}\frac{2\pi q}{N_{2}}\left(d^{-1}B^{3}A^{1}+d^{-1}A^{1}B^{3}\right)\right.\\
 & \left.+{\rm i}\int_{\gamma}A^{3}+\frac{1}{2}\frac{2\pi q}{N_{3}}\left(d^{-1}A^{1}A^{2}-d^{-1}A^{2}A^{1}\right)\right]\\
 & \times\delta\left(\int_{\sigma}B^{3}\right)\delta\left(\int_{\gamma}A^{1}\right)\delta\left(\int_{\gamma}A^{2}\right)
\end{alignedat}
$ & $\mathsf{L}_{010}^{001}=\mathsf{L}_{01n_{3}}^{c_{1}c_{2}1};c_{1},c_{2},n_{3}=0,1$\tabularnewline
\hline
$\mathbb{Z}_{N_{3}}$ & $\mathbb{Z}_{N_{1}}$ & $\mathsf{L}_{001}^{100}=\exp\left({\rm i}\int_{\gamma}A^{1}+{\rm i}\int_{\sigma}B^{3}\right)$ & -\tabularnewline
\hline
$\mathbb{Z}_{N_{3}}$ & $\mathbb{Z}_{N_{2}}$ & $\mathsf{L}_{001}^{010}=\exp\left({\rm i}\int_{\gamma}A^{2}+{\rm i}\int_{\sigma}B^{3}\right)$ & -\tabularnewline
\hline
$\mathbb{Z}_{N_{3}}$ & $\mathbb{Z}_{N_{1}},\mathbb{Z}_{N_{2}}$ & $\mathsf{L}_{001}^{110}=\exp\left({\rm i}\int_{\gamma}A^{1}+{\rm i}\int_{\gamma}A^{2}+{\rm i}\int_{\sigma}B^{3}\right)$ & -\tabularnewline
\hline
$\mathbb{Z}_{N_{3}}$ & $\mathbb{Z}_{N_{3}}$ & $\begin{alignedat}{1}\mathsf{L}_{001}^{001}= & 2\exp\left[{\rm i}\int_{\sigma}B^{3}+{\rm i}\int_{\gamma}A^{3}+\frac{1}{2}\frac{2\pi q}{N_{3}}\left(d^{-1}A^{1}A^{2}-d^{-1}A^{2}A^{1}\right)\right]\\
 & \times\delta\left(\int_{\gamma}A^{1}\right)\delta\left(\int_{\gamma}A^{2}\right)
\end{alignedat}
$ & $\mathsf{L}_{001}^{001}=\mathsf{L}_{001}^{c_{1}c_{2}1};c_{1},c_{2}=0,1$\tabularnewline
\hline
$\mathbb{Z}_{N_{1}},\mathbb{Z}_{N_{2}}$ & $\mathbb{Z}_{N_{1}}$ & $\begin{alignedat}{1}\mathsf{L}_{110}^{100}= & 2\exp\left[{\rm i}\int_{\gamma}A^{1}+{\rm i}\int_{\sigma}B^{1}+\frac{1}{2}\frac{2\pi q}{N_{1}}\left(d^{-1}A^{2}B^{3}+d^{-1}B^{3}A^{2}\right)\right.\\
 & +\left.{\rm i}\int_{\sigma}B^{2}-\frac{1}{2}\frac{2\pi q}{N_{2}}\left(d^{-1}B^{3}A^{1}+d^{-1}A^{1}B^{3}\right)\right]\\
 & \times\delta\left(\int_{\gamma}A^{2}-A^{1}\right)\delta\left(\int_{\sigma}B^{3}\right)
\end{alignedat}
$ & $\mathsf{L}_{110}^{100}=\mathsf{L}_{110}^{010}=\mathsf{L}_{111}^{110}=\mathsf{L}_{111}^{010}$\tabularnewline
\hline
$\mathbb{Z}_{N_{1}},\mathbb{Z}_{N_{2}}$ & $\mathbb{Z}_{N_{2}}$ & equivalent to $\mathsf{L}_{110}^{010}$ & $\mathsf{L}_{110}^{100}=\mathsf{L}_{110}^{010}=\mathsf{L}_{111}^{110}=\mathsf{L}_{111}^{010}$\tabularnewline
\hline
$\mathbb{Z}_{N_{1}},\mathbb{Z}_{N_{2}}$ & $\mathbb{Z}_{N_{1}},\mathbb{Z}_{N_{2}}$ & equivalent to $\mathsf{L}_{110}$ & $\mathsf{L}_{110}=\mathsf{L}_{110}^{110}=\mathsf{L}_{111}=\mathsf{L}_{111}^{110}$\tabularnewline
\hline
$\mathbb{Z}_{N_{1}},\mathbb{Z}_{N_{2}}$ & $\mathbb{Z}_{N_{3}}$ & $\begin{alignedat}{1}\mathsf{L}_{110}^{001}= & 4\exp\left[{\rm i}\int_{\sigma}B^{1}+\frac{1}{2}\frac{2\pi q}{N_{1}}\left(d^{-1}A^{2}B^{3}+d^{-1}B^{3}A^{2}\right)\right.\\
 & +{\rm i}\int_{\sigma}B^{2}-\frac{1}{2}\frac{2\pi q}{N_{2}}\left(d^{-1}B^{3}A^{1}+d^{-1}A^{1}B^{3}\right)\\
 & +\left.{\rm i}\int_{\gamma}A^{3}+\frac{1}{2}\frac{2\pi q}{N_{3}}\left(d^{-1}A^{1}A^{2}-d^{-1}A^{2}A^{1}\right)\right]\\
 & \times\delta\left(\int_{\gamma}A^{1}\right)\delta\left(\int_{\gamma}A^{2}\right)\delta\left(\int_{\sigma}B^{3}\right)
\end{alignedat}
$ & $\mathsf{L}_{110}^{001}=\mathsf{L}_{11n_{3}}^{c_{1}c_{2}1};c_{1},c_{2},n_{3}=0,1$\tabularnewline
\hline

\hline

\hline

\end{tabular}}
\end{table*}

%%\bigskip
%%\twocolumngrid

\section{Fusion rules and loop shrinking rules from path integrals\label{sec:fusion rule}}

\begin{figure}
  \centering
  \includegraphics[scale=0.35]{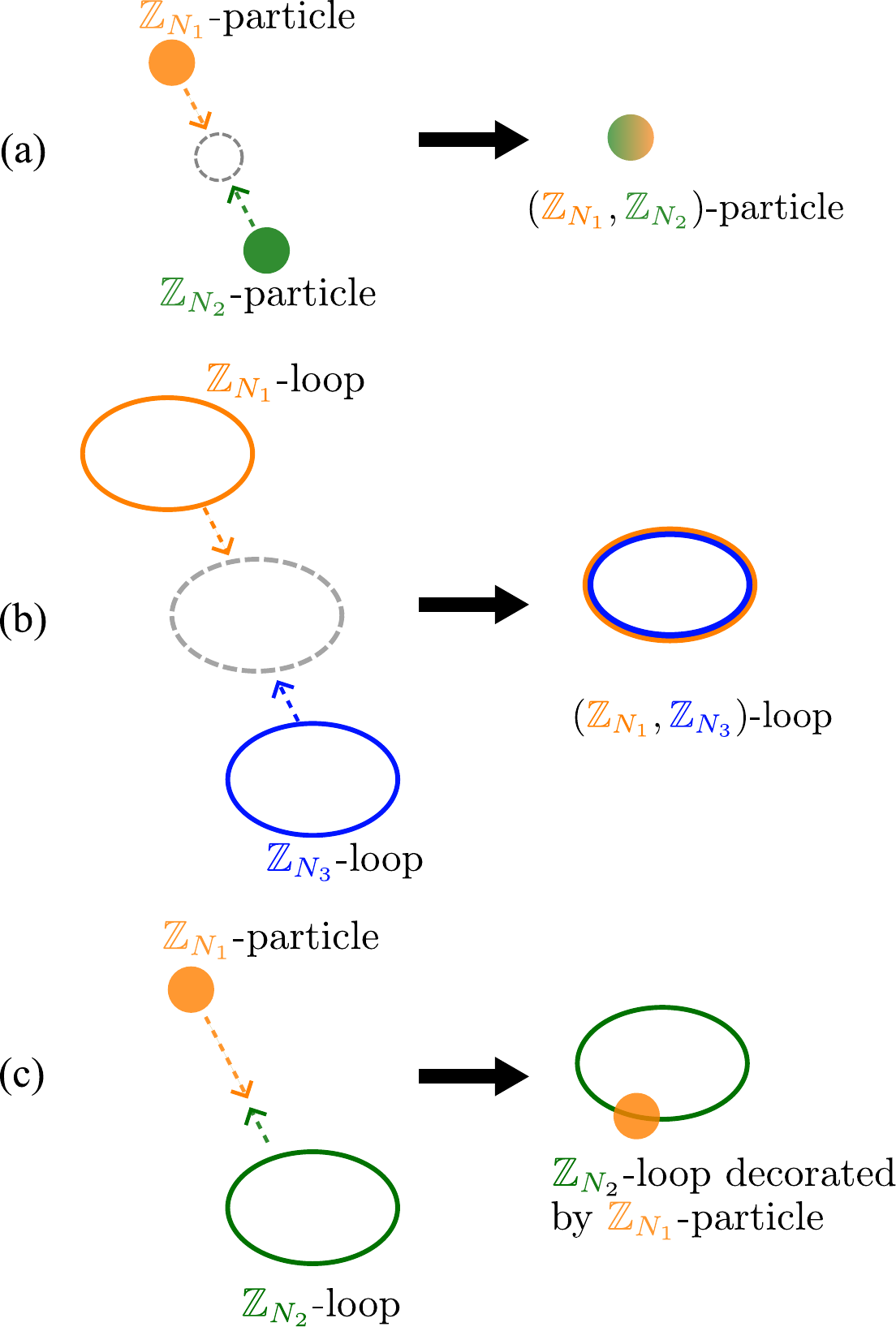}
  \caption{(a) Fusion of a $\Z_{N_1}$-particle and a $\Z_{N_2}$-particle. In continuous field theory, fusion of two particle excitations means that they moves towards each other until they meet at the same spatial location. This can be realized by making the world-lines of these two particles identical. The output of this fusion is a $\left(\Z_{N_1},\Z_{N_2}\right)$-particle which carries one unit of $\Z_{N_1}$ and $\Z_{N_2}$ gauge charge simultaneously. The $\Z_{N_1}$ and $\Z_{N_2}$ gauge charges in the output particle are represented by a mixed color of orange ($\Z_{N_1}$) and green ($\Z_{N_2}$). (b) Fusion of a $\Z_{N_1}$-loop and a $\Z_{N_3}$-loop. Similarly, fusion of two loops requires that they overlap at the same location. For this purpose, one can set their world-sheets identical in field theory. The result of this fusion is a loop carrying one unit of $\Z_{N_1}$ and $\Z_{N_3}$ flux respectively. These two fluxes are illustrated by two colors (orange and blue) circling along the loop. Notice that the boundary of two colors \emph{does not} indicate a separation of two different fluxes. According to Table~\ref{tab:AAB_pure_loop}, a $\left(\Z_{N_1},\Z_{N_3}\right)$-loop is equivalent to a $\Z_{N_1}$-loop. (c) Fusion of a $\Z_{N_1}$-particle and a $\Z_{N_2}$-loop. The outcome is a decorated loop: a $\Z_{N_2}$-loop decorated by a $\Z_{N_1}$-particle. In fact, this is the definition of fusion of particle and loop. In continuous field theory, this fusion is realized by making the world-line of particle live on the world-sheet of loop.}
  \label{fig:fusion_illustrate}
\end{figure}

In this section, we are going to calculate the fusion rules of excitations
for BR topological order with $G=\left(\mathbb{Z}_{2}\right)^{3}$.
These fusion rules, together with braiding phases, form a more complete
data set to characterize the BR topological order. Assume the
fusion of excitation $\mathsf{a}$ and $\mathsf{b}$ is
\begin{equation}
\mathsf{a}\otimes\mathsf{b}=\oplus_{i}N_{\mathsf{e}_{i}}^{\mathsf{a}\mathsf{b}}\mathsf{e}_{i}\label{eq:fusion_eg1}
\end{equation}
where $N_{\mathsf{e}_{i}}^{\mathsf{a}\mathsf{b}}$ is a non-zero integer
called fusion coefficient. Now we ask: how to represent this algebraic
fusion rule using field-theoretical language? If it is considered in
a lattice, the above fusion is that two excitations $\mathsf{a}$
and $\mathsf{b}$ are very close to each other such that they behave
like the superposition of other excitations $\mathsf{e}_{i}$. In
fact, if we consider the expectation value, the fusion rule (\ref{eq:fusion_eg1})
indicates that
\begin{equation}
\left\langle \mathsf{a}\otimes\mathsf{b}\right\rangle =\left\langle \oplus_{i}N_{\mathsf{e}_{i}}^{\mathsf{a}\mathsf{b}}\mathsf{e}_{i}\right\rangle =\oplus_{i}N_{\mathsf{e}_{i}}^{\mathsf{a}\mathsf{b}}\left\langle \mathsf{e}_{i}\right\rangle .
\end{equation}
If this fusion is considered in the scenario of continuous field theory, the excitations
should be replaced by gauge-invariant operators $\mathcal{O}_{\mathsf{e}_i}$. In addition, the correlation length in TQFT is zero which implies the infinite energy gap between the ground state and excited states. Any finite distance would be in fact infinitely larger than the correlation length. Therefore,
when discussing fusion in the framework of TQFT, we must set the topological excitations in the same spatial position strictly, as illustrated in Fig.~\ref{fig:fusion_illustrate}. In other words, the world-lines and/or world-sheets of two topological excitations in fusion should be identical. We can conclude that in the fusion
in TQFT, i.e., in terms of path integral, is given by
\begin{align}
\left\langle \mathsf{a}\otimes\mathsf{b}\right\rangle  & =\frac{1}{\mathcal{Z}}\int\mathcal{D}\left[A^{i}\right]\mathcal{D}\left[B^{i}\right]\exp\left({\rm i}S\right)\times\left(\mathcal{O}_{\mathsf{a}}\times\mathcal{O}_{\mathsf{b}}\right)\nonumber \\
 & =\frac{1}{\mathcal{Z}}\int\mathcal{D}\left[A^{i}\right]\mathcal{D}\left[B^{i}\right]\exp\left({\rm i}S\right)\times\left(\sum_{i}N_{\mathsf{e}_{i}}^{\mathsf{a}\mathsf{b}}\mathcal{O}_{\mathsf{e}_{i}}\right)\nonumber \\
 & =\left\langle \oplus_{i}N_{\mathsf{e}_{i}}^{\mathsf{a}\mathsf{b}}\mathsf{e}_{i}\right\rangle
\end{align}
in which $\mathcal{O}_{\mathsf{a}}$ and $\mathcal{O}_{\mathsf{b}}$ share the same world-line and/or world-sheet. In this way, we can read the fusion rule~(\ref{eq:fusion_eg1}) from
\begin{equation}
\mathcal{O}_{\mathsf{a}}\times\mathcal{O}_{\mathsf{b}}=\sum_{i}N_{\mathsf{e}_{i}}^{\mathsf{a}\mathsf{b}}\mathcal{O}_{\mathsf{e}_{i}}
\label{eq:operator_fusion_formula}
\end{equation}
which should be considered in the context of path integral. Below we first show several examples of computing
fusion rules through path integral. By exhausting all $19$ operators for topological excitations listed in Table~\ref{tab:AAB_particle_nonequivalent} (particles), Table~\ref{tab:AAB_pure_loop} (pure loops), and Table~\ref{tab:AAB_decorated_loop} (decorated loops), we can find out all
fusion rules for BR topological order with $G=\left(\Z_{2}\right)^3$. The complete fusion rules are shown in Table~\ref{tab:fusion_AAB_Z2Z2Z2_full} among which some are non-Abelian. Furthermore, the shrinking rules of loop excitations are studied in Sec.~\ref{subsec:shrinking_rule} and listed in Table~\ref{tab:shrinking rules_AAB_z2z2z2}.

\subsection{Examples of fusion rule calculation}\label{subsec:example_fusion_rules}

\begin{figure}
  \centering
  \includegraphics[scale=0.35]{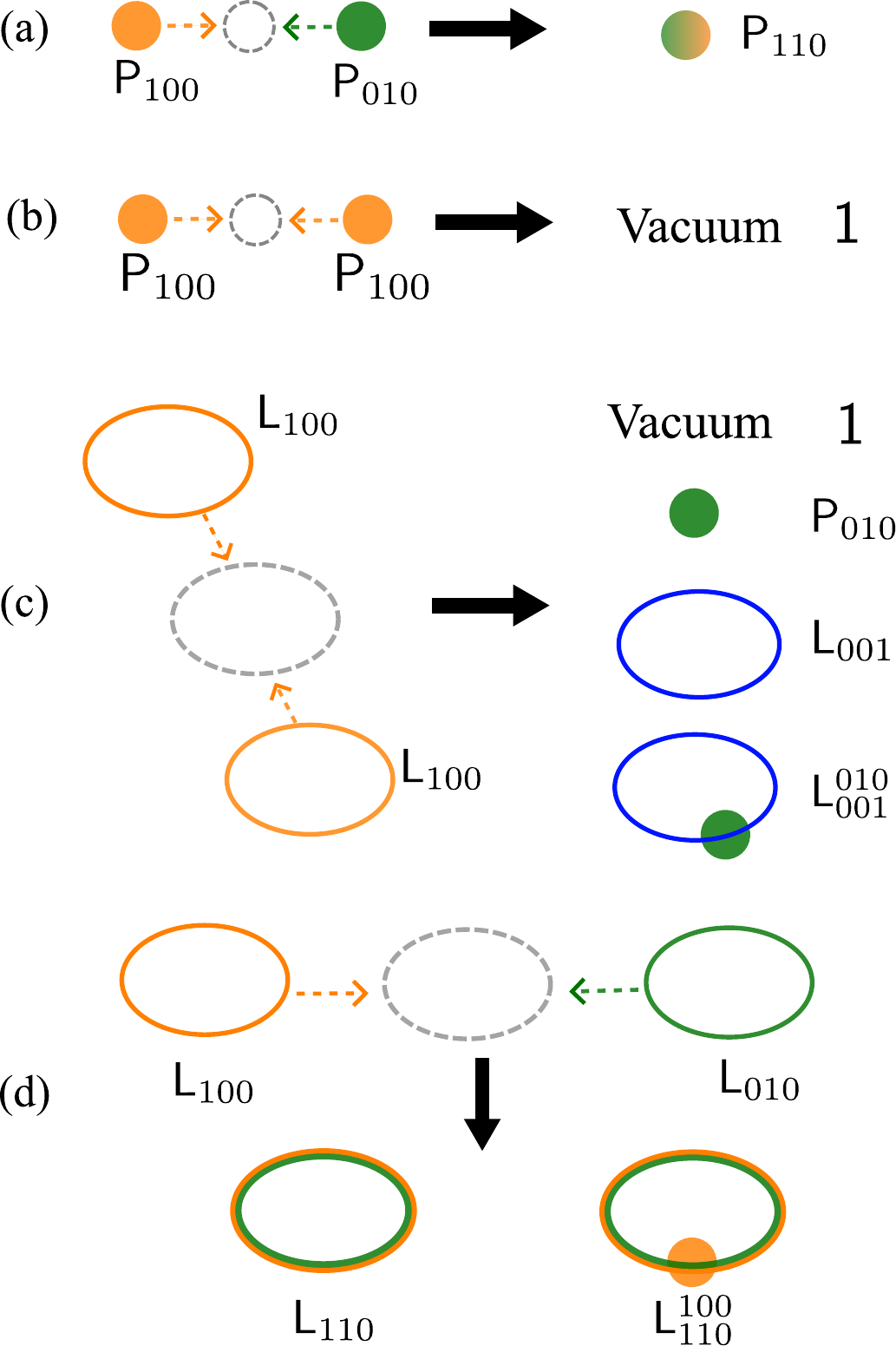}
  \caption{Illustrations of four fusion processes that are discussed in Sec.~\ref{subsec:example_fusion_rules} and detailed in Appendix~\ref{appendix:detail_fusion_example}. (a) Fusion: $\mathsf{P}_{100}\otimes\mathsf{P}_{010}=\mathsf{P}_{110}$. This is an Abelian fusion: fusing a $\Z_{N_1}$-particle and a $\Z_{N_2}$-particle results a $\left(\Z_{N_1},\Z_{N_2}\right)$-particle. (b) Fusion: $\mathsf{P}_{100}\otimes\mathsf{P}_{100}=\mathsf{1}$. This is also an Abelian fusion. This fusion rule indicates that in BR topological order with $G=\left(\Z_{2}\right)^{3}$ the anti-particle of $\Z_{N_1}$-particle is itself. This make sense since the $\Z_{N_1}$ gauge subgroup is a $\Z_{2}$ group. (c) Non-Abelian fusion: $\mathsf{L}_{100}\otimes\mathsf{L}_{100}=\mathsf{1}\oplus\mathsf{P}_{010}\oplus\mathsf{L}_{001}\oplus\mathsf{L}_{001}^{010}$. Fusion of two $\Z_{N_1}$-loops produces not a determined outcome, but a superposition of a vacuum $\mathsf{1}$, a particle $\mathsf{P}_{010}$, a pure loop $\mathsf{L}_{001}$, and a decorated loop $\mathsf{L}_{001}^{010}$. (d) Non-Abelian fusion: $\mathsf{L}_{100}\otimes\mathsf{L}_{010}=\mathsf{L}_{110}\oplus\mathsf{L}_{110}^{100}$. The outcome of fusing a $\Z_{N_1}$-loop and a $\Z_{N_2}$-loop is a superposition of a pure loop ($\mathsf{L}_{110}$) and a decorated loop ($\mathsf{L}_{110}^{100}$). Notice that $\mathsf{L}_{110}$ and $\mathsf{L}_{110}^{100}$ are nonequivalent according to Table.~\ref{tab:AAB_pure_loop} and Table.~\ref{tab:AAB_decorated_loop}.}
  \label{fig:4_fusion_examples}
\end{figure}

Now we explain how to exploit Eq.~(\ref{eq:operator_fusion_formula})
to obtain fusion rules of topological excitations by several examples. These examples of fusion are illustrated in Fig.~\ref{fig:4_fusion_examples} in which two are Abelian fusion and the others are non-Abelian.
The technical details can be found in Appendix~\ref{appendix:detail_fusion_example}.
The notations of operators ($\mathsf{P}_{n_{1}n_{2}n_{3}}$,
$\mathsf{L}_{n_{1}n_{2}n_{3}}$, and $\mathsf{L}_{n_{1}n_{2}n_{3}}^{c_{1}c_{2}c_{3}}$)
are also used to refer corresponding topological excitations in the context without causing ambiguity, e.g.,
$\mathsf{P}_{100}$ not only represents the operator of $\mathbb{Z}_{N_{1}}$-particle
but also denotes the $\mathbb{Z}_{N_{1}}$-particle excitation itself. When
we mention topological excitations in the fusion and loop shrinking
operations (discussed in Sec.~\ref{subsec:shrinking_rule}), we use
``$\otimes$'' and ``$\oplus$'' between the notations for direct
product and direct sum of fusion spaces. When Wilson operators in
path integrals are consider, their multiplication and addition are
indicated by ``$\times$'' and ``$+$''.
\subsubsection{$\mathbb{Z}_{N_{1}}$-particle and $\mathbb{Z}_{N_{2}}$-particle}

The \nth{1} example is the fusion of a $\mathbb{Z}_{N_{1}}$-particle
and a $\mathbb{Z}_{N_{2}}$-particle. Using Eq.~\ref{eq:operator_fusion_formula},
we can write down
\begin{align}
 & \left\langle \mathsf{P}_{100}\otimes\mathsf{P}_{010}\right\rangle \nonumber \\
= & \frac{1}{\mathcal{Z}}\int\mathcal{D}\left[A^{i}\right]\mathcal{D}\left[B^{i}\right]\exp\left({\rm i}S\right)\mathsf{P}_{100}\times\mathsf{P}_{010}\nonumber \\
= & \frac{1}{\mathcal{Z}}\int\mathcal{D}\left[A^{i}\right]\mathcal{D}\left[B^{i}\right]\exp\left({\rm i}S\right)\exp\left({\rm i}\int_{\gamma}A^{1}+A^{2}\right)\nonumber \\
= & \frac{1}{\mathcal{Z}}\int\mathcal{D}\left[A^{i}\right]\mathcal{D}\left[B^{i}\right]\exp\left({\rm i}S\right)\mathsf{P}_{110}\nonumber \\
= & \left\langle \mathsf{P}_{110}\right\rangle
\end{align}
and find that
\begin{equation}
\mathsf{P}_{100}\otimes\mathsf{P}_{010}=\mathsf{P}_{110}.
\end{equation}
This result indicates that by fusing two particles carrying $\mathbb{Z}_{N_{1}}$
and $\mathbb{Z}_{N_{2}}$ gauge charges respectively we obtain a single
particle that carries both $\mathbb{Z}_{N_{1}}$ and $\mathbb{Z}_{N_{2}}$
gauge charges.

\subsubsection{Two $\mathbb{Z}_{N_{1}}$-particles}

The \nth{2} example is the fusion of two $\mathbb{Z}_{N_{1}}$-particles:
\begin{align}
\left\langle \mathsf{P}_{100}\otimes\mathsf{P}_{100}\right\rangle = & \frac{1}{\mathcal{Z}}\int\mathcal{D}\left[A^{i}\right]\mathcal{D}\left[B^{i}\right]\exp\left({\rm i}S\right)\exp\left({\rm i}2\int_{\gamma}A^{1}\right).
\end{align}
Integrating out $B^{1}$, $B^{2}$, and $A^{3}$ and we obtain constraints
for $A^{1}$, $A^{2}$, and $B^{3}$ respectively:
\begin{equation}
\oint A^{1}=\frac{2\pi m_{1}}{N_{1}},
\end{equation}
\begin{equation}
\oint A^{2}=\frac{2\pi m_{2}}{N_{2}},
\end{equation}
\begin{equation}
\oint B^{3}=\frac{2\pi m_{3}}{N_{3}},
\end{equation}
where $m_{1,2,3}\in\mathbb{Z}$. Notice that gauge group is $G=\prod_{i=1}^{3}\mathbb{Z}_{N_{i}}=\left(\mathbb{Z}_{2}\right)^{3}$,
we have
\begin{equation}
\left\langle \mathsf{P}_{100}\otimes\mathsf{P}_{100}\right\rangle =1=\left\langle \mathsf{1}\right\rangle ,
\end{equation}
i.e.,
\begin{equation}
\mathsf{P}_{100}\otimes\mathsf{P}_{100}=\mathsf{1}.
\end{equation}
This result tells us that $\mathsf{P}_{100}$ is the anti-particle
of itself , which is reasonable since $\mathsf{P}_{100}$ carries
one unit of $\mathbb{Z}_{N_{1}}=\mathbb{Z}_{2}$ gauge charge.

\subsubsection{Two $\mathbb{Z}_{N_{1}}$-loops}

In the \nth{3} example, we consider the fusion of two $\mathbb{Z}_{N_{1}}$-loops.
We start with
\begin{align}
\left\langle \mathsf{L}_{100}\otimes\mathsf{L}_{100}\right\rangle = & \frac{1}{\mathcal{Z}}\int\mathcal{D}\left[A^{i}\right]\mathcal{D}\left[B^{i}\right]\exp\left({\rm i}S\right)\times\mathsf{L}_{100}\times\mathsf{L}_{100}.
\end{align}
Using Table.~\ref{tab:AAB_pure_loop}, we plug in the expression of $\mathsf{L}_{100}$ and obtain (details are collected in Appendix~\ref{appendix:detail_fusion_example}):
\begin{align}
 & \left\langle \mathsf{L}_{100}\otimes\mathsf{L}_{100}\right\rangle \nonumber \\
= & 1+\exp\left(\frac{{\rm i}2\pi m_{2}}{2}\right)+\exp\left(\frac{{\rm i}2\pi m_{3}}{2}\right)+\exp\left[\frac{{\rm i}2\pi\left(m_{2}+m_{3}\right)}{2}\right].
\end{align}
We can immediately find that
\begin{align}
 & \left\langle \mathsf{L}_{100}\otimes\mathsf{L}_{100}\right\rangle \nonumber \\
= & \frac{1}{\mathcal{Z}}\int\mathcal{D}\left[A^{i}\right]\mathcal{D}\left[B^{i}\right]\exp\left({\rm i}S\right)\times\left(\mathsf{1}+\mathsf{P}_{010}+\mathsf{L}_{001}+\mathsf{L}_{001}^{010}\right)
\end{align}
thus we can conclude with
\begin{equation}
\mathsf{L}_{100}\otimes\mathsf{L}_{100}=\mathsf{1}\oplus\mathsf{P}_{010}\oplus\mathsf{L}_{001}\oplus\mathsf{L}_{001}^{010}.
\end{equation}
This is a non-Abelian fusion rule which tells us that if we fuse two
$\mathbb{Z}_{N_{1}}$-loops we would obtain the superposition of a
vacuum, a $\mathbb{Z}_{N_{2}}$-particle, a $\mathbb{Z}_{N_{3}}$-loop,
and a $\mathbb{Z}_{N_{3}}$-loop decorated by a $\mathbb{Z}_{N_{2}}$-particle.

\subsubsection{$\mathbb{Z}_{N_{1}}$-loop and $\mathbb{Z}_{N_{2}}$-loop}

In the \nth{4} example, we continue to consider $\mathsf{L}_{100}\otimes\mathsf{L}_{010}$:
\begin{align}
 & \left\langle \mathsf{L}_{100}\otimes\mathsf{L}_{010}\right\rangle \nonumber\\
 = & \frac{1}{\mathcal{Z}}\int\mathcal{D}\left[A^{i}\right]\mathcal{D}\left[B^{i}\right]\exp\left({\rm i}S\right)\times\mathsf{L}_{100}\times\mathsf{L}_{010}.
\end{align}
For simplicity, we denote $\mathsf{L}_{100}\times\mathsf{L}_{010}$
as
\begin{align}
\mathsf{L}_{100}\times\mathsf{L}_{010}= & 4\exp\left({\rm i}f_{1}+{\rm i}f_{2}\right)\delta\left(\int_{\gamma}A^{2}\right)\delta\left(\int_{\sigma}B^{3}\right)\nonumber \\
 & \times\delta\left(\int_{\gamma}A^{1}\right)\delta\left(\int_{\sigma}B^{3}\right)
\end{align}
where $f_{1}=\int_{\sigma}B^{1}+\frac{1}{2}\frac{2\pi q}{N_{1}}\left(d^{-1}A^{2}B^{3}+d^{-1}B^{3}A^{2}\right)$
and $f_{2}=\int_{\sigma}B^{2}-\frac{1}{2}\frac{2\pi q}{N_{2}}\left(d^{-1}B^{3}A^{1}+d^{-1}A^{1}B^{3}\right)$.
For the complete expression of $\mathsf{L}_{100}$ and $\mathsf{L}_{010}$,
one can refer to Table~\ref{tab:AAB_pure_loop}. Since the Kronecker
delta functions can be re-written as
\begin{align}
\delta\left(\int_{\sigma}B^{3}\right)\delta\left(\int_{\sigma}B^{3}\right)= & \delta\left(\int_{\sigma}B^{3}\right),\\
\delta\left(\int_{\gamma}A^{2}\right)\delta\left(\int_{\gamma}A^{1}\right)= & \delta\left(\int_{\gamma}A^{2}-A^{1}\right)\delta\left(\int_{\gamma}A^{1}\right),
\end{align}
we can write $\mathsf{L}_{100}\times\mathsf{L}_{010}$ as
\begin{align}
\mathsf{L}_{100}\times\mathsf{L}_{010}= & 4\exp\left({\rm i}f_{1}+{\rm i}f_{2}\right)\delta\left(\int_{\gamma}A^{1}\right)\nonumber \\
 & \times\delta\left(\int_{\gamma}A^{2}-A^{1}\right)\delta\left(\int_{\sigma}B^{3}\right).
\end{align}
The Kronecker delta function $\delta\left(\int_{\gamma}A^{1}\right)$
can be expressed as $\delta\left(\int_{\gamma}A^{1}\right)=\frac{1}{2}\left[1+\exp\left({\rm i}\int_{\gamma}A^{1}\right)\right]$ in path integral.
Therefore, in the sense of expectation value,
\begin{align}
\mathsf{L}_{100}\times\mathsf{L}_{010}= & 2\exp\left({\rm i}f_{1}+{\rm i}f_{2}+{\rm i}\int_{\gamma}A^{1}\right)\nonumber \\
 & \times\delta\left(\int_{\gamma}A^{2}-A^{1}\right)\delta\left(\int_{\sigma}B^{3}\right).
\end{align}
By checking the all $19$ operators, we find (here $\mathsf{L}$ denotes operators of loops)
\begin{align}
\left\langle\mathsf{L}_{100}\times\mathsf{L}_{010}\right\rangle= & \left\langle\mathsf{L}_{110}+\mathsf{L}_{110}^{100}\right\rangle
\end{align}
which indicates the following fusion rule of topological excitations (here $\mathsf{L}$ denotes loop excitations):
\begin{equation}
\mathsf{L}_{100}\otimes\mathsf{L}_{010}=\mathsf{L}_{110}\oplus\mathsf{L}_{110}^{100}.
\end{equation}
This is another non-Abelian fusion rule. The output of fusion of a
$\mathbb{Z}_{N_{1}}$-loop and a $\mathbb{Z}_{N_{2}}$-loop is the
superposition of a pure $\left(\mathbb{Z}_{N_{1}},\mathbb{Z}_{N_{2}}\right)$-loop,
$\mathsf{L}_{110}$ and a $\left(\mathbb{Z}_{N_{1}},\mathbb{Z}_{N_{2}}\right)$-loop
decorated by a $\mathbb{Z}_{N_{1}}$-particle, $\mathsf{L}_{110}^{100}$.
One should notice the following equivalence relation as indicated
in Table.~\ref{tab:AAB_decorated_loop}: $\mathsf{L}_{110}=\mathsf{L}_{110}^{110}$
and $\mathsf{L}_{110}^{100}=\mathsf{L}_{110}^{010}$. The above results are obtained through a calculation
of path integral though the formulas are written in a simplified manner.
The detailed derivation can be found in Appendix~\ref{appendix:detail_fusion_example}.

\subsection{Fusion table for BR topological order with $G=\left(\mathbb{Z}_{2}\right)^{3}$}

The above examples show how to calculate fusion rules from path integral.
By exhausting all combinations of two excitations, we obtain the complete
fusion rules and quantum dimension $d$ of excitations for Borromean
rings topological order with $G=\left(\mathbb{Z}_{2}\right)^{3}$
as shown in Table~\ref{tab:fusion_AAB_Z2Z2Z2_full} and Table~\ref{tab:qu_dim_AAB_Z2Z2Z2}.
The fusion rules satisfy the properties of commutativity and associativity,
i.e., $\mathsf{a}\otimes \mathsf{b}=\mathsf{b}\otimes \mathsf{a}$ and $\left(\mathsf{a}\otimes \mathsf{b}\right)\otimes \mathsf{c}=\mathsf{a}\otimes\left(\mathsf{b}\otimes \mathsf{c}\right)$,
which is automatically guaranteed by the path integral calculation of Abelian
gauge fields.

In the BR topological order with $G=\left(\mathbb{Z}_{2}\right)^{3}$,
the excitations can be divided as follows: vacuum, $\mathsf{1}$;
$4$ nonequivalent particles, $\mathsf{P}_{100}$, $\mathsf{P}_{010}$,
$\mathsf{P}_{110}$, and $\mathsf{P_{001}}$; $4$ nonequivalent \emph{pure}
loops, $\mathsf{L}_{001}$, $\mathsf{L}_{100}$, $\mathsf{L}_{010}$,
and $\mathsf{L}_{110}$; $10$ nonequivalent loops decorated with
particle, $\mathsf{L}_{001}^{100}$, $\mathsf{L}_{001}^{010}$, $\mathsf{L}_{001}^{110}$,
$\mathsf{L}_{100}^{100}$, $\mathsf{L}_{010}^{010}$, $\mathsf{L}_{001}^{001}$,
$\mathsf{L}_{110}^{100}$, $\mathsf{L}_{100}^{001}$, $\mathsf{L}_{010}^{001}$,
and $\mathsf{L}_{110}^{001}$. All these excitations, no matter particles
or loops, can be thought as combinations of gauge charges and gauge
fluxes. Since the gauge group is $G=\left(\mathbb{Z}_{2}\right)^{3}$,
one may find there are $\left(2^{3}\right)^{2}=64$ different combinations
that exceeds the number of excitations listed in Table~\ref{tab:fusion_AAB_Z2Z2Z2_full}.
In fact, among all $64$ combinations of gauge charges and fluxes,
some of them behave without any difference thus collected to the same
equivalence class, as shown in Sec. \ref{subsec:Operators_equivalence class} and Table~\ref{tab:AAB_decorated_loop}.

In the following lines, we make some explanation about the Table~\ref{tab:fusion_AAB_Z2Z2Z2_full} of fusion rules. First, there are $8$ \emph{Abelian excitations} ($\mathsf{1}$, $\mathsf{P}_{100}$, $\mathsf{P}_{010}$, $\mathsf{L}_{001}$, $\mathsf{P}_{110}$, $\mathsf{L}_{001}^{100}$, $\mathsf{L}_{001}^{010}$, and $\mathsf{L}_{001}^{110}$; labeled by number
$1$ to $8$ in Table~\ref{tab:fusion_AAB_Z2Z2Z2_full}) whose fusion rules with any other excitations are always Abelian,
i.e., single fusion channel. All other excitations are called \emph{non-Abelian excitations}.
This fusion table is obtained in the case of $G=\left(\mathbb{Z}_{2}\right)^{3}$
and we may expect that the fusion of an excitation and itself would
produce a vacuum due to the $\mathbb{Z}_{2}$ cyclic nature. Nevertheless,
for $\mathsf{L}_{100}^{001}$, a loop carrying $\mathbb{Z}_{N_{1}}$
flux and decorated by a $\mathbb{Z}_{N_{3}}$-particle, the fusion
of two $\mathsf{L}_{100}^{001}$ produces two copies of the direct sum of all $8$ Abelian
excitations that is denoted as $\mathbf{Ab}$ (see Table~\ref{tab:fusion_AAB_Z2Z2Z2_full}). This means that $\mathsf{L}_{100}^{001}\otimes\mathsf{L}_{100}^{001}$
generates a direct sum of \emph{two} vacuums. Meanwhile,
$\mathsf{L}_{010}^{001}$ and $\mathsf{L}_{110}^{001}$ also have
this property. Field-theoretical calculation for this result can be found in Appendix~\ref{appendix:two_vacuum_example}. For other excitations, the fusion of its two copies just
produce a single vacuum.

From this fusion table, we can obtain all fusion coefficients $N_{k}^{ij}$'s
and matrices $N_{i}$ whose element is $\left(N_{i}\right)_{kj}=N_{k}^{ij}$,
where $i,j,k$ are integers ranging from $1$ to $19$ to label the topological excitations. The largest
eigenvalue of $N_{i}$ is the quantum dimension of corresponding excitations, as shown in Table~\ref{tab:qu_dim_AAB_Z2Z2Z2}.
We notice that the quantum dimension of topological excitation is exactly
the coefficient in the front of corresponding gauge-invariant operator. This fact may imply a connection between the Wilson operator and the fusion space of a topological excitation.
Finally, we note that, non-Abelian fusion
rules are also found in $\left(2+1\right)$D DW gauge theory with Abelian gauge group~\citep{propitius1995topological,PhysRevB.95.035131}, where it is found that for a gauge group $G=\left(\mathbb{Z}_{2}\right)^{3}$,
the fusion rules for particles can be captured by the so-called twisted quantum
double model $D^{\omega_3}(G)$. Yet the present situation in $\left(3+1\right)$D
is different: we still consider a $G=\left(\mathbb{Z}_{2}\right)^{3}$
gauge group as a simple illustration, the algebra of fusion rules is apparently different from that of
$D^{\omega_3}(G)$.

\begin{table*}
\centering
\caption{Complete fusion rules of excitations for
BR topological order with $G=\left(\mathbb{Z}_{2}\right)^{3}$. Numbers $1$ to $19$ are used to label the nineteen different excitations. For excitations labeled by $1$ to $8$, their fusion rules with any other excitations are always Abelian so we call them Abelian excitations. The remaining excitations are dubbed non-Abelian excitations since among all fusion rules only the ones involving them are non-Abelian.
$\mathbf{Ab}$ denotes the direct sum of all the Abelian excitations: $\mathbf{Ab}\equiv\mathsf{1}\oplus\mathsf{P}_{100}\oplus\mathsf{P}_{010}\oplus\mathsf{L}_{001}\oplus\mathsf{P}_{110}\oplus\mathsf{L}_{001}^{100}\oplus\mathsf{L}_{001}^{010}\oplus\mathsf{L}_{001}^{110}$.
All these fusion rules are obtained using field-theoretical approach. Some of them are explained as examples in Sec.~\ref{subsec:example_fusion_rules} and Appendix~\ref{appendix:detail_fusion_example}.
\label{tab:fusion_AAB_Z2Z2Z2_full}}
\resizebox{\textwidth}{!}{
%%\begin{tabular*}{\textwidth}{@{\extracolsep{\fill}}|c|c|c|c|c|c|c|c|c|c|c|c|c|c|c|c|c|c|c|c|c|}
\begin{tabular}{|c|c|c|c|c|c|c|c|c|c|c|c|c|c|c|c|c|c|c|c|c|}
\hline

\hline

\hline
  &  & \multicolumn{8}{c|}{\color{blue}\textbf{Abelian excitations}} & \multicolumn{11}{c|}{\color{blue}\textbf{non-Abelian excitations}}\tabularnewline
\hline
 & $\otimes$ & {\color{blue}$\mathsf{1}$} & {\color{blue}$\mathsf{P}_{100}$} & {\color{blue}$\mathsf{P}_{010}$} & {\color{blue}$\mathsf{L}_{001}$} & {\color{blue}$\mathsf{P}_{110}$} & {\color{blue}$\mathsf{L}_{001}^{100}$} & {\color{blue}$\mathsf{L}_{001}^{010}$} &{\color{blue} $\mathsf{L}_{001}^{110}$} & {\color{blue}$\mathsf{L}_{100}$ }& {\color{blue}$\mathsf{L}_{100}^{100}$ }& {\color{blue}$\mathsf{L}_{010}$} &{\color{blue} $\mathsf{L}_{010}^{010}$ }&{\color{blue} $\mathsf{P}_{001}$ }&{\color{blue} $\mathsf{L}_{001}^{001}$ }&{\color{blue} $\mathsf{L}_{110}$ }&{\color{blue} $\mathsf{L}_{110}^{100}$ }& {\color{blue}$\mathsf{L}_{100}^{001}$ }&{\color{blue} $\mathsf{L}_{010}^{001}$ }&{\color{blue} $\mathsf{L}_{110}^{001}$}\tabularnewline
\hline
1 & {\color{blue}$\mathsf{1}$} & $\mathsf{1}$ & $\mathsf{P}_{100}$ & $\mathsf{P}_{010}$ & $\mathsf{L}_{001}$ & $\mathsf{P}_{110}$ & $\mathsf{L}_{001}^{100}$ & $\mathsf{L}_{001}^{010}$ & $\mathsf{L}_{001}^{110}$ & $\mathsf{L}_{100}$ & $\mathsf{L}_{100}^{100}$ & $\mathsf{L}_{010}$ & $\mathsf{L}_{010}^{010}$ & $\mathsf{P}_{001}$ & $\mathsf{L}_{001}^{001}$ & $\mathsf{L}_{110}$ & $\mathsf{L}_{110}^{100}$ & $\mathsf{L}_{100}^{001}$ & $\mathsf{L}_{010}^{001}$ & $\mathsf{L}_{110}^{001}$\tabularnewline
\hline
2 & {\color{blue}$\mathsf{P}_{100}$} & $\mathsf{P}_{100}$ & $\mathsf{1}$ & $\mathsf{P}_{110}$ & $\mathsf{L}_{001}^{100}$ & $\mathsf{P}_{010}$ & $\mathsf{L}_{001}$ & $\mathsf{L}_{001}^{110}$ & $\mathsf{L}_{001}^{010}$ & $\mathsf{L}_{100}^{100}$ & $\mathsf{L}_{100}$ & $\mathsf{L}_{010}$ & $\mathsf{L}_{010}^{010}$ & $\mathsf{P}_{001}$ & $\mathsf{L}_{001}^{001}$ & $\mathsf{L}_{110}^{100}$ & $\mathsf{L}_{110}$ & $\mathsf{L}_{100}^{001}$ & $\mathsf{L}_{010}^{001}$ & $\mathsf{L}_{110}^{001}$\tabularnewline
\hline
3 & {\color{blue}$\mathsf{P}_{010}$}& $\mathsf{P}_{010}$ & $\mathsf{P}_{110}$ & $\mathsf{1}$ & $\mathsf{L}_{001}^{010}$ & $\mathsf{P}_{100}$ & $\mathsf{L}_{001}^{110}$ & $\mathsf{L}_{001}$ & $\mathsf{L}_{001}^{100}$ & $\mathsf{L}_{100}$ & $\mathsf{L}_{100}^{100}$ & $\mathsf{L}_{010}^{010}$ & $\mathsf{L}_{010}$ & $\mathsf{P}_{001}$ & $\mathsf{L}_{001}^{001}$ & $\mathsf{L}_{110}^{100}$ & $\mathsf{L}_{110}$ & $\mathsf{L}_{100}^{001}$ & $\mathsf{L}_{010}^{001}$ & $\mathsf{L}_{110}^{001}$\tabularnewline
\hline
4 & {\color{blue}$\mathsf{L}_{001}$}& $\mathsf{L}_{001}$ & $\mathsf{L}_{001}^{100}$ & $\mathsf{L}_{001}^{010}$ & $\mathsf{1}$ & $\mathsf{L}_{001}^{110}$ & $\mathsf{P}_{100}$ & $\mathsf{P}_{010}$ & $\mathsf{P}_{110}$ & $\mathsf{L}_{100}$ & $\mathsf{L}_{100}^{100}$ & $\mathsf{L}_{010}$ & $\mathsf{L}_{010}^{010}$ & $\mathsf{L}_{001}^{001}$ & $\mathsf{P}_{001}$ & $\mathsf{L}_{110}$ & $\mathsf{L}_{110}^{100}$ & $\mathsf{L}_{100}^{001}$ & $\mathsf{L}_{010}^{001}$ & $\mathsf{L}_{110}^{001}$\tabularnewline
\hline
5 & {\color{blue}$\mathsf{P}_{110}$}& $\mathsf{P}_{110}$ & $\mathsf{P}_{010}$ & $\mathsf{P}_{100}$ & $\mathsf{L}_{001}^{110}$ & $\mathsf{1}$ & $\mathsf{L}_{001}^{010}$ & $\mathsf{L}_{001}^{100}$ & $\mathsf{L}_{001}$ & $\mathsf{L}_{100}^{100}$ & $\mathsf{L}_{100}$ & $\mathsf{L}_{010}^{010}$ & $\mathsf{L}_{010}$ & $\mathsf{P}_{001}$ & $\mathsf{L}_{001}^{001}$ & $\mathsf{L}_{110}$ & $\mathsf{L}_{110}^{100}$ & $\mathsf{L}_{100}^{001}$ & $\mathsf{L}_{010}^{001}$ & $\mathsf{L}_{110}^{001}$\tabularnewline
\hline
6 & {\color{blue}$\mathsf{L}_{001}^{100}$} & $\mathsf{L}_{001}^{100}$ & $\mathsf{L}_{001}$ & $\mathsf{L}_{001}^{110}$ & $\mathsf{P}_{100}$ & $\mathsf{L}_{001}^{010}$ & $\mathsf{1}$ & $\mathsf{P}_{110}$ & $\mathsf{P}_{010}$ & $\mathsf{L}_{100}^{100}$ & $\mathsf{L}_{100}$ & $\mathsf{L}_{010}$ & $\mathsf{L}_{010}^{010}$ & $\mathsf{L}_{001}^{001}$ & $\mathsf{P}_{001}$ & $\mathsf{L}_{110}^{100}$ & $\mathsf{L}_{110}$ & $\mathsf{L}_{100}^{001}$ & $\mathsf{L}_{010}^{001}$ & $\mathsf{L}_{110}^{001}$\tabularnewline
\hline
7 & {\color{blue}$\mathsf{L}_{001}^{010}$} & $\mathsf{L}_{001}^{010}$ & $\mathsf{L}_{001}^{110}$ & $\mathsf{L}_{001}$ & $\mathsf{P}_{010}$ & $\mathsf{L}_{001}^{100}$ & $\mathsf{P}_{110}$ & $\mathsf{1}$ & $\mathsf{P}_{100}$ & $\mathsf{L}_{100}$ & $\mathsf{L}_{100}^{100}$ & $\mathsf{L}_{010}^{010}$ & $\mathsf{L}_{010}$ & $\mathsf{L}_{001}^{001}$ & $\mathsf{P}_{001}$ & $\mathsf{L}_{110}^{100}$ & $\mathsf{L}_{110}$ & $\mathsf{L}_{100}^{001}$ & $\mathsf{L}_{010}^{001}$ & $\mathsf{L}_{110}^{001}$\tabularnewline
\hline
8 & {\color{blue}$\mathsf{L}_{001}^{110}$} & $\mathsf{L}_{001}^{110}$ & $\mathsf{L}_{001}^{010}$ & $\mathsf{L}_{001}^{100}$ & $\mathsf{P}_{110}$ & $\mathsf{L}_{001}$ & $\mathsf{P}_{010}$ & $\mathsf{P}_{100}$ & $\mathsf{1}$ & $\mathsf{L}_{100}^{100}$ & $\mathsf{L}_{100}$ & $\mathsf{L}_{010}^{010}$ & $\mathsf{L}_{010}$ & $\mathsf{L}_{001}^{001}$ & $\mathsf{P}_{001}$ & $\mathsf{L}_{110}$ & $\mathsf{L}_{110}^{100}$ & $\mathsf{L}_{100}^{001}$ & $\mathsf{L}_{010}^{001}$ & $\mathsf{L}_{110}^{001}$\tabularnewline
\hline
9 & {\color{blue}$\mathsf{L}_{100}$} & $\mathsf{L}_{100}$ & $\mathsf{L}_{100}^{100}$ & $\mathsf{L}_{100}$ & $\mathsf{L}_{100}$ & $\mathsf{L}_{100}^{100}$ & $\mathsf{L}_{100}^{100}$ & $\mathsf{L}_{100}$ & $\mathsf{L}_{100}^{100}$ & $\begin{aligned} & \mathsf{1}\\
\oplus & \mathsf{P}_{010}\\
\oplus & \mathsf{L}_{001}\\
\oplus & \mathsf{L}_{001}^{010}
\end{aligned}
$ & $\begin{aligned} & \mathsf{P}_{100}\\
\oplus & \mathsf{P}_{110}\\
\oplus & \mathsf{L}_{001}^{100}\\
\oplus & \mathsf{L}_{001}^{110}
\end{aligned}
$ & $\begin{aligned} & \mathsf{L}_{110}\\
\oplus & \mathsf{L}_{110}^{100}
\end{aligned}
$ & $\begin{aligned} & \mathsf{L}_{110}\\
\oplus & \mathsf{L}_{110}^{100}
\end{aligned}
$ & $\mathsf{L}_{100}^{001}$ & $\mathsf{L}_{100}^{001}$ & $\begin{aligned} & \mathsf{L}_{010}\\
\oplus & \mathsf{L}_{010}^{010}
\end{aligned}
$ & $\begin{aligned} & \mathsf{L}_{010}\\
\oplus & \mathsf{L}_{010}^{010}
\end{aligned}
$ & $\begin{aligned} & 2\cdot\mathsf{P}_{001}\\
\oplus & 2\cdot\mathsf{L}_{001}^{001}
\end{aligned}
$ & $2\cdot\mathsf{L}_{110}^{001}$ & $2\cdot\mathsf{L}_{010}^{001}$\tabularnewline
\hline
10 & {\color{blue}$\mathsf{L}_{100}^{100}$} & $\mathsf{L}_{100}^{100}$ & $\mathsf{L}_{100}$ & $\mathsf{L}_{100}^{100}$ & $\mathsf{L}_{100}^{100}$ & $\mathsf{L}_{100}$ & $\mathsf{L}_{100}$ & $\mathsf{L}_{100}^{100}$ & $\mathsf{L}_{100}$ & $\begin{aligned} & \mathsf{P}_{100}\\
\oplus & \mathsf{P}_{110}\\
\oplus & \mathsf{L}_{001}^{100}\\
\oplus & \mathsf{L}_{001}^{110}
\end{aligned}
$ & $\begin{aligned} & \mathsf{1}\\
\oplus & \mathsf{P}_{010}\\
\oplus & \mathsf{L}_{001}\\
\oplus & \mathsf{L}_{001}^{010}
\end{aligned}
$ & $\begin{aligned} & \mathsf{L}_{110}\\
\oplus & \mathsf{L}_{110}^{100}
\end{aligned}
$ & $\begin{aligned} & \mathsf{L}_{110}\\
\oplus & \mathsf{L}_{110}^{100}
\end{aligned}
$ & $\mathsf{L}_{100}^{001}$ & $\mathsf{L}_{100}^{001}$ & $\begin{aligned} & \mathsf{L}_{010}\\
\oplus & \mathsf{L}_{010}^{010}
\end{aligned}
$ & $\begin{aligned} & \mathsf{L}_{010}\\
\oplus & \mathsf{L}_{010}^{010}
\end{aligned}
$ & $\begin{aligned} & 2\cdot\mathsf{P}_{001}\\
\oplus & 2\cdot\mathsf{L}_{001}^{001}
\end{aligned}
$ & $2\cdot\mathsf{L}_{110}^{001}$ & $2\cdot\mathsf{L}_{010}^{001}$\tabularnewline
\hline
11 & {\color{blue}$\mathsf{L}_{010}$} & $\mathsf{L}_{010}$ & $\mathsf{L}_{010}$ & $\mathsf{L}_{010}^{010}$ & $\mathsf{L}_{010}$ & $\mathsf{L}_{010}^{010}$ & $\mathsf{L}_{010}$ & $\mathsf{L}_{010}^{010}$ & $\mathsf{L}_{010}^{010}$ & $\begin{aligned} & \mathsf{L}_{110}\\
\oplus & \mathsf{L}_{110}^{100}
\end{aligned}
$ & $\begin{aligned} & \mathsf{L}_{110}\\
\oplus & \mathsf{L}_{110}^{100}
\end{aligned}
$ & $\begin{aligned} & \mathsf{1}\\
\oplus & \mathsf{L}_{001}\\
\oplus & \mathsf{P}_{100}\\
\oplus & \mathsf{L}_{001}^{100}
\end{aligned}
$ & $\begin{aligned} & \mathsf{P}_{010}\\
\oplus & \mathsf{L}_{001}^{010}\\
\oplus & \mathsf{P}_{110}\\
\oplus & \mathsf{L}_{001}^{110}
\end{aligned}
$ & $\mathsf{L}_{010}^{001}$ & $\mathsf{L}_{010}^{001}$ & $\begin{aligned} & \mathsf{L}_{100}\\
\oplus & \mathsf{L}_{100}^{100}
\end{aligned}
$ & $\begin{aligned} & \mathsf{L}_{100}\\
\oplus & \mathsf{L}_{100}^{100}
\end{aligned}
$ & $2\cdot\mathsf{L}_{110}^{001}$ & $\begin{aligned} & 2\cdot\mathsf{P}_{001}\\
\oplus & 2\cdot\mathsf{L}_{001}^{001}
\end{aligned}
$ & $2\cdot\mathsf{L}_{100}^{001}$\tabularnewline
\hline
12 & {\color{blue}$\mathsf{L}_{010}^{010}$} & $\mathsf{L}_{010}^{010}$ & $\mathsf{L}_{010}^{010}$ & $\mathsf{L}_{010}$ & $\mathsf{L}_{010}^{010}$ & $\mathsf{L}_{010}$ & $\mathsf{L}_{010}^{010}$ & $\mathsf{L}_{010}$ & $\mathsf{L}_{010}$ & $\begin{aligned} & \mathsf{L}_{110}\\
\oplus & \mathsf{L}_{110}^{100}
\end{aligned}
$ & $\begin{aligned} & \mathsf{L}_{110}\\
\oplus & \mathsf{L}_{110}^{100}
\end{aligned}
$ & $\begin{aligned} & \mathsf{P}_{010}\\
\oplus & \mathsf{L}_{001}^{010}\\
\oplus & \mathsf{P}_{110}\\
\oplus & \mathsf{L}_{001}^{110}
\end{aligned}
$ & $\begin{aligned} & \mathsf{1}\\
\oplus & \mathsf{L}_{001}\\
\oplus & \mathsf{P}_{100}\\
\oplus & \mathsf{L}_{001}^{100}
\end{aligned}
$ & $\mathsf{L}_{010}^{001}$ & $\mathsf{L}_{010}^{001}$ & $\begin{aligned} & \mathsf{L}_{100}\\
\oplus & \mathsf{L}_{100}^{100}
\end{aligned}
$ & $\begin{aligned} & \mathsf{L}_{100}\\
\oplus & \mathsf{L}_{100}^{100}
\end{aligned}
$ & $2\cdot\mathsf{L}_{110}^{001}$ & $\begin{aligned} & 2\cdot\mathsf{P}_{001}\\
\oplus & 2\cdot\mathsf{L}_{001}^{001}
\end{aligned}
$ & $2\cdot\mathsf{L}_{100}^{001}$\tabularnewline
\hline
13 & {\color{blue}$\mathsf{P}_{001}$} & $\mathsf{P}_{001}$ & $\mathsf{P}_{001}$ & $\mathsf{P}_{001}$ & $\mathsf{L}_{001}^{001}$ & $\mathsf{P}_{001}$ & $\mathsf{L}_{001}^{001}$ & $\mathsf{L}_{001}^{001}$ & $\mathsf{L}_{001}^{001}$ & $\mathsf{L}_{100}^{001}$ & $\mathsf{L}_{100}^{001}$ & $\mathsf{L}_{010}^{001}$ & $\mathsf{L}_{010}^{001}$ & $\begin{aligned} & \mathsf{1}\\
\oplus & \mathsf{P}_{100}\\
\oplus & \mathsf{P}_{010}\\
\oplus & \mathsf{P}_{110}
\end{aligned}
$ & $\begin{aligned} & \mathsf{L}_{001}\\
\oplus & \mathsf{L}_{001}^{100}\\
\oplus & \mathsf{L}_{001}^{010}\\
\oplus & \mathsf{L}_{001}^{110}
\end{aligned}
$ & $\mathsf{L}_{110}^{001}$ & $\mathsf{L}_{110}^{001}$ & $\begin{aligned} & 2\cdot\mathsf{L}_{100}\\
\oplus & 2\cdot\mathsf{L}_{100}^{100}
\end{aligned}
$ & $\begin{aligned} & 2\cdot\mathsf{L}_{010}\\
\oplus & 2\cdot\mathsf{L}_{010}^{010}
\end{aligned}
$ & $\begin{aligned} & 2\cdot\mathsf{L}_{110}\\
\oplus & 2\cdot\mathsf{L}_{110}^{100}
\end{aligned}
$\tabularnewline
\hline
14 & {\color{blue}$\mathsf{L}_{001}^{001}$} & $\mathsf{L}_{001}^{001}$ & $\mathsf{L}_{001}^{001}$ & $\mathsf{L}_{001}^{001}$ & $\mathsf{P}_{001}$ & $\mathsf{L}_{001}^{001}$ & $\mathsf{P}_{001}$ & $\mathsf{P}_{001}$ & $\mathsf{P}_{001}$ & $\mathsf{L}_{100}^{001}$ & $\mathsf{L}_{100}^{001}$ & $\mathsf{L}_{010}^{001}$ & $\mathsf{L}_{010}^{001}$ & $\begin{aligned} & \mathsf{L}_{001}\\
\oplus & \mathsf{L}_{001}^{100}\\
\oplus & \mathsf{L}_{001}^{010}\\
\oplus & \mathsf{L}_{001}^{110}
\end{aligned}
$ & $\begin{aligned} & \mathsf{1}\\
\oplus & \mathsf{P}_{100}\\
\oplus & \mathsf{P}_{010}\\
\oplus & \mathsf{P}_{110}
\end{aligned}
$ & $\mathsf{L}_{110}^{001}$ & $\mathsf{L}_{110}^{001}$ & $\begin{aligned} & 2\cdot\mathsf{L}_{100}\\
\oplus & 2\cdot\mathsf{L}_{100}^{100}
\end{aligned}
$ & $\begin{aligned} & 2\cdot\mathsf{L}_{010}\\
\oplus & 2\cdot\mathsf{L}_{010}^{010}
\end{aligned}
$ & $\begin{aligned} & 2\cdot\mathsf{L}_{110}\\
\oplus & 2\cdot\mathsf{L}_{110}^{100}
\end{aligned}
$\tabularnewline
\hline
15 & {\color{blue}$\mathsf{L}_{110}$} & $\mathsf{L}_{110}$ & $\mathsf{L}_{110}^{100}$ & $\mathsf{L}_{110}^{100}$ & $\mathsf{L}_{110}$ & $\mathsf{L}_{110}$ & $\mathsf{L}_{110}^{100}$ & $\mathsf{L}_{110}^{100}$ & $\mathsf{L}_{110}$ & $\begin{aligned} & \mathsf{L}_{010}\\
\oplus & \mathsf{L}_{010}^{010}
\end{aligned}
$ & $\begin{aligned} & \mathsf{L}_{010}\\
\oplus & \mathsf{L}_{010}^{010}
\end{aligned}
$ & $\begin{aligned} & \mathsf{L}_{100}\\
\oplus & \mathsf{L}_{100}^{100}
\end{aligned}
$ & $\begin{aligned} & \mathsf{L}_{100}\\
\oplus & \mathsf{L}_{100}^{100}
\end{aligned}
$ & $\mathsf{L}_{110}^{001}$ & $\mathsf{L}_{110}^{001}$ & $\begin{aligned} & \mathsf{1}\\
\oplus & \mathsf{P}_{110}\\
\oplus & \mathsf{L}_{001}\\
\oplus & \mathsf{L}_{001}^{110}
\end{aligned}
$ & $\begin{aligned} & \mathsf{P}_{100}\\
\oplus & \mathsf{P}_{010}\\
\oplus & \mathsf{L}_{001}^{100}\\
\oplus & \mathsf{L}_{001}^{010}
\end{aligned}
$ & $2\cdot\mathsf{L}_{010}^{001}$ & $2\cdot\mathsf{L}_{100}^{001}$ & $\begin{aligned} & 2\cdot\mathsf{P}_{001}\\
\oplus & 2\cdot\mathsf{L}_{001}^{001}
\end{aligned}
$\tabularnewline
\hline
16 & {\color{blue}$\mathsf{L}_{110}^{100}$} & $\mathsf{L}_{110}^{100}$ & $\mathsf{L}_{110}$ & $\mathsf{L}_{110}$ & $\mathsf{L}_{110}^{100}$ & $\mathsf{L}_{110}^{100}$ & $\mathsf{L}_{110}$ & $\mathsf{L}_{110}$ & $\mathsf{L}_{110}^{100}$ & $\begin{aligned} & \mathsf{L}_{010}\\
\oplus & \mathsf{L}_{010}^{010}
\end{aligned}
$ & $\begin{aligned} & \mathsf{L}_{010}\\
\oplus & \mathsf{L}_{010}^{010}
\end{aligned}
$ & $\begin{aligned} & \mathsf{L}_{100}\\
\oplus & \mathsf{L}_{100}^{100}
\end{aligned}
$ & $\begin{aligned} & \mathsf{L}_{100}\\
\oplus & \mathsf{L}_{100}^{100}
\end{aligned}
$ & $\mathsf{L}_{110}^{001}$ & $\mathsf{L}_{110}^{001}$ & $\begin{aligned} & \mathsf{P}_{100}\\
\oplus & \mathsf{P}_{010}\\
\oplus & \mathsf{L}_{001}^{100}\\
\oplus & \mathsf{L}_{001}^{010}
\end{aligned}
$ & $\begin{aligned} & \mathsf{1}\\
\oplus & \mathsf{P}_{110}\\
\oplus & \mathsf{L}_{001}\\
\oplus & \mathsf{L}_{001}^{110}
\end{aligned}
$ & $2\cdot\mathsf{L}_{010}^{001}$ & $2\cdot\mathsf{L}_{100}^{001}$ & $\begin{aligned} & 2\cdot\mathsf{P}_{001}\\
\oplus & 2\cdot\mathsf{L}_{001}^{001}
\end{aligned}
$\tabularnewline
\hline
17 & {\color{blue}$\mathsf{L}_{100}^{001}$} & $\mathsf{L}_{100}^{001}$ & $\mathsf{L}_{100}^{001}$ & $\mathsf{L}_{100}^{001}$ & $\mathsf{L}_{100}^{001}$ & $\mathsf{L}_{100}^{001}$ & $\mathsf{L}_{100}^{001}$ & $\mathsf{L}_{100}^{001}$ & $\mathsf{L}_{100}^{001}$ & $\begin{aligned} & 2\cdot\mathsf{P}_{001}\\
\oplus & 2\cdot\mathsf{L}_{001}^{001}
\end{aligned}
$ & $\begin{aligned} & 2\cdot\mathsf{P}_{001}\\
\oplus & 2\cdot\mathsf{L}_{001}^{001}
\end{aligned}
$ & $2\cdot\mathsf{L}_{110}^{001}$ & $2\cdot\mathsf{L}_{110}^{001}$ & $\begin{aligned} & 2\cdot\mathsf{L}_{100}\\
\oplus & 2\cdot\mathsf{L}_{100}^{100}
\end{aligned}
$ & $\begin{aligned} & 2\cdot\mathsf{L}_{100}\\
\oplus & 2\cdot\mathsf{L}_{100}^{100}
\end{aligned}
$ & $2\cdot\mathsf{L}_{010}^{001}$ & $2\cdot\mathsf{L}_{010}^{001}$ & $2\cdot\mathbf{Ab}$ & $\begin{aligned} & 4\cdot\mathsf{L}_{110}\\
\oplus & 4\cdot\mathsf{L}_{110}^{100}
\end{aligned}
$ & $\begin{aligned} & 4\cdot\mathsf{L}_{010}\\
\oplus & 4\cdot\mathsf{L}_{010}^{010}
\end{aligned}
$\tabularnewline
\hline
18 & {\color{blue}$\mathsf{L}_{010}^{001}$} & $\mathsf{L}_{010}^{001}$ & $\mathsf{L}_{010}^{001}$ & $\mathsf{L}_{010}^{001}$ & $\mathsf{L}_{010}^{001}$ & $\mathsf{L}_{010}^{001}$ & $\mathsf{L}_{010}^{001}$ & $\mathsf{L}_{010}^{001}$ & $\mathsf{L}_{010}^{001}$ & $2\cdot\mathsf{L}_{110}^{001}$ & $2\cdot\mathsf{L}_{110}^{001}$ & $\begin{aligned} & 2\cdot\mathsf{P}_{001}\\
\oplus & 2\cdot\mathsf{L}_{001}^{001}
\end{aligned}
$ & $\begin{aligned} & 2\cdot\mathsf{P}_{001}\\
\oplus & 2\cdot\mathsf{L}_{001}^{001}
\end{aligned}
$ & $\begin{aligned} & 2\cdot\mathsf{L}_{010}\\
\oplus & 2\cdot\mathsf{L}_{010}^{010}
\end{aligned}
$ & $\begin{aligned} & 2\cdot\mathsf{L}_{010}\\
\oplus & 2\cdot\mathsf{L}_{010}^{010}
\end{aligned}
$ & $2\cdot\mathsf{L}_{100}^{001}$ & $2\cdot\mathsf{L}_{100}^{001}$ & $\begin{aligned} & 4\cdot\mathsf{L}_{110}\\
\oplus & 4\cdot\mathsf{L}_{110}^{100}
\end{aligned}
$ & $2\cdot\mathbf{Ab}$ & $\begin{aligned} & 4\cdot\mathsf{L}_{100}\\
\oplus & 4\cdot\mathsf{L}_{100}^{100}
\end{aligned}
$\tabularnewline
\hline
19 & {\color{blue}$\mathsf{L}_{110}^{001}$} & $\mathsf{L}_{110}^{001}$ & $\mathsf{L}_{110}^{001}$ & $\mathsf{L}_{110}^{001}$ & $\mathsf{L}_{110}^{001}$ & $\mathsf{L}_{110}^{001}$ & $\mathsf{L}_{110}^{001}$ & $\mathsf{L}_{110}^{001}$ & $\mathsf{L}_{110}^{001}$ & $2\cdot\mathsf{L}_{010}^{001}$ & $2\cdot\mathsf{L}_{010}^{001}$ & $2\cdot\mathsf{L}_{100}^{001}$ & $2\cdot\mathsf{L}_{100}^{001}$ & $\begin{aligned} & 2\cdot\mathsf{L}_{110}\\
\oplus & 2\cdot\mathsf{L}_{110}^{100}
\end{aligned}
$ & $\begin{aligned} & 2\cdot\mathsf{L}_{110}\\
\oplus & 2\cdot\mathsf{L}_{110}^{100}
\end{aligned}
$ & $\begin{aligned} & 2\cdot\mathsf{P}_{001}\\
\oplus & 2\cdot\mathsf{L}_{001}^{001}
\end{aligned}
$ & $\begin{aligned} & 2\cdot\mathsf{P}_{001}\\
\oplus & 2\cdot\mathsf{L}_{001}^{001}
\end{aligned}
$ & $\begin{aligned} & 4\cdot\mathsf{L}_{010}\\
\oplus & 4\cdot\mathsf{L}_{010}^{010}
\end{aligned}
$ & $\begin{aligned} & 4\cdot\mathsf{L}_{100}\\
\oplus & 4\cdot\mathsf{L}_{100}^{100}
\end{aligned}
$ & $2\cdot\mathbf{Ab}$\tabularnewline
\hline

\hline

\hline

\end{tabular}}

\end{table*}

\begin{table*}
\caption{Quantum dimensions of operators in BR topological order
with $G=\left(\mathbb{Z}_{2}\right)^{3}$. By definition, quantum
dimension is the largest eigenvalue of the matrix $N_{i}$ whose element
is given by $\left(N_{i}\right)_{kj}=N_{k}^{ij}$. Topological excitations with quantum dimension larger than $1$ are non-Abelian, as indicated by multi-channel fusion rules in Table.~\ref{tab:fusion_AAB_Z2Z2Z2_full}. \label{tab:qu_dim_AAB_Z2Z2Z2}}
\centering
\begin{tabular*}{\textwidth}{@{\extracolsep{\fill}}cccccccccccccccccccc}
\hline

\hline

\hline
\textbf{Wilson} \textbf{Operators} & $\mathsf{1}$ & $\mathsf{P}_{100}$ & $\mathsf{P}_{010}$ & $\mathsf{L}_{001}$ & $\mathsf{P}_{110}$ & $\mathsf{L}_{001}^{100}$ & $\mathsf{L}_{001}^{010}$ & $\mathsf{L}_{001}^{110}$ & $\mathsf{L}_{100}$ & $\mathsf{L}_{100}^{100}$ & $\mathsf{L}_{010}$ & $\mathsf{L}_{010}^{010}$ & $\mathsf{P}_{001}$ & $\mathsf{L}_{001}^{001}$ & $\mathsf{L}_{110}$ & $\mathsf{L}_{110}^{100}$ & $\mathsf{L}_{100}^{001}$ & $\mathsf{L}_{010}^{001}$ & $\mathsf{L}_{110}^{001}$\tabularnewline
\hline
\textbf{Quantum dimension} & $1$ & $1$ & $1$ & $1$ & $1$ & $1$ & $1$ & $1$ & $2$ & $2$ & $2$ & $2$ & $2$ & $2$ & $2$ & $2$ & $4$ & $4$ & $4$\tabularnewline
\hline

\hline

\hline
\end{tabular*}
\end{table*}

\subsection{Loop-shrinking rules, consistency, and anomaly}\label{subsec:shrinking_rule}

Since the loop excitations we consider in this paper are not linked
with other loops, a loop can be shrunk to a point that turns out to correspond a (or several) particle excitation. This feature is absent in $\left(2+1\right)$D topological orders yet is important in higher-dimensional cases. The loop shrinking operation may be important when we consider the dimension reduction of topological order.
In this section, we show that the loop shrinking operation
can   be represented in the framework of TQFT. The Wilson operators studied in Sec.~\ref{sec:operator} help to provide a general algorithm to understand shrinking operation
of loop excitations in $3$D space. The loop shrinking rules may also be an important
characterization for $3$D topological orders.

\begin{figure}
  \centering
  \includegraphics[scale=0.4]{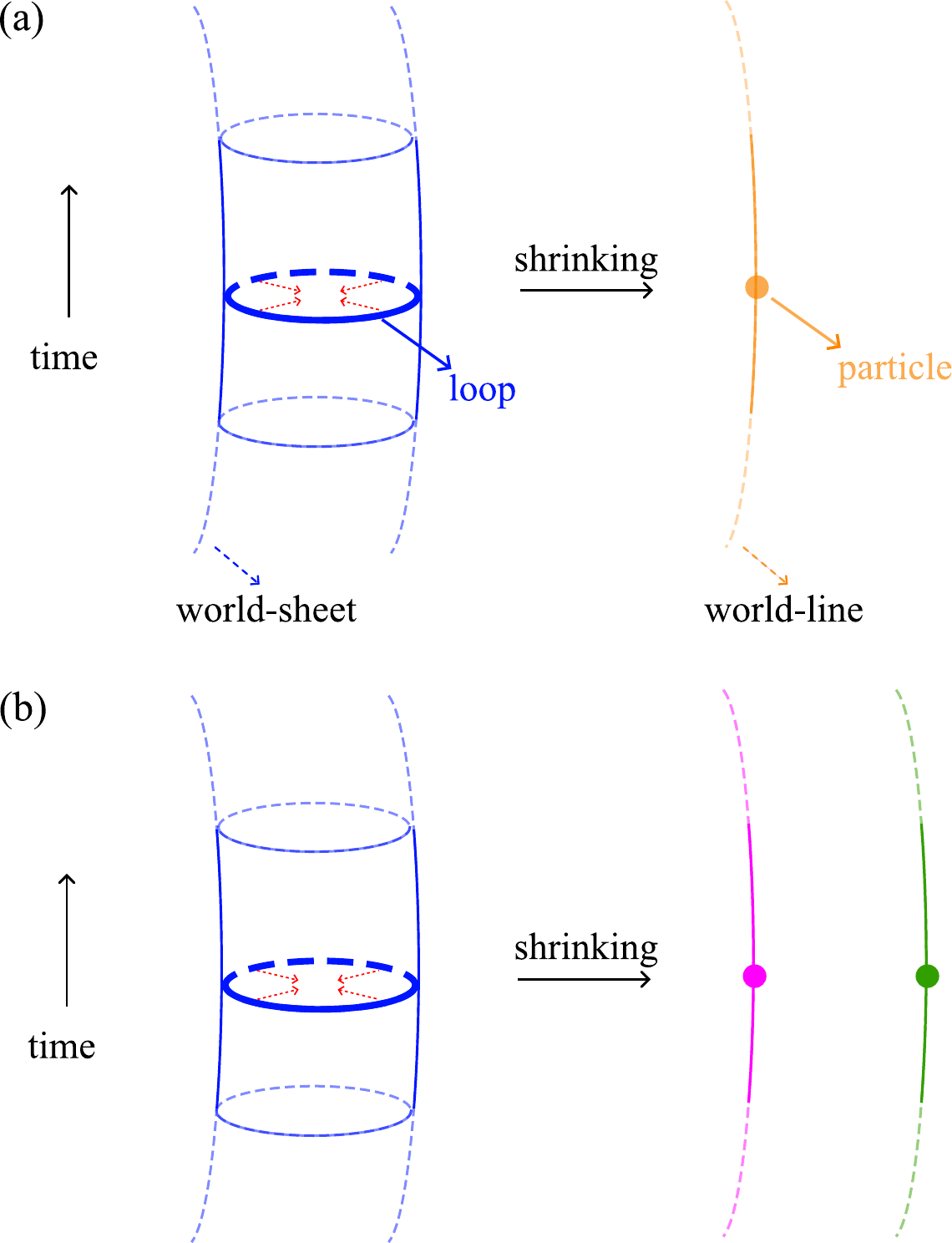}
  \caption{Illustration of the loop-shrinking operation. When a loop is shrunk to a particle, its world-sheet is shrunk to a line which turns out to be the world-line of the particle. In this situation, the integral of $2$-form gauge field in Wilson operator for loop excitation naturally vanishes, see Eq.~(\ref{eq:loop_shrink_formula}) and the main text. (a) The outcome of loop shrinking operation can be a single particle. As shown in Table~\ref{tab:shrinking rules_AAB_z2z2z2}, shrinking an Abelian loop always results in one particle excitation. (b) The outcome of loop shrinking operation can also be a composite formed by  multiple particles, e.g., the example of Eq.~(\ref{eq:shrinking_L_100_example}) discussed in the main text.}
  \label{fig:loop_shrinking_pic}
\end{figure}

Back to our work, how to represent this shrinking operation in terms
of gauge-invariant operators and path integral? Since   the
world-sheet of a loop would contract to a world-line after the loop shrinking operation,
we conjecture that the shrinking operation can be represented in the
path integral by shrinking the world-sheet to a closed curve that
can be viewed as a world-line of particle, as illustrated in Fig.~\ref{fig:loop_shrinking_pic}. In details, the world-sheet
$\sigma$ is a $2$-torus $T^{2}$, the shrinking operation is taking
the limit $T^{2}\rightarrow S^{1}$ where $S^{1}$ is the non-contractible
path circling along time direction on $T^{2}$. Let $\mathcal{S}$
be the shrinking operation for loop excitations. For example, if we
consider to shrink a $\mathbb{Z}_{N_{1}}$-loop, $\mathsf{L}_{100}$,
we can write down (we still consider $G=\left(\mathbb{Z}_{2}\right)^{3}$)
\begin{widetext}
\begin{align}
\left\langle \mathcal{S}\left(\mathsf{L}_{100}\right)\right\rangle =\left\langle \lim_{\sigma\rightarrow S^{1}}\mathsf{L}_{100}\right\rangle = & \lim_{\sigma\rightarrow S^{1}}\frac{1}{\mathcal{Z}}\int\mathcal{D}\left[A^{1}\right]\mathcal{D}\left[B^{i}\right]\exp\left({\rm i}S\right)\nonumber \\
 & \times2\exp\left[{\rm i}\int_{\sigma}B^{1}+\frac{1}{2}\frac{2\pi q}{N_{1}}\left(d^{-1}A^{2}B^{3}+d^{-1}B^{3}A^{2}\right)\right]\delta\left(\int_{\gamma}A^{2}\right)\delta\left(\int_{\sigma}B^{3}\right)\nonumber \\
= & \frac{1}{\mathcal{Z}}\int\mathcal{D}\left[A^{1}\right]\mathcal{D}\left[B^{i}\right]\exp\left({\rm i}S\right)\times2\exp\left({\rm i}0\right)\times\delta\left(\int_{\gamma}A^{2}\right)\delta\left(0\right)\nonumber \\
= & \frac{1}{\mathcal{Z}}\int\mathcal{D}\left[A^{1}\right]\mathcal{D}\left[B^{i}\right]\exp\left({\rm i}S\right)\times2\times\delta\left(\int_{\gamma}A^{2}\right)\times 1\nonumber \\
= & \frac{1}{\mathcal{Z}}\int\mathcal{D}\left[A^{1}\right]\mathcal{D}\left[B^{i}\right]\exp\left({\rm i}S\right)\times\left[1+\exp\left({\rm i}\int_{\gamma}A^{2}\right)\right]\nonumber \\
= & \left\langle \mathsf{1}\oplus\mathsf{P}_{010}\right\rangle.
\label{eq:loop_shrink_formula}
\end{align}
\end{widetext}
So we can claim that the $\mathbb{Z}_{N_{1}}$-loop can be shrunk
into the superposition of a trivial particle (vacuum) and a $\mathbb{Z}_{N_{2}}$-particle:
\begin{equation}
\mathcal{S}\left(\mathsf{L}_{100}\right)=\mathsf{1}\oplus\mathsf{P}_{010}.
\label{eq:shrinking_L_100_example}
\end{equation}
This loop shrinking rule (\ref{eq:shrinking_L_100_example}) indicates that one would obtain a superposition of a trivial particle and a particle carrying one unit of $\Z_{N_2}$ gauge charge after shrinking the loop $\mathsf{L}_{100}$.
Similarly, we can obtain shrinking rules for all loop excitations, as shown
in Table~\ref{tab:shrinking rules_AAB_z2z2z2}.

\begin{table*}
\centering
\caption{Shrinking rules of topological excitations for BR topological
order with $G=\left(\mathbb{Z}_{2}\right)^{3}$. The loop excitations are classified as Abelian and non-Abelian ones, depending on whether they have Abelian or non-Abelian fusion rules with other topological excitations. All these shrinking rules respect fusion rules, i.e., $\mathcal{S}\left(a\otimes b\right)=\mathcal{S}\left(a\right)\otimes\mathcal{S}\left(b\right)$,
as explained in Sec.~\ref{subsec:shrinking_rule}. Shrinking an Abelian loop always results in an Abelian particle. On the other hand, shrinking a non-Abelian loop leads to either a non-Abelian particle or a composite particle (superposition of multiple simple particles).
\label{tab:shrinking rules_AAB_z2z2z2}}
\begin{tabular*}{\textwidth}{@{\extracolsep{\fill}}ccccccccccccccc}
\hline

\hline

\hline
\textbf{Abelian loops} & $\mathsf{L}_{001}$ & $\mathsf{L}_{001}^{100}$ & $\mathsf{L}_{001}^{010}$ & $\mathsf{L}_{001}^{110}$ &  &  &  &  &  & \tabularnewline
\tabularnewline
$\mathcal{S}\left(\textbf{Abelian loop}\right)$ & $\mathsf{1}$ & $\mathsf{P}_{100}$ & $\mathsf{P}_{010}$ & $\mathsf{P}_{110}$ &  &  &  &  &  & \tabularnewline

\hline
\textbf{non-Abelian loops} & $\mathsf{L}_{100}$ & $\mathsf{L}_{100}^{100}$ & $\mathsf{L}_{010}$ & $\mathsf{L}_{010}^{010}$ & $\mathsf{L}_{001}^{001}$ & $\mathsf{L}_{110}$ & $\mathsf{L}_{110}^{100}$ & $\mathsf{L}_{100}^{001}$ & $\mathsf{L}_{010}^{001}$ & $\mathsf{L}_{110}^{001}$
\tabularnewline
\tabularnewline

$ \mathcal{S}\left(\textbf{non-Abelian loop}\right)$ & $\mathsf{1}\oplus\mathsf{P}_{010}$ & $\mathsf{P}_{100}\oplus\mathsf{P}_{110}$ & $\mathsf{1}\oplus\mathsf{P}_{100}$ & $\mathsf{P}_{010}\oplus\mathsf{P}_{110}$ & $\mathsf{P}_{001}$ & $\mathsf{1}\oplus\mathsf{P}_{110}$ & $\mathsf{P}_{100}\oplus\mathsf{P}_{010}$ & $2\cdot\mathsf{P}_{001}$ & $2\cdot\mathsf{P}_{001}$ & $2\cdot\mathsf{P}_{001}$\tabularnewline
\hline

\hline

\hline
\end{tabular*}
\end{table*}
Physically, 
one may expect that the shrinking operation should respect the fusion
rules:
\begin{equation}
\mathcal{S}\left(\mathsf{a}\otimes \mathsf{b}\right)=\mathcal{S}\left(\mathsf{a}\right)\otimes\mathcal{S}\left(\mathsf{b}\right)
\label{eq:shrinking_condition}
\end{equation}
where $\mathsf{a}$ and $\mathsf{b}$ are excitations. It is natural to set $\mathcal{S}\left(\mathsf{a}\right)=\mathsf{a}$
if $\mathsf{a}$ is a particle excitation. Analogous to fusion rules, we can
write the shrinking rules in the form of
\begin{equation}
\mathcal{S}\left(\mathsf{a}\right)=\oplus_{\mathsf{c}}\mathrm{S}_{\mathsf{c}}^{\mathsf{a}}\cdot\mathsf{c}
\end{equation}
where the non-zero integer $\mathrm{S}_{\mathsf{c}}^{\mathsf{a}}$
behaves as the ``shrinking coefficient''. Using this notation, we
have
\begin{align}
\mathcal{S}\left(\mathsf{a}\otimes\mathsf{b}\right)= & \mathcal{S}\left(\oplus_{\mathsf{c}}N_{\mathsf{c}}^{\mathsf{a}\mathsf{b}}\cdot\mathsf{c}\right)\nonumber \\
= & \sum_{\mathsf{c}}N_{\mathsf{c}}^{\mathsf{a}\mathsf{b}}\cdot\mathcal{S}\left(\mathsf{c}\right)\nonumber \\
= & \sum_{\mathsf{c}}N_{\mathsf{c}}^{\mathsf{a}\mathsf{b}}\cdot\oplus_{\mathsf{d}}\mathrm{S}_{\mathsf{d}}^{\mathsf{c}}\cdot\mathsf{d}\nonumber \\
= & \oplus_{\mathsf{d}}\left(\sum_{\mathsf{c}}N_{\mathsf{c}}^{\mathsf{a}\mathsf{b}}\mathrm{S}_{\mathsf{d}}^{\mathsf{c}}\right)\cdot\mathsf{d}
\end{align}
and
\begin{align}
\mathcal{S}\left(\mathsf{a}\right)\otimes\mathcal{S}\left(\mathsf{b}\right)= & \left(\oplus_{\mathsf{k_{1}}}\mathrm{S}_{\mathsf{k_{1}}}^{\mathsf{a}}\cdot\mathsf{k_{1}}\right)\otimes\left(\oplus_{\mathsf{k_{2}}}\mathrm{S}_{\mathsf{k_{2}}}^{\mathsf{b}}\cdot\mathsf{k_{2}}\right)\nonumber \\
= & \oplus_{\mathsf{k_{1}}}\mathrm{S}_{\mathsf{k_{1}}}^{\mathsf{a}}\cdot\oplus_{\mathsf{k_{2}}}\mathrm{S}_{\mathsf{k_{2}}}^{\mathsf{b}}\cdot\left(\mathsf{k_{1}}\otimes\mathsf{k_{2}}\right)\nonumber \\
= & \sum_{\mathsf{k_{1}}}\sum_{\mathsf{k_{2}}}\mathrm{S}_{\mathsf{k_{1}}}^{\mathsf{a}}\mathrm{S}_{\mathsf{k_{2}}}^{\mathsf{b}}\left(\oplus_{\mathsf{d}}N_{\mathsf{d}}^{\mathsf{k_{1}}\mathsf{k_{2}}}\cdot\mathsf{d}\right)\nonumber \\
= & \oplus_{\mathsf{d}}\left(\sum_{\mathsf{k_{1}},\mathsf{k_{2}}}\mathrm{S}_{\mathsf{k_{1}}}^{\mathsf{a}}\mathrm{S}_{\mathsf{k_{2}}}^{\mathsf{b}}N_{\mathsf{d}}^{\mathsf{k_{1}}\mathsf{k_{2}}}\right)\cdot\mathsf{d}.
\end{align}
By calculating the $N_{k}^{ij}$ and $\mathrm{S}_{c}^{a}$ data from
Table~\ref{tab:fusion_AAB_Z2Z2Z2_full} and Table~\ref{tab:shrinking rules_AAB_z2z2z2}, we confirm that for arbitrary excitations $\mathsf{a}$ and $\mathsf{b}$
\begin{equation}
\sum_{\mathsf{c}}N_{\mathsf{c}}^{\mathsf{a}\mathsf{b}}\mathrm{S}_{\mathsf{d}}^{\mathsf{c}}=\sum_{\mathsf{k_{1}},\mathsf{k_{2}}}\mathrm{S}_{\mathsf{k_{1}}}^{\mathsf{a}}\mathrm{S}_{\mathsf{k_{2}}}^{\mathsf{b}}N_{\mathsf{d}}^{\mathsf{k_{1}}\mathsf{k_{2}}}
\end{equation}
is always satisfied, i.e., the shrinking rules respect the fusion
rules as Eq.~(\ref{eq:shrinking_condition}).

Furthermore, by taking a closer look at the loop shrinking rules in Table~\ref{tab:shrinking rules_AAB_z2z2z2}, we can conclude the following facts from our field-theoretical analysis:
\begin{enumerate}
\item The quantum dimensions of topological excitations are conserved under
loop shrinking operation. This can be checked by referring to Table~\ref{tab:qu_dim_AAB_Z2Z2Z2}.
\item An Abelian loop is always shrunk into an Abelian particle. On the
other hand, a non-Abelian loop is shrunk into either a non-Abelian
particle or a composite particle.
\item The loop shrinking rules is consistent with fusion rules, i.e., $\mathcal{S}\left(\mathsf{a}\otimes\mathsf{b}\right)=\mathcal{S}\left(\mathsf{a}\right)\otimes\mathcal{S}\left(\mathsf{b}\right)$.
\end{enumerate}
All these facts indicate the consistency of fusion rules and loop
shrinking rules. We believe that this consistency of fusion rules and loop shrinking rules play an important role in establishing an anomaly-free topological order. A quantum anomaly may occur if the loop shrinking rules conflict with fusion rules, which is an interesting future direction for field theory study.

\section{Fusion rules of topological orders with compatible braidings in $3$D
space\label{sec:compatible_fusion} }

Though braiding processes in topological order can be described by TQFT in a unified framework, not all of them can be supported in one system without incompatibility \citep{zhang2021compatible}. For a given gauge group, different topological order can be obtained according to different combinations of compatible braiding processes. In this section we would like to answer this question: how the fusion rules of topological order \emph{differ} depending on the combination of compatible braidings.

In $3$D space, nontrivial braiding processes in $3$D
space include particle-loop braiding, multi-loop braiding, and Borromean
rings braiding. Yet not arbitrary combination of these braiding processes
are compatible. For example, when the gauge group is $G=\prod_{i=1}^{3}\mathbb{Z}_{N_{i}}$,
the Borromean rings braiding can be compatible with three-loop braiding,
if the loops in $3$-loop braiding only carry two kinds of $\mathbb{Z}_{N_{i}}$
fluxes. If these three loops carry \emph{three} kinds\emph{ }of $\mathbb{Z}_{N_{i}}$
fluxes, the $3$-loop braiding is not compatible with the Borromean
rings braiding. The origin of this incompatibility is that we cannot
construct a gauge-invariant TQFT action for these two braiding processes~\citep{zhang2021compatible}.

Below, we study the fusion rules for topological order with particle-loop
braiding only, with particle-loop braiding and $3$-loop braiding,
and with all three kinds of braiding processes, respectively. We find
that for topological orders with particle-loop braidings and $3$-loop
braidings only, the fusion rules are Abelian. However, once we introduce
Borromean rings braiding, the fusion rules become non-Abelian.

\subsection{Topological order with particle-loop braiding only: $G=\left(\mathbb{Z}_{2}\right)^{2}$}\label{subsec:fusion_BF_z2z2}

The TQFT action for topological order with particle-loop braiding
only is
\begin{equation}
S=S_{BF}=\int\sum_{i=1}^{2}\frac{N_{i}}{2\pi}B^{i}dA^{i}.\label{eq:BdA_z2z2}
\end{equation}
Since $G=\left(\mathbb{Z}_{2}\right)^{2}$, $N_{1}=N_{2}=2$. The
gauge transformations are
\begin{align}
A^{i}\rightarrow & A^{i}+d\chi^{i},\\
B^{i}\rightarrow & B^{i}+dV^{i}.\nonumber
\end{align}
 Particle excitations are represented by operators
\begin{equation}
\mathsf{P}_{ij}=\exp\left({\rm i}i\int_{\gamma}A^{1}+{\rm i}j\int_{\gamma}A^{2}\right)
\end{equation}
with $i,j=0,1$. Loop excitations are represented by
\begin{equation}
\mathsf{L}_{10}^{ij}=\exp\left({\rm i}\int_{\sigma}B^{1}+{\rm i}i\int_{\gamma}A^{1}+{\rm i}j\int_{\gamma}A^{2}\right),
\end{equation}
\begin{equation}
\mathsf{L}_{01}^{ij}=\exp\left({\rm i}\int_{\sigma}B^{2}+{\rm i}i\int_{\gamma}A^{1}+{\rm i}j\int_{\gamma}A^{2}\right),
\end{equation}
\begin{equation}
\mathsf{L}_{11}^{ij}=\exp\left({\rm i}\int_{\sigma}B^{1}+{\rm i}\int_{\sigma}B^{2}+{\rm i}i\int_{\gamma}A^{1}+{\rm i}j\int_{\gamma}A^{2}\right),
\end{equation}
with $i,j=0,1$. The fusion rules for these excitations are summarized
in Table~\ref{tab:fusion_BdA_Z2Z2}. These fusion rules, being Abelian, form a $\left(\mathbb{Z}_{2}\right)^{4}$ group.

\begin{table*}
\centering
\caption{Fusion table for $S=S_{BF}$, $S=S_{BF}+S_{A^1 A^2 dA^2}$, or $S=S_{BF}+S_{A^1 A^2 dA^2}+S_{A^2 A^1 dA^1}$ with $G=\mathbb{Z}_{2}\times\mathbb{Z}_{2}$. For different TQFT actions (see Sec.~\ref{subsec:fusion_BF_z2z2},~\ref{subsec:fusion_BF_AAdA_z2z2} and~\ref{subsec:fusion_two_AAdA_z2z2}), the same notation $\mathsf{P}_{n_1 n_2}$ or $\mathsf{L}_{n_1 n_2}^{c_1 c_2}$ represents the same excitation though the expression of Wilson operator depends on the gauge transformations. For example, $\mathsf{L}_{10}$ denotes the loop carrying $\Z_{N_1}$ flux yet its explicit operator expression varies for different TQFT actions, as shown in Sec.~\ref{subsec:fusion_BF_z2z2},~\ref{subsec:fusion_BF_AAdA_z2z2} and~\ref{subsec:fusion_two_AAdA_z2z2}. All fusion rules are Abelian hence the quantum dimension of each excitation
is $1$. The fusions of all $16$ excitations form a $\left(\mathbb{Z}_{2}\right)^{4}$ group.
\label{tab:fusion_BdA_Z2Z2}}
\begin{tabular*}{\textwidth}{@{\extracolsep{\fill}}|c|c|c|c|c|c|c|c|c|c|c|c|c|c|c|c|c|}
\hline

\hline

\hline

 & $\mathsf{P}_{00}\equiv\mathsf{1}$ & $\mathsf{P}_{10}$ & $\mathsf{P}_{01}$ & $\mathsf{P}_{11}$ & $\mathsf{L}_{10}$ & $\mathsf{L}_{01}$ & $\mathsf{L}_{11}$ & $\mathsf{L}_{10}^{10}$ & $\mathsf{L}_{01}^{10}$ & $\mathsf{L}_{11}^{10}$ & $\mathsf{L}_{10}^{01}$ & $\mathsf{L}_{01}^{01}$ & $\mathsf{L}_{11}^{01}$ & $\mathsf{L}_{10}^{11}$ & $\mathsf{L}_{01}^{11}$ & $\mathsf{L}_{11}^{11}$\tabularnewline
\hline
$\mathsf{1}$ & $\mathsf{1}$ & $\mathsf{P}_{10}$ & $\mathsf{P}_{01}$ & $\mathsf{P}_{11}$ & $\mathsf{L}_{10}$ & $\mathsf{L}_{01}$ & $\mathsf{L}_{11}$ & $\mathsf{L}_{10}^{10}$ & $\mathsf{L}_{01}^{10}$ & $\mathsf{L}_{11}^{10}$ & $\mathsf{L}_{10}^{01}$ & $\mathsf{L}_{01}^{01}$ & $\mathsf{L}_{11}^{01}$ & $\mathsf{L}_{10}^{11}$ & $\mathsf{L}_{01}^{11}$ & $\mathsf{L}_{11}^{11}$\tabularnewline
\hline
$\mathsf{P}_{10}$ & $\mathsf{P}_{10}$ & $\mathsf{1}$ & $\mathsf{P}_{11}$ & $\mathsf{P}_{01}$ & $\mathsf{L}_{10}^{10}$ & $\mathsf{L}_{01}^{10}$ & $\mathsf{L}_{11}^{10}$ & $\mathsf{L}_{10}$ & $\mathsf{L}_{01}$ & $\mathsf{L}_{11}$ & $\mathsf{L}_{10}^{11}$ & $\mathsf{L}_{01}^{11}$ & $\mathsf{L}_{11}^{11}$ & $\mathsf{L}_{10}^{01}$ & $\mathsf{L}_{01}^{01}$ & $\mathsf{L}_{11}^{01}$\tabularnewline
\hline
$\mathsf{P}_{01}$ & $\mathsf{P}_{01}$ & $\mathsf{P}_{11}$ & $\mathsf{1}$ & $\mathsf{P}_{10}$ & $\mathsf{L}_{10}^{01}$ & $\mathsf{L}_{01}^{01}$ & $\mathsf{L}_{11}^{01}$ & $\mathsf{L}_{10}^{11}$ & $\mathsf{L}_{01}^{11}$ & $\mathsf{L}_{11}^{11}$ & $\mathsf{L}_{10}$ & $\mathsf{L}_{01}$ & $\mathsf{L}_{11}$ & $\mathsf{L}_{10}^{10}$ & $\mathsf{L}_{01}^{10}$ & $\mathsf{L}_{11}^{10}$\tabularnewline
\hline
$\mathsf{P}_{11}$ & $\mathsf{P}_{11}$ & $\mathsf{P}_{01}$ & $\mathsf{P}_{10}$ & $\mathsf{1}$ & $\mathsf{L}_{10}^{11}$ & $\mathsf{L}_{01}^{11}$ & $\mathsf{L}_{11}^{11}$ & $\mathsf{L}_{10}^{01}$ & $\mathsf{L}_{01}^{01}$ & $\mathsf{L}_{11}^{01}$ & $\mathsf{L}_{10}^{10}$ & $\mathsf{L}_{01}^{10}$ & $\mathsf{L}_{11}^{10}$ & $\mathsf{L}_{10}$ & $\mathsf{L}_{01}$ & $\mathsf{L}_{11}$\tabularnewline
\hline
$\mathsf{L}_{10}$ & $\mathsf{L}_{10}$ & $\mathsf{L}_{10}^{10}$ & $\mathsf{L}_{10}^{01}$ & $\mathsf{L}_{10}^{11}$ & $\mathsf{1}$ & $\mathsf{L}_{11}$ & $\mathsf{L}_{01}$ & $\mathsf{P}_{10}$ & $\mathsf{L}_{11}^{10}$ & $\mathsf{L}_{01}^{10}$ & $\mathsf{P}_{01}$ & $\mathsf{L}_{11}^{01}$ & $\mathsf{L}_{01}^{01}$ & $\mathsf{P}_{11}$ & $\mathsf{L}_{11}^{11}$ & $\mathsf{L}_{01}^{11}$\tabularnewline
\hline
$\mathsf{L}_{01}$ & $\mathsf{L}_{01}$ & $\mathsf{L}_{01}^{10}$ & $\mathsf{L}_{01}^{01}$ & $\mathsf{L}_{01}^{11}$ & $\mathsf{L}_{11}$ & $\mathsf{1}$ & $\mathsf{L}_{10}$ & $\mathsf{L}_{11}^{10}$ & $\mathsf{P}_{10}$ & $\mathsf{L}_{10}^{10}$ & $\mathsf{L}_{11}^{01}$ & $\mathsf{P}_{01}$ & $\mathsf{L}_{10}^{01}$ & $\mathsf{L}_{11}^{11}$ & $\mathsf{P}_{11}$ & $\mathsf{L}_{10}^{11}$\tabularnewline
\hline
$\mathsf{L}_{11}$ & $\mathsf{L}_{11}$ & $\mathsf{L}_{11}^{10}$ & $\mathsf{L}_{11}^{01}$ & $\mathsf{L}_{11}^{11}$ & $\mathsf{L}_{01}$ & $\mathsf{L}_{10}$ & $\mathsf{1}$ & $\mathsf{L}_{01}^{10}$ & $\mathsf{L}_{10}^{10}$ & $\mathsf{P}_{10}$ & $\mathsf{L}_{01}^{01}$ & $\mathsf{L}_{10}^{01}$ & $\mathsf{P}_{01}$ & $\mathsf{L}_{01}^{11}$ & $\mathsf{L}_{10}^{11}$ & $\mathsf{P}_{11}$\tabularnewline
\hline
$\mathsf{L}_{10}^{10}$ & $\mathsf{L}_{10}^{10}$ & $\mathsf{L}_{10}$ & $\mathsf{L}_{10}^{11}$ & $\mathsf{L}_{10}^{01}$ & $\mathsf{P}_{10}$ & $\mathsf{L}_{11}^{10}$ & $\mathsf{L}_{01}^{10}$ & $\mathsf{1}$ & $\mathsf{L}_{11}$ & $\mathsf{L}_{01}$ & $\mathsf{P}_{11}$ & $\mathsf{L}_{11}^{11}$ & $\mathsf{L}_{01}^{11}$ & $\mathsf{P}_{01}$ & $\mathsf{L}_{11}^{01}$ & $\mathsf{L}_{01}^{01}$\tabularnewline
\hline
$\mathsf{L}_{01}^{10}$ & $\mathsf{L}_{01}^{10}$ & $\mathsf{L}_{01}$ & $\mathsf{L}_{01}^{11}$ & $\mathsf{L}_{01}^{01}$ & $\mathsf{L}_{11}^{10}$ & $\mathsf{P}_{10}$ & $\mathsf{L}_{10}^{10}$ & $\mathsf{L}_{11}$ & $\mathsf{1}$ & $\mathsf{L}_{10}$ & $\mathsf{L}_{11}^{11}$ & $\mathsf{P}_{11}$ & $\mathsf{L}_{10}^{11}$ & $\mathsf{L}_{11}^{01}$ & $\mathsf{P}_{01}$ & $\mathsf{L}_{10}^{01}$\tabularnewline
\hline
$\mathsf{L}_{11}^{10}$ & $\mathsf{L}_{11}^{10}$ & $\mathsf{L}_{11}$ & $\mathsf{L}_{11}^{11}$ & $\mathsf{L}_{11}^{01}$ & $\mathsf{L}_{01}^{10}$ & $\mathsf{L}_{10}^{10}$ & $\mathsf{P}_{10}$ & $\mathsf{L}_{01}$ & $\mathsf{L}_{10}$ & $\mathsf{1}$ & $\mathsf{L}_{01}^{11}$ & $\mathsf{L}_{10}^{11}$ & $\mathsf{P}_{11}$ & $\mathsf{L}_{01}^{01}$ & $\mathsf{L}_{10}^{01}$ & $\mathsf{P}_{01}$\tabularnewline
\hline
$\mathsf{L}_{10}^{01}$ & $\mathsf{L}_{10}^{01}$ & $\mathsf{L}_{10}^{11}$ & $\mathsf{L}_{10}$ & $\mathsf{L}_{10}^{10}$ & $\mathsf{P}_{01}$ & $\mathsf{L}_{11}^{01}$ & $\mathsf{L}_{01}^{01}$ & $\mathsf{P}_{11}$ & $\mathsf{L}_{11}^{11}$ & $\mathsf{L}_{01}^{11}$ & $\mathsf{1}$ & $\mathsf{L}_{11}$ & $\mathsf{L}_{01}$ & $\mathsf{P}_{10}$ & $\mathsf{L}_{11}^{10}$ & $\mathsf{L}_{01}^{10}$\tabularnewline
\hline
$\mathsf{L}_{01}^{01}$ & $\mathsf{L}_{01}^{01}$ & $\mathsf{L}_{01}^{11}$ & $\mathsf{L}_{01}$ & $\mathsf{L}_{01}^{10}$ & $\mathsf{L}_{11}^{01}$ & $\mathsf{P}_{01}$ & $\mathsf{L}_{10}^{01}$ & $\mathsf{L}_{11}^{11}$ & $\mathsf{P}_{11}$ & $\mathsf{L}_{10}^{11}$ & $\mathsf{L}_{11}$ & $\mathsf{1}$ & $\mathsf{L}_{10}$ & $\mathsf{L}_{11}^{10}$ & $\mathsf{P}_{10}$ & $\mathsf{L}_{10}^{10}$\tabularnewline
\hline
$\mathsf{L}_{11}^{01}$ & $\mathsf{L}_{11}^{01}$ & $\mathsf{L}_{11}^{11}$ & $\mathsf{L}_{11}$ & $\mathsf{L}_{11}^{10}$ & $\mathsf{L}_{01}^{01}$ & $\mathsf{L}_{10}^{01}$ & $\mathsf{P}_{01}$ & $\mathsf{L}_{01}^{11}$ & $\mathsf{L}_{10}^{11}$ & $\mathsf{P}_{11}$ & $\mathsf{L}_{01}$ & $\mathsf{L}_{10}$ & $\mathsf{1}$ & $\mathsf{L}_{01}^{10}$ & $\mathsf{L}_{10}^{10}$ & $\mathsf{P}_{10}$\tabularnewline
\hline
$\mathsf{L}_{10}^{11}$ & $\mathsf{L}_{10}^{11}$ & $\mathsf{L}_{10}^{01}$ & $\mathsf{L}_{10}^{10}$ & $\mathsf{L}_{10}$ & $\mathsf{P}_{11}$ & $\mathsf{L}_{11}^{11}$ & $\mathsf{L}_{01}^{11}$ & $\mathsf{P}_{01}$ & $\mathsf{L}_{11}^{01}$ & $\mathsf{L}_{01}^{01}$ & $\mathsf{P}_{10}$ & $\mathsf{L}_{11}^{10}$ & $\mathsf{L}_{01}^{10}$ & $\mathsf{1}$ & $\mathsf{L}_{11}$ & $\mathsf{L}_{01}$\tabularnewline
\hline
$\mathsf{L}_{01}^{11}$ & $\mathsf{L}_{01}^{11}$ & $\mathsf{L}_{01}^{01}$ & $\mathsf{L}_{01}^{10}$ & $\mathsf{L}_{01}$ & $\mathsf{L}_{11}^{11}$ & $\mathsf{P}_{11}$ & $\mathsf{L}_{10}^{11}$ & $\mathsf{L}_{11}^{01}$ & $\mathsf{P}_{01}$ & $\mathsf{L}_{10}^{01}$ & $\mathsf{L}_{11}^{10}$ & $\mathsf{P}_{10}$ & $\mathsf{L}_{10}^{10}$ & $\mathsf{L}_{11}$ & $\mathsf{1}$ & $\mathsf{L}_{10}$\tabularnewline
\hline
$\mathsf{L}_{11}^{11}$ & $\mathsf{L}_{11}^{11}$ & $\mathsf{L}_{11}^{01}$ & $\mathsf{L}_{11}^{10}$ & $\mathsf{L}_{11}$ & $\mathsf{L}_{01}^{11}$ & $\mathsf{L}_{10}^{11}$ & $\mathsf{P}_{11}$ & $\mathsf{L}_{01}^{01}$ & $\mathsf{L}_{10}^{01}$ & $\mathsf{P}_{01}$ & $\mathsf{L}_{01}^{10}$ & $\mathsf{L}_{10}^{10}$ & $\mathsf{P}_{10}$ & $\mathsf{L}_{01}$ & $\mathsf{L}_{10}$ & $\mathsf{1}$\tabularnewline
\hline

\hline

\hline

\end{tabular*}
\end{table*}

\subsection{Topological order with particle-loop braiding and $3$-loop braiding:
$G=\left(\mathbb{Z}_{2}\right)^{2}$}\label{subsec:fusion_BF_AAdA_z2z2}

The TQFT action for topological order with particle-loop braiding and $3$-loop braiding is
\begin{align}
S= & S_{BF}+S_{A^{1}A^{2}dA^{2}}\nonumber \\
= & \sum_{i=1}^{2}\frac{N_{i}}{2\pi}B^{i}dA^{i}+\frac{pN_{1}N_{2}}{\left(2\pi\right)^{2}N_{12}}A^{1}A^{2}dA^{2}\label{eq:AAdA_z2z2}
\end{align}
with $N_{1}=N_{2}=2$, $N_{12}\equiv\gcd\left(N_{1},N_{2}\right)=2$
and $p\in\mathbb{Z}_{N_{12}}$. For a nontrivial action, we can set $p=1$.
The gauge transformations are
\begin{align}
A^{i}\rightarrow & A^{i}+d\chi^{i},\\
B^{1}\rightarrow & B^{1}+dV^{1}+\frac{pN_{2}}{2\pi N_{12}}d\chi^{2}A^{2},\\
B^{2}\rightarrow & B^{2}+dV^{2}-\frac{pN_{1}}{2\pi N_{12}}d\chi^{1}A^{2}.
\end{align}
 Particle excitations are represented by operators
\begin{equation}
\mathsf{P}_{ij}=\exp\left({\rm i}i\int_{\gamma}A^{1}+{\rm i}j\int_{\gamma}A^{2}\right)
\end{equation}
with $i,j=0,1$. Loop excitations are represented by
\begin{align}
\mathsf{L}_{10}^{ij}= & \exp\left[{\rm i}\left(\int_{\sigma}B^{1}+\frac{pN_{2}}{2\pi N_{12}}\int_{\Omega}A^{2}dA^{2}\right)\right.\nonumber \\
 & +\left.{\rm i}i\int_{\gamma}A^{1}+{\rm i}j\int_{\gamma}A^{2}\right],
\end{align}
\begin{align}
\mathsf{L}_{01}^{ij}= & \exp\left[{\rm i}\left(\int_{\sigma}B^{2}-\frac{pN_{1}}{2\pi N_{12}}\int_{\Omega}A^{1}dA^{2}\right)\right.\nonumber \\
 & +\left.{\rm i}i\int_{\gamma}A^{1}+{\rm i}j\int_{\gamma}A^{2}\right],
\end{align}
\begin{align}
\mathsf{L}_{11}^{ij}= & \exp\left[{\rm i}\left(\int_{\sigma}B^{1}+\frac{pN_{2}}{2\pi N_{12}}\int_{\Omega}A^{2}dA^{2}\right)\right.\nonumber \\
 & +{\rm i}\left(\int_{\sigma}B^{2}-\frac{pN_{1}}{2\pi N_{12}}\int_{\Omega}A^{1}dA^{2}\right)\nonumber \\
 & +\left.{\rm i}i\int_{\gamma}A^{1}+{\rm i}j\int_{\gamma}A^{2}\right],
\end{align}
with $i,j=0,1$. There are $2^{4}=16$ nonequivalent excitations in total.
Using the field-theoretical approach developed in Sec.~\ref{sec:fusion rule}, we find that the fusion rules in this case are the same as those of $S=S_{BF}$ with $G=\Z_2 \times \Z_2$. In other words, the fusion rules in this case are also shown in Table.~\ref{tab:fusion_BdA_Z2Z2}.

\subsection{Topological order with particle-loop braiding and two different $3$-loop braidings: $G=\left(\mathbb{Z}_{2}\right)^{2}$}\label{subsec:fusion_two_AAdA_z2z2}

The TQFT action $S=S_{BF}+S_{A^{1}A^{2}dA^{2}}+S_{A^{2}A^{1}dA^{1}}$
describes the topological order with particle-loop braiding and two different but compatible $3$-loop braidings.
The action is
\begin{align}
S= & \int\sum_{i=1}^{2}\frac{N_{i}}{2\pi}B^{i}dA^{i}\nonumber \\
 & +\frac{p_{1}N_{1}N_{2}}{\left(2\pi\right)^{2}N_{12}}A^{1}A^{2}dA^{2}+\frac{p_{2}N_{1}N_{2}}{\left(2\pi\right)^{2}N_{12}}A^{2}A^{1}dA^{1}
\end{align}
with $N_{1}=N_{2}=2$, $N_{12}\equiv\gcd\left(N_{1},N_{2}\right)=2$
and $p_{1}p_{2}\in\mathbb{Z}_{N_{12}}$. We can view this action as
the stack of $S=S_{BF}+S_{A^{1}A^{2}dA^{2}}$ and $S=S_{BF}+S_{A^{2}A^{1}dA^{1}}$.
The gauge transformations are
\begin{align}
A^{i}\rightarrow & A^{i}+d\chi^{i},\\
B^{1}\rightarrow & B^{1}+dV^{1}+\frac{p_{1}N_{2}}{2\pi N_{12}}d\chi^{2}A^{2}-\frac{p_{2}N_{2}}{2\pi N_{12}}d\chi^{2}A^{1},\\
B^{2}\rightarrow & B^{2}+dV^{2}-\frac{p_{1}N_{1}}{2\pi N_{12}}d\chi^{1}A^{2}+\frac{p_{2}N_{1}}{2\pi N_{12}}d\chi^{1}A^{1}.
\end{align}
The particle and loop (including pure loop and decorated loop) excitations
are represented by the following operators:
\begin{equation}
\mathsf{P}_{ij}=\exp\left({\rm i}i\int_{\gamma}A^{1}+{\rm i}j\int_{\gamma}A^{2}\right),
\end{equation}
\begin{align}
 & \mathsf{L}_{10}^{ij}\nonumber \\
= & \exp\left[{\rm i}\left(\int_{\sigma}B^{1}+\frac{p_{1}N_{2}}{2\pi N_{12}}\int_{\Omega}A^{2}dA^{2}-\frac{p_{2}N_{2}}{2\pi N_{12}}\int_{\Omega}A^{2}dA^{1}\right)\right.\nonumber \\
 & +\left.{\rm i}i\int_{\gamma}A^{1}+{\rm i}j\int_{\gamma}A^{2}\right],
\end{align}
\begin{align}
 & \mathsf{L}_{01}^{ij}\nonumber \\
= & \exp\left[{\rm i}\left(\int_{\sigma}B^{2}-\frac{p_{1}N_{1}}{2\pi N_{12}}\int_{\Omega}A^{1}dA^{2}+\frac{p_{2}N_{1}}{2\pi N_{12}}\int_{\Omega}A^{1}dA^{1}\right)\right.\nonumber \\
 & +\left.{\rm i}i\int_{\gamma}A^{1}+{\rm i}j\int_{\gamma}A^{2}\right],
\end{align}
\begin{align}
 & \mathsf{L}_{11}^{ij}\nonumber \\
= & \exp\left[{\rm i}\left(\int_{\sigma}B^{1}+\frac{pN_{2}}{2\pi N_{12}}\int_{\Omega}A^{2}dA^{2}-\frac{p_{2}N_{2}}{2\pi N_{12}}\int_{\Omega}A^{2}dA^{1}\right)\right.\nonumber \\
 & +{\rm i}\left(\int_{\sigma}B^{2}-\frac{pN_{1}}{2\pi N_{12}}\int_{\Omega}A^{1}dA^{2}+\frac{p_{2}N_{1}}{2\pi N_{12}}\int_{\Omega}A^{1}dA^{1}\right)\nonumber \\
 & +\left.{\rm i}i\int_{\gamma}A^{1}+{\rm i}j\int_{\gamma}A^{2}\right],
\end{align}
where $i,j=0,1$. The number of all nonequivalent excitations is $16$.
Again, we find fusion rules in this case same as those of $S=S_{BF}$
with $G=\mathbb{Z}_{2}\times\mathbb{Z}_{2}$, i.e., shown in Table.~\ref{tab:fusion_BdA_Z2Z2}. Combining the discussion
in Sec.~\ref{subsec:fusion_BF_z2z2},~\ref{subsec:fusion_BF_AAdA_z2z2} and~\ref{subsec:fusion_two_AAdA_z2z2}, we can conclude that when $G=\mathbb{Z}_{2}\times\mathbb{Z}_{2}$
the fusion rules of excitations are unchanged, forming a $\left(\mathbb{Z}_{2}\right)^{4}$
group, no matter the TQFT action contains twisted terms or not. In
other words, with $G=\mathbb{Z}_{2}\times\mathbb{Z}_{2}$, the fusion
rules of different topologically ordered systems which support different
but mutually compatible braidings are same.

\subsection{Topological order with particle-loop braiding and $3$-loop braiding:
$G=\left(\mathbb{Z}_{2}\right)^{3}$}\label{subsec:fusion_BF_AAdA_z2z2z2}

When $G=\prod_{i=1}^{3}\mathbb{Z}_{N_{i}}$, the TQFT action for topological order with particle-loop braiding and $3$-loop braiding can be
\begin{align}
S & =S_{BF}+S_{A^{1}A^{2}dA^{3}}\nonumber \\
 & =\int\sum_{i=1}^{3}\frac{N_{i}}{2\pi}B^{i}dA^{i}+\frac{pN_{1}N_{2}}{\left(2\pi\right)^{2}N_{12}}A^{1}A^{2}dA^{3}\label{eq:AAdA_z1z2z3}
\end{align}
with $N_{1}=N_{2}=N_{3}=2$, $N_{12}=2$, and $p\in\mathbb{Z}_{N_{12}}$,
i.e., $p\in\mathbb{Z}_{2}$. We set $p=1$ so that the action is nontrivial:
\begin{equation}
S=\int\sum_{i=1}^{3}\frac{N_{i}}{2\pi}B^{i}dA^{i}+\frac{2}{\left(2\pi\right)^{2}}A^{1}A^{2}dA^{3}.
\end{equation}
The gauge transformations are
\begin{align}
A^{i}\rightarrow & A^{i}+d\chi^{i},\\
B^{1}\rightarrow & B^{1}+dV^{1}+\frac{pN_{2}}{2\pi N_{12}}d\chi^{2}A^{3},\\
B^{2}\rightarrow & B^{2}+dV^{2}-\frac{pN_{1}}{2\pi N_{12}}d\chi^{1}A^{3},\\
B^{3}\rightarrow & B^{3}+dV^{3}.
\end{align}
In this case, the particle excitations are represented by
\begin{equation}
\mathsf{P}_{ijl}=\exp\left({\rm i}i\int_{\gamma}A^{1}+{\rm i}j\int_{\gamma}A^{2}+{\rm i}k\int_{\gamma}A^{3}\right)
\end{equation}
with $i,j,k=0,1$. The loop (pure loop and decorated loop) excitations
are represented by the following Wilson operators:
\begin{align}
\mathsf{L}_{100}^{ijl}= & \exp\left({\rm i}\int_{\sigma}B^{1}+{\rm i}\frac{pN_{2}}{2\pi N_{12}}\int_{\Omega}A^{2}dA^{3}\right.\nonumber \\
 & \left.+{\rm i}i\int_{\gamma}A^{1}+{\rm i}j\int_{\gamma}A^{2}+{\rm i}k\int_{\gamma}A^{3}\right),
\end{align}
\begin{align}
\mathsf{L}_{010}^{ijl}= & \exp\left({\rm i}\int_{\sigma}B^{1}+{\rm i}\frac{pN_{2}}{2\pi N_{12}}\int_{\Omega}A^{2}dA^{3}\right.\nonumber \\
 & \left.+{\rm i}i\int_{\gamma}A^{1}+{\rm i}j\int_{\gamma}A^{2}+{\rm i}k\int_{\gamma}A^{3}\right),
\end{align}
\begin{align}
\mathsf{L}_{001}^{ijl}= & \exp\left({\rm i}\int_{\sigma}B^{3}\right.\nonumber \\
 & \left.+{\rm i}i\int_{\gamma}A^{1}+{\rm i}j\int_{\gamma}A^{2}+{\rm i}k\int_{\gamma}A^{3}\right),
\end{align}
where $i,j,k=0,1$. Similarly, we find that the fusion rules of these
excitations form a $\left(\mathbb{Z}_{2}\right)^{6}$ group. In addition, $S=S_{BF}+S_{AAdA}$
with arbitrary $AAdA$ twisted term produces identical fusion rules as $S=S_{BF}$ when $G=\left(\mathbb{Z}_{2}\right)^{^{3}}$. This is also true when $G=\mathbb{Z}_{2}\times\mathbb{Z}_{2}$ (see Sec.~\ref{subsec:fusion_BF_z2z2},~\ref{subsec:fusion_BF_AAdA_z2z2} and~\ref{subsec:fusion_two_AAdA_z2z2}). These results lead to the general
conclusion: once given the gauge group $G=\prod_{i}\mathbb{Z}_{N_{i}}$, for different topologically ordered systems which support particle-loop braidings and/or $3$-loop braidings \emph{only}, the fusion rules are same: they are Abelian and constitute a $\left(\prod_{i}\Z_{N_i}\right)^{2}$ group.

\subsection{Topological order with particle-loop braiding, $3$-loop braiding, and BR braiding: $G=\left(\mathbb{Z}_{2}\right)^{3}$}\label{subsec:fusion_BF_AAdA_AAB_z2z2z2}

The Borromean rings braiding described by $S_{A^{1}A^{2}B^{3}}$ is
compatible with the $3$-loop braiding described by $S_{A^{1}A^{2}dA^{2}}$ \citep{zhang2021compatible}.
The TQFT action is given by
\begin{align}
S= & \int\sum_{i=1}^{3}\frac{N_{i}}{2\pi}B^{i}dA^{i}\nonumber \\
 & +\frac{p_{1}N_{1}N_{2}}{\left(2\pi\right)^{2}N_{12}}A^{1}A^{2}dA^{2}+\frac{p_{2}N_{1}N_{2}N_{3}}{\left(2\pi\right)^{2}N_{123}}A^{1}A^{2}B^{3}
\end{align}
with $p_{1}\in\mathbb{Z}_{N_{12}}$ and $p_{2}\in\mathbb{Z}_{N_{123}}$.
We set $p_{1}=p_{2}=1$. The gauge transformations are
\begin{align}
A^{1}\rightarrow & A^{1}+d\chi^{1},\\
A^{2}\rightarrow & A^{2}+d\chi^{2},\\
A^{3}\rightarrow & A^{3}+d\chi^{3}\nonumber \\
 & -\frac{p_{2}N_{1}N_{2}}{2\pi N_{123}}\left(\chi^{1}A^{2}+\frac{1}{2}\chi^{1}d\chi^{2}\right)\nonumber \\
 & +\frac{p_{2}N_{1}N_{2}}{2\pi N_{123}}\left(\chi^{2}A^{1}+\frac{1}{2}\chi^{2}d\chi^{1}\right),\\
B^{1}\rightarrow & B^{1}+dV^{1}\nonumber \\
 & +\frac{p_{1}N_{2}}{2\pi N_{12}}d\chi^{2}A^{2}\nonumber \\
 & -\frac{p_{2}N_{2}N_{3}}{2\pi N_{123}}\left(\chi^{2}B^{3}-A^{2}V^{3}+\chi^{2}dV^{3}\right),\\
B^{2}\rightarrow & B^{2}+dV^{2}\nonumber \\
 & -\frac{p_{2}N_{1}}{2\pi N_{12}}d\chi^{1}A^{2}\nonumber \\
 & +\frac{p_{2}N_{1}N_{3}}{2\pi N_{123}}\left(\chi^{1}B^{3}-A^{1}V^{3}+\chi^{1}dV^{3}\right),\\
B^{3}\rightarrow & B^{3}+dV^{3}.
\end{align}
The loop and particle excitations are represented by operators shown
in Table.~\ref{tab:AAB+AAdA_loop_nonequivalent} and Table.~\ref{tab:AAB+AAdA_particle_nonequivalent}.
We find these operators have a similar expression of those for $S=S_{BF}+S_{A^{1}A^{2}B^{3}}$, i.e., Eq.~(\ref{eq:action_AAB}).
In Sec. \ref{sec:fusion rule} we have seen that non-Abelian fusion
can be traced back to the Kronecker delta function in
operators. By performing similar calculation, we find that the operators listed in Table~\ref{tab:AAB+AAdA_particle_nonequivalent} and Table~\ref{tab:AAB+AAdA_loop_nonequivalent} obey the same fusion
rules of $S=S_{BF}+S_{A^{1}A^{2}B^{3}}$, i.e., those shown in Table~\ref{tab:fusion_AAB_Z2Z2Z2_full}. This result is different from those of $S=S_{BF}$ and $S=S_{BF}+S_{AAdA}$ aforementioned. As pointed out in Ref.~\citep{zhang2021compatible}, BR braiding is not always compatible with multi-loop braidings. If a BR braiding is introduced compatibly to a system that only supports particle-loop braiding and/or multi-loop braiding only, the formerly Abelian fusion rules would be dramatically changed to be non-Abelian.

\section{Discussion and outlook\label{sec:discussion_outlook}}

In this work,   we perform field-theoretical analysis on   Wilson operators (i.e., the excitation contents), fusion rules, and loop-shrinking rules  in  three dimensional topological orders.  
Let us briefly review this paper.
First, gauge-invariant Wilson operators are written down for nonequivalent
topological excitations. The number
of particle excitations and pure loop excitations agrees with that
calculated from a lattice cocycle model. Next, fusion rules are represented
in terms of path integral of TQFT and we find out all fusion rules as well as quantum dimensions.
Some of the fusion rules are non-Abelian though the input gauge group
for this topological order is Abelian. %The origin of such difference is still unknown.
Beside the fusion rules, we also study the shrinking rules of loop
excitations, which is a very interesting topological property of spatially nonlocal topological excitations. We propose a field-theoretical framework to perform the shrinking operation
in terms of operators and path integral, i.e, shrinking the loop's
world-sheet to a world-line. The loop shrinking rules obtained are consistent with fusion rules, i.e., they respect fusion rules and conserve the quantum dimensions through the shrinking process. The consistency between fusion rules and loop-shrinking rules is critical in establishing an anomaly-free topological order in 3D. Motivated by the present work, we expect to explore  the following topics in the near future.
   
\textbf{i.}---We expect more field-theoretical calculations may give a hint on   the consistency among braiding data, fusion rules and shrinking rules in general 3D topological orders. Once inconsistency happens, the corresponding topological orders might be potentially anomalous and only realizable on the boundary of some 4D topological phases of matter.

\textbf{ii.}---It will be  interesting to attempt to understand the algebraic structure behind the fusion rules of Borromean rings
topological order and all topological orders with compatible braidings discussed in this paper. Considering that the BR topological
order is beyond the usual DW gauge theory~\citep{string3,lantian3dto1,Bullivant:2019fmk,Bullivant:2020xhy,2022JMP....63h1901B} classification,  it may be described by the generalized Drinfel'd center (a braided monoidal 2-category) of a 2-group (a special kind of fusion 2-category).
In addition, our theory finds that, fusing a loop and its anti-loop may generate a fusion channel with two vacua. This is very unusual since in    2D topological orders, fusing a particle and its anti-particle must only have one vacuum. We conjecture this phenomenon may be related to the incorporation of ``2-group'' structure in our field theory, which is absent in field-theory of 2D topological order. In summary, the goal of this paper is to construct a field-theoretical study, more precisely, the path-integral calculation on topological invariants; the corresponding algebraic description is   also important, which will be one of future directions.

\textbf{iii.}---It will be important to generalize the classification of Abelian symmetry fractionalization in Ref.~\cite{ning2022fractionalizing} to BR topological order as well as all other topological orders with compatible braidings, which leads to a more complete field-theoretical understanding on Symmetry Enriched Topological phases in 3D~\cite{ning2022fractionalizing,ye16_set,2016arXiv161008645Y} and thus generalize Table~I of Ref.~\cite{ning2022fractionalizing} to non-Abelian fractionalization.

\textbf{iv.}---Just like the study of non-Abelian anyons in 2D topological systems, braiding, fusion, and shrinking are topological invariants are vital in   theory of TQC of higher dimensional stabilizer codes.  So, we expect that our field-theoretical study will be helpful along this line of efforts, especially on the roles of loop-like excitations (errors/defects).

\textbf{v.}---It will be curious to ask what is the physical consequence of knotted loops or mutually linked loops when performing fusion or shrinking operations.  Our field-theoretical study has provided  concrete procedures for computation of unknotted loops, which in principle can be applied to more complicated loops. We leave these exciting questions for future exploration.
 %\cite{Vijay2015,Prem2018,PhysRevB.99.245135,PhysRevB.97.144106,Pretko2017,rxotic2021Nathan,Slagle21foliatedfield,Pai2019,2020PhRvR2c3124G,You2019,Pai2018,PhysRevB.100.125150,Shirley2018,2020PhRvB.102t5106P,ye19a,PhysRevB.104.235127,Fractons2019Nandkishore,Fracton2016Vijay,Ma2018,Ma2017,Prem2017,2019arXiv190913879W,Pai2019,PhysRevA.83.042330,2020PhRvR...2d3219W,Slagle2017a,Chamon05fracton,Liu21fracton,Schmidt20fractonpcut,Hermele21fracton,Pretko2018,Aasen20tqftnetwork,Ma2018a,2020IJMPA..3530003P,PhysRevX.9.031035,Fractonic2020Yuan,Fractonic2021Chen,Higher2022Yuan,Quantum2022Yuan,Renormalization2021Li,2022arXiv220300015Z,PhysRevResearch.4.033111,Localization2019Pai,Emergent2017Pretko}

\acknowledgements
 We thank X. G. Wen, A. Tiwari, and C. Delcamp for helpful communications on this work. This work was supported by Guangdong Basic and Applied Basic Research Foundation under Grant No.~2020B1515120100, NSFC Grant (No.~11847608 \& No.~12074438). The work was performed in part on resources provided by the Guangdong Provincial Key Laboratory of Magnetoelectric Physics and Devices (LaMPad).

\begin{table*}
\centering
\caption{Operators for nonequivalent loop excitations in $S=S_{BF}+S_{A^{1}A^{2}dA^{2}}+S_{A^{1}A^{2}B^{3}}$
with $G=\left(\mathbb{Z}_{2}\right)^{3}$ (see Sec.~\ref{subsec:fusion_BF_AAdA_AAB_z2z2z2}). Among them, there are $1$ trivial loop, $4$ nontrivial pure loops and $10$ decorated loops. The ``$0$ or $\Z_{N_i}$'' charge decoration means that the operator of pure loop (no particle attached to it) is equivalent to that of the loop decorated with a particle carrying $\Z_{N_i}$ gauge charge. The equivalence operators are explained in Sec.~\ref{subsec:Operators_equivalence class}.   \label{tab:AAB+AAdA_loop_nonequivalent}}
\resizebox{\textwidth}{!}{
%%\begin{tabular*}{\textwidth}{@{\extracolsep{\fill}}|c|c|c|c|}
\begin{tabular}{cccc}
\hline

\hline

\hline

\textbf{Fluxes} & \textbf{Charge decoration} & \textbf{Operators for loop excitations} &\textbf{ Equivalent operators}\tabularnewline
\hline
$0$ & $0$ & $\mathsf{L}_{000}=\mathsf{1}=\exp\left({\rm i}0\right)=1$ & -\tabularnewline
\hline
$\mathbb{Z}_{N_{1}}$ & $0$ or $\mathbb{Z}_{N_{2}}$ & $\begin{alignedat}{1}\mathsf{L}_{100}= & 2\exp\left[{\rm i}\int_{\sigma}B^{1}+\frac{1}{2}\frac{2\pi q}{N_{1}}\left(d^{-1}A^{2}B^{3}+d^{-1}B^{3}A^{2}\right)+{\rm i}\int_{\Omega}\frac{p_{1}N_{2}}{2\pi N_{12}}A^{2}dA^{2}\right]\\
 & \times\delta\left(\int_{\gamma}A^{2}\right)\delta\left(\int_{\sigma}B^{3}\right)
\end{alignedat}
$ & $\mathsf{L}_{100}=\mathsf{L}_{10n_{3}}^{0c_{2}0};c_{2},n_{3}=0,1$\tabularnewline
\hline
$\mathbb{Z}_{N_{1}}$ & $\mathbb{Z}_{N_{1}}$ & $\begin{alignedat}{1}\mathsf{L}_{100}^{100}= & 2\exp\left[{\rm i}\int_{\gamma}A^{1}+{\rm i}\int_{\sigma}B^{1}+\frac{1}{2}\frac{2\pi q}{N_{1}}\left(d^{-1}A^{2}B^{3}+d^{-1}B^{3}A^{2}\right)+{\rm i}\int_{\Omega}\frac{p_{1}N_{2}}{2\pi N_{12}}A^{2}dA^{2}\right]\\
 & \times\delta\left(\int_{\gamma}A^{2}\right)\delta\left(\int_{\sigma}B^{3}\right)
\end{alignedat}
$ & $\mathsf{L}_{100}^{100}=\mathsf{L}_{10n_{3}}^{1c_{2}0};c_{2},n_{3}=0,1$\tabularnewline
\hline
$\mathbb{Z}_{N_{1}}$ & $\mathbb{Z}_{N_{3}}$ & $\begin{alignedat}{1}\mathsf{L}_{100}^{001}= & 4\exp\left[{\rm i}\int_{\sigma}B^{1}+\frac{1}{2}\frac{2\pi q}{N_{1}}\left(d^{-1}A^{2}B^{3}+d^{-1}B^{3}A^{2}\right)+{\rm i}\int_{\Omega}\frac{p_{1}N_{2}}{2\pi N_{12}}A^{2}dA^{2}\right.\\
 & \left.+{\rm i}\int_{\gamma}A^{3}+\frac{1}{2}\frac{2\pi q}{N_{3}}\left(d^{-1}A^{1}A^{2}-d^{-1}A^{2}A^{1}\right)\right]\\
 & \times\delta\left(\int_{\gamma}A^{2}\right)\delta\left(\int_{\sigma}B^{3}\right)\delta\left(\int_{\gamma}A^{1}\right)
\end{alignedat}
$ & $\mathsf{L}_{100}^{001}=\mathsf{L}_{10n_{3}}^{c_{1}c_{2}1};c_{1},c_{2},n_{3}=0,1$\tabularnewline
\hline
$\mathbb{Z}_{N_{2}}$ & $0$ or $\mathbb{Z}_{N_{1}}$ & $\begin{alignedat}{1}\mathsf{L}_{010}= & 2\exp\left[{\rm i}\int_{\sigma}B^{2}-\frac{1}{2}\frac{2\pi q}{N_{2}}\left(d^{-1}B^{3}A^{1}+d^{-1}A^{1}B^{3}\right)-\frac{pN_{1}}{2\pi N_{12}}\int_{\Omega}A^{1}dA^{2}\right]\\
 & \times\delta\left(\int_{\sigma}B^{3}\right)\delta\left(\int_{\gamma}A^{1}\right)
\end{alignedat}
$ & $\mathsf{L}_{010}=\mathsf{L}_{01n_{3}}^{c_{1}00};c_{1},n_{3}=0,1$\tabularnewline
\hline
$\mathbb{Z}_{N_{2}}$ & $\mathbb{Z}_{N_{2}}$ & $\begin{alignedat}{1}\mathsf{L}_{010}^{010}= & 2\exp\left[{\rm i}\int_{\gamma}A^{2}+{\rm i}\int_{\sigma}B^{2}-\frac{1}{2}\frac{2\pi q}{N_{2}}\left(d^{-1}B^{3}A^{1}+d^{-1}A^{1}B^{3}\right)-{\rm i}\int_{\Omega}\frac{pN_{1}}{2\pi N_{12}}A^{1}dA^{2}\right]\\
 & \times\delta\left(\int_{\sigma}B^{3}\right)\delta\left(\int_{\gamma}A^{1}\right)
\end{alignedat}
$ & $\mathsf{L}_{010}^{010}=\mathsf{L}_{01n_{3}}^{c_{1}10};c_{1},n_{3}=0,1$\tabularnewline
\hline
$\mathbb{Z}_{N_{2}}$ & $\mathbb{Z}_{N_{3}}$ & $\begin{alignedat}{1}\mathsf{L}_{010}^{001}= & 4\exp\left[{\rm i}\int_{\sigma}B^{2}-\frac{1}{2}\frac{2\pi q}{N_{2}}\left(d^{-1}B^{3}A^{1}+d^{-1}A^{1}B^{3}\right)-{\rm i}\int_{\Omega}\frac{pN_{1}}{2\pi N_{12}}A^{1}dA^{2}\right.\\
 & \left.+{\rm i}\int_{\gamma}A^{3}+\frac{1}{2}\frac{2\pi q}{N_{3}}\left(d^{-1}A^{1}A^{2}-d^{-1}A^{2}A^{1}\right)\right]\\
 & \times\delta\left(\int_{\sigma}B^{3}\right)\delta\left(\int_{\gamma}A^{1}\right)\delta\left(\int_{\gamma}A^{2}\right)
\end{alignedat}
$ & $\mathsf{L}_{010}^{001}=\mathsf{L}_{01n_{3}}^{c_{1}c_{2}1};c_{1},c_{2},n_{3}=0,1$\tabularnewline
\hline
$\mathbb{Z}_{N_{3}}$ & $0$ & $\mathsf{L}_{001}=\exp\left({\rm i}\int_{\sigma}B^{3}\right)$ & -\tabularnewline
\hline
$\mathbb{Z}_{N_{3}}$ & $\mathbb{Z}_{N_{1}}$ & $\mathsf{L}_{001}^{100}=\exp\left({\rm i}\int_{\gamma}A^{1}+{\rm i}\int_{\sigma}B^{3}\right)$ & -\tabularnewline
\hline
$\mathbb{Z}_{N_{3}}$ & $\mathbb{Z}_{N_{2}}$ & $\mathsf{L}_{001}^{010}=\exp\left({\rm i}\int_{\gamma}A^{2}+{\rm i}\int_{\sigma}B^{3}\right)$ & -\tabularnewline
\hline
$\mathbb{Z}_{N_{3}}$ & $\mathbb{Z}_{N_{1}},\mathbb{Z}_{N_{2}}$ & $\mathsf{L}_{001}^{110}=\exp\left({\rm i}\int_{\gamma}A^{1}+{\rm i}\int_{\gamma}A^{2}+{\rm i}\int_{\sigma}B^{3}\right)$ & -\tabularnewline
\hline
$\mathbb{Z}_{N_{3}}$ & $\mathbb{Z}_{N_{3}}$ & $\begin{alignedat}{1}\mathsf{L}_{001}^{001}= & 2\exp\left[{\rm i}\int_{\sigma}B^{3}+{\rm i}\int_{\gamma}A^{3}+\frac{1}{2}\frac{2\pi q}{N_{3}}\left(d^{-1}A^{1}A^{2}-d^{-1}A^{2}A^{1}\right)\right]\\
 & \times\delta\left(\int_{\gamma}A^{1}\right)\delta\left(\int_{\gamma}A^{2}\right)
\end{alignedat}
$ & $\mathsf{L}_{001}^{001}=\mathsf{L}_{001}^{c_{1}c_{2}1};c_{1},c_{2}=0,1$\tabularnewline
\hline
$\mathbb{Z}_{N_{1}},\mathbb{Z}_{N_{2}}$ & $0$ or $\left(\mathbb{Z}_{N_{1}},\mathbb{Z}_{N_{2}}\right)$ & $\begin{alignedat}{1}\mathsf{L}_{110}= & 2\exp\left[{\rm i}\int_{\sigma}B^{1}+\frac{1}{2}\frac{2\pi q}{N_{1}}\left(d^{-1}A^{2}B^{3}+d^{-1}B^{3}A^{2}\right)+{\rm i}\int_{\Omega}\frac{p_{1}N_{2}}{2\pi N_{12}}A^{2}dA^{2}\right.\\
 & +\left.{\rm i}\int_{\sigma}B^{2}-\frac{1}{2}\frac{2\pi q}{N_{2}}\left(d^{-1}B^{3}A^{1}+d^{-1}A^{1}B^{3}\right)-{\rm i}\int_{\Omega}\frac{pN_{1}}{2\pi N_{12}}A^{1}dA^{2}\right]\\
 & \times\delta\left(\int_{\gamma}A^{2}-A^{1}\right)\delta\left(\int_{\sigma}B^{3}\right)
\end{alignedat}
$ & $\mathsf{L}_{110}=\mathsf{L}_{110}^{110}=\mathsf{L}_{111}=\mathsf{L}_{111}^{110}$\tabularnewline
\hline
$\mathbb{Z}_{N_{1}},\mathbb{Z}_{N_{2}}$ & $\mathbb{Z}_{N_{1}}$ or $\mathbb{Z}_{N_{2}}$ & $\begin{alignedat}{1}\mathsf{L}_{110}^{100}= & 2\exp\left[{\rm i}\int_{\gamma}A^{1}+{\rm i}\int_{\sigma}B^{1}+\frac{1}{2}\frac{2\pi q}{N_{1}}\left(d^{-1}A^{2}B^{3}+d^{-1}B^{3}A^{2}\right)+\frac{p_{1}N_{2}}{2\pi N_{12}}\int_{\Omega}A^{2}dA^{2}\right.\\
 & +\left.{\rm i}\int_{\sigma}B^{2}-\frac{1}{2}\frac{2\pi q}{N_{2}}\left(d^{-1}B^{3}A^{1}+d^{-1}A^{1}B^{3}\right)\right]\\
 & \times\delta\left(\int_{\gamma}A^{2}-A^{1}\right)\delta\left(\int_{\sigma}B^{3}\right)
\end{alignedat}
$ & $\mathsf{L}_{110}^{100}=\mathsf{L}_{110}^{010}=\mathsf{L}_{111}^{110}=\mathsf{L}_{111}^{010}$\tabularnewline
\hline
$\mathbb{Z}_{N_{1}},\mathbb{Z}_{N_{2}}$ & $\mathbb{Z}_{N_{3}}$ & $\begin{alignedat}{1}\mathsf{L}_{110}^{001}= & 4\exp\left[{\rm i}\int_{\sigma}B^{1}+\frac{1}{2}\frac{2\pi q}{N_{1}}\left(d^{-1}A^{2}B^{3}+d^{-1}B^{3}A^{2}\right)+{\rm i}\int_{\Omega}\frac{p_{1}N_{2}}{2\pi N_{12}}A^{2}dA^{2}\right.\\
 & +{\rm i}\int_{\sigma}B^{2}-\frac{1}{2}\frac{2\pi q}{N_{2}}\left(d^{-1}B^{3}A^{1}+d^{-1}A^{1}B^{3}\right)-{\rm i}\int_{\Omega}\frac{pN_{1}}{2\pi N_{12}}A^{1}dA^{2}\\
 & +\left.{\rm i}\int_{\gamma}A^{3}+\frac{1}{2}\frac{2\pi q}{N_{3}}\left(d^{-1}A^{1}A^{2}-d^{-1}A^{2}A^{1}\right)\right]\\
 & \times\delta\left(\int_{\gamma}A^{1}\right)\delta\left(\int_{\gamma}A^{2}\right)\delta\left(\int_{\sigma}B^{3}\right)
\end{alignedat}
$ & $\mathsf{L}_{110}^{001}=\mathsf{L}_{11n_{3}}^{c_{1}c_{2}1};c_{1},c_{2},n_{3}=0,1$\tabularnewline
\hline

\hline

\hline

\end{tabular}}
\end{table*}

\begin{table*}
\centering
\caption{Operators for nonequivalent particle excitations in $S=S_{BF}+S_{A^{1}A^{2}dA^{2}}+S_{A^{1}A^{2}B^{3}}$
with $G=\left(\mathbb{Z}_{2}\right)^{3}$ (see Sec.~\ref{subsec:fusion_BF_AAdA_AAB_z2z2z2}). The operators for particle excitations share the same expression as those of $S=S_{BF}+S_{A^{1}A^{2}B^{3}}$ (listed in Table.~\ref{tab:AAB_particle_nonequivalent}). The equivalence operators are explained in Sec.~\ref{subsec:Operators_equivalence class}. \label{tab:AAB+AAdA_particle_nonequivalent}}
\begin{tabular*}{\textwidth}{@{\extracolsep{\fill}}ccc}
%%\begin{tabular}{|c|c|c|}
\hline

\hline

\hline

\textbf{Charges} & \textbf{Operator for particle excitations} & \textbf{Equivalent operators}\tabularnewline
\hline
$0$ & $\mathsf{P}_{000}=\mathsf{1}=\exp\left({\rm i}0\right)=1$ & -\tabularnewline
\hline
$\mathbb{Z}_{N_{1}}$ & $\mathsf{P}_{100}=\exp\left({\rm i}\int_{\gamma}A^{1}\right)$ & -\tabularnewline
\hline
$\mathbb{Z}_{N_{2}}$ & $\mathsf{P}_{010}=\exp\left({\rm i}\int_{\gamma}A^{2}\right)$ & -\tabularnewline
\hline
$\mathbb{Z}_{N_{3}}$ & $\mathsf{P}_{001}=2\exp\left[{\rm i}\int_{\gamma}A^{3}+\frac{1}{2}\frac{2\pi q}{N_{3}}\left(d^{-1}A^{1}A^{2}-d^{-1}A^{2}A^{1}\right)\right]\delta\left(\int_{\gamma}A^{1}\right)\delta\left(\int_{\gamma}A^{2}\right)$ & $\mathsf{P}_{001}=\mathsf{P}_{101}=\mathsf{P}_{011}=\mathsf{P}_{111}$\tabularnewline
\hline
$\mathbb{Z}_{N_{1}},\mathbb{Z}_{N_{2}}$ & $\mathsf{P}_{110}=\exp\left({\rm i}\int_{\gamma}A^{1}+{\rm i}\int_{\gamma}A^{2}\right)$ & -\tabularnewline
\hline

\hline

\hline

\end{tabular*}
\end{table*}

\appendix
\onecolumngrid
\section{Derivation of normalization factors of operators\label{appendix:derivation_normalization_factor}}

The normalization factors of operators are derived following principles.
First, if a particle or pure loop fuses with its anti-particle/anti-loop,
there should be a \emph{single} vacuum after fusion. Second, the fusion
result of excitations should be positive integer combinations of excitations.
For simplicity, in the following calculation we neglect the notation of expectation value but we should keep in mind that the following
formulas are in fact discussed in the context of path integrals.

For example, consider a particle with $\mathbb{Z}_{N_{1}}$ gauge charge,
its operator is
\begin{equation}
\mathsf{P}_{100}=\mathcal{N}_{000}^{100}\exp\left({\rm i}\int_{\gamma}A^{1}\right)
\end{equation}
with $\mathcal{N}_{000}^{100}$ is the normalization factor to be determined. Since $\mathbb{Z}_{N_{1}}=\mathbb{Z}_{2}$,
it is expected that
\begin{equation}
\mathsf{P}_{100}\otimes\mathsf{P}_{100}=\mathsf{1}\oplus\cdots.
\end{equation}
By comparing the coefficient, we obtain that $\mathcal{N}_{000}^{100}=1$,
i.e.,
\begin{equation}
\mathsf{P}_{100}=\exp\left({\rm i}\int_{\gamma}A^{1}\right).
\end{equation}

Next, consider a pure loop with $\mathbb{Z}_{N_{1}}$ flux, the operator is
\begin{equation}
\mathsf{L}_{100}=\mathcal{N}_{100}^{000}\exp\left[{\rm i}\int_{\sigma}B^{1}+\frac{1}{2}\frac{2\pi q}{N_{1}}\left(d^{-1}A^{2}B^{3}+d^{-1}B^{3}A^{2}\right)\right]\delta\left(\int_{\gamma}A^{2}\right)\delta\left(\int_{\sigma}B^{3}\right).
\end{equation}
Similarly, for $\mathbb{Z}_{N_{i}}=\mathbb{Z}_{2}$ ($i=1,2,3$), we expect
\begin{equation}
\mathsf{L}_{100}\otimes\mathsf{L}_{100}=\mathsf{1}\oplus\cdots.
\end{equation}
We calculate this fusion:
\begin{align}
 & \mathsf{L}_{100}\times\mathsf{L}_{100}\nonumber \\
= & \mathcal{N}_{100}^{000}\exp\left[{\rm i}\int_{\sigma}B^{1}+\frac{1}{2}\frac{2\pi q}{N_{1}}\left(d^{-1}A^{2}B^{3}+d^{-1}B^{3}A^{2}\right)\right]\delta\left(\int_{\gamma}A^{2}\right)\delta\left(\int_{\sigma}B^{3}\right)\nonumber \\
 & \times\mathcal{N}_{100}^{000}\exp\left[{\rm i}\int_{\sigma}B^{1}+\frac{1}{2}\frac{2\pi q}{N_{1}}\left(d^{-1}A^{2}B^{3}+d^{-1}B^{3}A^{2}\right)\right]\delta\left(\int_{\gamma}A^{2}\right)\delta\left(\int_{\sigma}B^{3}\right)\nonumber \\
= & \left(\mathcal{N}_{100}^{000}\right)^{2}\times\exp\left[{\rm i}2\int_{\sigma}B^{1}+\frac{1}{2}\frac{2\pi q}{N_{1}}\left(d^{-1}A^{2}B^{3}+d^{-1}B^{3}A^{2}\right)\right]\left[\delta\left(\int_{\gamma}A^{2}\right)\right]^{2}\left[\delta\left(\int_{\sigma}B^{3}\right)\right]^{2}\nonumber \\
= & \left(\mathcal{N}_{100}^{000}\right)^{2}\times1\times\delta\left(\int_{\gamma}A^{2}\right)\delta\left(\int_{\sigma}B^{3}\right)\nonumber \\
= & \left(\mathcal{N}_{100}^{000}\right)^{2}\times\frac{1}{2}\left(\mathsf{1}+\exp\left({\rm i}\int_{\gamma}A^{2}\right)\right)\times\frac{1}{2}\left(\mathsf{1}+\exp\left({\rm i}\int_{\sigma}B^{3}\right)\right)\nonumber \\
= & \left(\mathcal{N}_{100}^{000}\right)^{2}\times\frac{1}{4}\left(\mathsf{1}+\mathsf{P}_{010}+\mathsf{L}_{001}+\mathsf{L}_{001}^{010}\right),
\end{align}
where we have used
\begin{equation}
\left\langle \exp\left[{\rm i}2\int_{\sigma}B^{1}+\frac{1}{2}\frac{2\pi q}{N_{1}}\left(d^{-1}A^{2}B^{3}+d^{-1}B^{3}A^{2}\right)\right]\right\rangle =\left(\pm1\right)^{2}=1.
\end{equation}
The first principle mentioned above requires that
\begin{equation}
\left(\mathcal{N}_{100}^{000}\right)^{2}\times\frac{1}{4}=1
\end{equation}
Therefore, we have
\begin{equation}
\mathcal{N}_{100}^{000}=2.
\end{equation}
Following similar consideration, we can the fix normalization factor for operators of all particle and pure loop excitations. For operators of decorated loops, their factors are obtained from the fusion of corresponding pure loops and particles.

\section{Lattice cocycle model and emergent 2-group gauge theory}\label{appendix:cocycle_model}

In this section, we define lattice cocycle models~\citep{PhysRevB.95.205142} to realize
the TQFT in Eq.~(\ref{eq:action_AAB}). By extracting the topological part of the partition
function that is independent of the system volume, we can calculate
the ground state degeneracies of different 3D spacial manifolds. In
particular, the number of nonequivalent point-like and pure loop-like
topological excitations can be obtained in this lattice model. All
the results agree with the field theory analysis in previous sections.

\subsection{Topological partition functions}

After integrating out the Lagrange multipliers $B^{1}$, $B^{2}$
and $A^{3}$ in Eq.~(\ref{eq:action_AAB}), the remaining fields
$A^{1}$, $A^{2}$ and $B^{3}$ take values in $\Z_{N_{i}}$. It motivates
us to define the following lattice model with both 1-form and 2-form
cocycles on arbitrary 4D spacetime manifold triangulation $M_{4}$:
\begin{equation}
\mathcal{Z}_{k}(M_{4})=\sum_{\substack{a_{1},a_{2}\in Z^{1}(M_{4},\Z_{N})\\
b\in Z^{2}(M_{4},\Z_{N})
}
}e^{2\pi\ii\frac{k}{N}\int_{M_{4}}a_{1}a_{2}b}.\label{ZkM}
\end{equation}
For simplicity, we assumed $N_{i}=N$ ($i=1,2,3$). We have two kinds
of $\Z_{N}$ degrees of freedom defined on links and triangles of
$M_{4}$. The degrees of freedom $a_{1}$ and $a_{2}$ are two 1-cochains
of $M_{4}$, which map each link $\langle ij\rangle\in M_{4}$ to
$a_{ij}\in\Z_{N}$. Moreover, $a_{1}$ and $a_{2}$ satisfy the cocycle
(flat connection) condition $(\di a)_{ijk}:=a_{jk}-a_{ik}+a_{ij}=0$
on each triangle $\langle ijk\rangle\in M_{4}$. These cocycles form
a subgroup of the cochain group. Similarly, $b$ is a 2-cochain that
maps each triangle $\langle ijk\rangle\in M_{4}$ to $b_{ijk}\in\Z_{N}$.
It is also a 2-cocycle satisfying the cocycle (flat connection) condition
$(\di b)_{ijkl}:=b_{jkl}-b_{ikl}+b_{ijl}-b_{ijk}=0$ on each tetrahedron
$\langle ijkl\rangle\in M_{4}$. The sets of 1- and 2-cochains on
$M_{4}$ are denoted as $C^{1}(M_{4},\Z_{N})$ and $C^{2}(M_{4},\Z_{N})$.
And the sets of 1- and 2-cocycles on $M_{4}$ are denoted as $Z^{1}(M_{4},\Z_{N})$
and $Z^{2}(M_{4},\Z_{N})$. The integral $\int_{M_{4}}a_{1}a_{2}b$
is the analogous notation of discrete summation on triangulation $M_{4}$:
\begin{equation}
\int_{M_{4}}a_{1}a_{2}b=\sum_{\langle ijklm\rangle\in M_{4}}(a_{1})_{ij}(a_{2})_{jk}b_{klm}.
\end{equation}
The summation on the right-hand side is the cup product of $a_{1},a_{2}$
and $b$, which is the discrete version of wedge product of differential
forms.

The cocycle model Eq.~(\ref{ZkM}) is a local boson model. After
appropriate normalization, the topological part of it will be equivalent
to a 2-group gauge theory. To begin with, let us consider first $k=0$.
In this case, the action amplitude is always one, and Eq.~(\ref{ZkM})
becomes
\begin{equation}
\mathcal{Z}_{0}(M_{4})=|Z^{1}|^{2}|Z^{2}|=\frac{|H^{1}|\cdot|H^{2}|}{|H^{0}|}|C^{0}|\cdot|C^{1}|.
\end{equation}
In the last step we used $|Z^{i}|=|H^{i}|\cdot|C^{i-1}|/|Z^{i-1}|$
to relate the order of cocycle group $Z^{i}(M_{4},\Z_{N})$, the cochain
group $C^{i}(M_{4},\Z_{N})$ and the cohomology group $H^{i}(M_{4},\Z_{N})=Z^{i}(M_{4},\Z_{N})/B^{i}(M_{4},\Z_{N})$,
where $B^{i}(M_{4},\Z_{N}):=\{\di a_{i-1}|a\in C^{i-1}(M_{4},\Z_{N})\}$.
From the $k=0$ partition function, we see that the terms $|C^{0}|$
and $|C^{1}|$ are the numbers of vertices and links of the system,
which are volume-dependent. And the topological part of the partition
function is simply $|H^{1}|\cdot|H^{2}|/|H^{0}|=(|H^{1}|^{2}\cdot|H^{2}|)/(|H^{0}|\cdot|H^{1}|)$.
Therefore, we normalize and define the topological partition function
of the cocycle model Eq.~(\ref{ZkM}) to be
\begin{equation}
\mathcal{Z}_{k}^{\mathrm{top}}(M_{4})=\frac{1}{|H^{0}|\cdot|H^{1}|}\sum_{\substack{a_{1},a_{2}\in H^{1}(M_{4},\Z_{N})\\
b\in H^{2}(M_{4},\Z_{N})
}
}e^{2\pi\ii\frac{k}{N}\int_{M_{4}}a_{1}a_{2}b},\label{ZkMtop}
\end{equation}
where the summation is over cohomology classes $H^{i}(M_{4},\Z_{N})$,
rather than cocycles $Z^{i}(M_{4},\Z_{N})$. In this sense, the topological
cocycle model is a 2-group lattice gauge theory, because the gauge
equivalent configurations (coboundaries) are mod out as nonphysical
states. The three cocycle fields $a_{1}$, $a_{2}$ and $b$ correspond
to 1-form and 2-form gauge fields in the continuum.

We believe that the above topological cocycle model is equivalent
to the TQFT defined in Eq.~(\ref{eq:action_AAB}) in the continuum limit. In particular,
they should share the same universal properties such as ground state
degeneracies, number of nonequivalent excitations and their fusion
rules and braidings, etc.

\subsection{Number of topological excitations}

We can extract physical properties of the topological cocycle model
Eq.~(\ref{ZkMtop}) by calculating the partition function on different
spacetime 4-manifolds. If we choose the 4-manifold to be $M_{4}=S^{1}\times M_{3}$,
where $S^{1}$ is the time circle, the partition function is a trace
of identity operator in the ground state subspace. Therefore, it equals
to the ground state degeneracy on the space 3-manifold $M_{3}$:
\begin{align}
\mathrm{GSD}_{k}(M_{3})=\mathcal{Z}_{k}^{\mathrm{top}}(S^{1}\times M_{3}).
\end{align}
The ground state degeneracy is ultimately related to the topological
excitations in the system, as we can wrap around the nontrivial cycles
of $M_{3}$ by creation operator of point-like or loop-like excitations
to transform one ground state to another.

In particular, we can choose $M_{3}$ to be the three dimensional
sphere $S^{3}$. Since both the first and second homotopy groups of
$S^{3}$ are trivial, there is no nontrivial string or membrane operator
wrapping around $S^{3}$. Therefore, the ground state degeneracy should
always be one. In fact, one can also show directly that
\begin{equation}
\mathrm{GSD}_{k}(S^{3})=\mathcal{Z}_{k}^{\mathrm{top}}(S^{1}\times S^{3})=\frac{1}{|H^{0}|\cdot|H^{1}|}|H^{1}|^{2}\cdot|H^{2}|=\frac{N^{2}}{N^{2}}=1
\end{equation}
where we used $H^{0}(S^{1}\times S^{3},\Z_{N})=H^{1}(S^{1}\times S^{3},\Z_{N})=\Z_{N}$
and $H^{2}(S^{1}\times S^{3},\Z_{N})=0$.

If the space manifold is $M_{3}=S^{1}\times S^{2}$, we can use a
string operator to create a pair of point-like excitations, transport
one of them around the $S^{1}$ and finally annihilate them. On the
other hand, we can use a membrane operator to create a pure loop-like
excitations, warp it around the $S^{2}$ and finally shrink it to
vacuum. The ground states created in the above two procedures are
not independent, because the point-like and loop-like excitations
has nontrivial mutual statistics. In summary, the ground state degeneracy
on space manifold $M_{3}=S^{1}\times S^{2}$ equals to the number
of point-like excitations, and the number of pure loop-like excitations~\citep{PhysRevB.95.205142}.

Now let us calculate $\mathrm{GSD}_{k}(S^{1}\times S^{2})$ for the
topological cocycle model Eq.~(\ref{ZkM}). The path integral involves
the following cohomology groups of $M_{4}=S^{1}\times M_{3}=T^{2}\times S^{2}$:
\begin{align}
H^{0}(T^{2}\times S^{2},\Z_{N}) & =\Z_{N},\\
H^{1}(T^{2}\times S^{2},\Z_{N}) & =H^{1}(T^{2},\Z_{N})=H^{1}(S^{1},\Z_{N})\times H^{1}(S^{1},\Z_{N})=\Z_{N}\times\Z_{N}=\langle\alpha_{1},\alpha_{2}\rangle,\\
H^{2}(T^{2}\times S^{2},\Z_{N}) & =H^{2}(T^{2},\Z_{N})\times H^{2}(S^{2},\Z_{N})=\Z_{N}\times\Z_{N}=\langle\alpha_{1}\alpha_{2},\beta\rangle,
\end{align}
where $\alpha_{1},\alpha_{2}\in H^{1}(M_{4},\Z_{N})=(\Z_{N})^{2}$
are two 1-cocycle generators associated with the temporal $S^{1}$
and the spacial $S^{1}$ in $M_{4}$. There cup product $\alpha_{1}\alpha_{2}$
is one of the 2-cocycle generators for $H^{2}(M_{4},\Z_{N})=(\Z_{N})^{2}$.
Another 2-cocycle generator $\beta$ is associated with the spatial
$S^{2}$ in $M_{4}$. The cup product of $\alpha_{i}$ itself is trivial:
$(\alpha_{i})^{2}=0$ for $i=1,2$. The pairing between the fundamental
class of $M_{4}$ and the 4-cocycle $\alpha_{1}\alpha_{2}\beta$ gives
us the integral
\begin{equation}
\int_{M_{4}}\alpha_{1}\alpha_{2}\beta=1.
\end{equation}
Using these results, we can decompose the cocycles $a_{1},a_{2}$
and $b$ in terms of the cohomology generators as
\begin{align}
a_{1} & =\mu_{11}\alpha_{1}+\mu_{12}\alpha_{2},\\
a_{2} & =\mu_{21}\alpha_{1}+\mu_{22}\alpha_{2},\\
b & =\mu_{31}\alpha_{1}\alpha_{2}+\mu_{32}\beta,
\end{align}
where $\mu_{ij}\in\Z_{N}$ are the coefficients. And the action integral
becomes $\int_{M_{4}}a_{1}a_{1}b=(\mu_{11}\mu_{22}-\mu_{12}\mu_{21})\mu_{32}\int_{M_{4}}\alpha_{1}\alpha_{2}\beta=(\mu_{11}\mu_{22}-\mu_{12}\mu_{21})\mu_{32}$,
since $(\alpha_{i})^{2}=0$. Therefore, the path integral in the partition
function $Z_{k}^{\mathrm{top}}(M_{4})$ is now a finite summation
over $\mu_{ij}$:
\begin{align}
\mathrm{GSD}_{k}(S^{1}\times S^{2}) & =\mathcal{Z}_{k}^{\mathrm{top}}(T^{2}\times S^{2})\nonumber \\
 & =\frac{1}{|H^{0}|\cdot|H^{1}|}\sum_{\substack{a_{1},a_{2}\in H^{1}(M_{4},\Z_{N})\\
b\in H^{2}(M_{4},\Z_{N})
}
}e^{2\pi\ii\frac{k}{N}\int_{M_{4}}a_{1}a_{2}b}\nonumber \\
 & =\frac{1}{N^{3}}\sum_{\{\mu_{ij}\}}e^{2\pi\ii\frac{k}{N}(\mu_{11}\mu_{22}-\mu_{12}\mu_{21})\mu_{32}}\nonumber \\
 & =\sum_{\mu_{32}\in\Z_{N}}\gcd(k\mu_{32},N)^{2}\nonumber \\
 & =\gcd(k,N)^{3}\cdot g\left(\frac{N}{\gcd(k,N)}\right),
\end{align}
where the gcd-square-sum function is defined as $g(n)=\sum_{\mu=0}^{n-1}\gcd(\mu,n)^{2}$.
For the theory of $N=2$ and $k=1$, we have $\mathrm{GSD}(S^{1}\times S^{2})=g(2)=2^{2}+1^{2}=5$.
This is the number of nonequivalent particles and pure loop excitations.
The results agree with the field theory calculations in the main text.

Similarly, if the spacial manifold is 3-torus $T^{3}$, the GSD can
be calculated as
\begin{align}
\mathrm{GSD}_{k}(T^{3}) & =\mathcal{Z}_{k}^{\mathrm{top}}(S^{1}\times T^{3})\nonumber \\
 & =\frac{1}{|H^{0}|\cdot|H^{1}|}\sum_{\substack{a_{1},a_{2}\in H^{1}(M_{4},\Z_{N})\\
b\in H^{2}(M_{4},\Z_{N})
}
}e^{2\pi\ii\frac{k}{N}\int_{M_{4}}a_{1}a_{2}b}\nonumber \\
 & =\frac{1}{N^{5}}\sum_{\{\mu_{i},\nu_{j},\lambda_{kl}\}}e^{2\pi\ii\frac{k}{N}\sum_{1\leq i,j\leq4}\sum_{1\leq k<l\leq4}\mathrm{sgn}(ijkl)\mu_{i}\nu_{j}\lambda_{kl}}.
\end{align}
The summation in the exponent is over all $\mu_{i},\mu_{j},\lambda_{kl}\in\Z_{N}$
for $1\leq i,j\leq4$ and $1\leq k<l\leq4$. For $N=2$, the above
formula gives us $\mathrm{GSD}_{0}(T^{3})=N^{9}=512$ and $\mathrm{GSD}_{1}(T^{3})=92$.

\section{Detailed calculation for examples of fusion rules in the main text}\label{appendix:detail_fusion_example}
In this appendix, we derive the several fusion rules mentioned in Sec.~\ref{subsec:example_fusion_rules} in details.
\subsection{$\mathbb{Z}_{N_{1}}$-particle and $\mathbb{Z}_{N_{2}}$-particle}

The first example is the fusion of a $\mathbb{Z}_{N_{1}}$-particle
and a $\mathbb{Z}_{N_{2}}$-particle. We can write down
\begin{align}
\left\langle \mathsf{P}_{100}\otimes\mathsf{P}_{010}\right\rangle = & \frac{1}{\mathcal{Z}}\int\mathcal{D}\left[A^{i}\right]\mathcal{D}\left[B^{i}\right]\exp\left({\rm i}S\right)\exp\left({\rm i}\int_{\gamma}A^{1}\right)\times\exp\left({\rm i}\int_{\gamma}A^{2}\right)\nonumber \\
= & \frac{1}{\mathcal{Z}}\int\mathcal{D}\left[A^{i}\right]\mathcal{D}\left[B^{i}\right]\exp\left({\rm i}S\right)\exp\left({\rm i}\int_{\gamma}A^{1}+A^{2}\right)\nonumber \\
= & \frac{1}{\mathcal{Z}}\int\mathcal{D}\left[A^{i}\right]\mathcal{D}\left[B^{i}\right]\exp\left({\rm i}S\right)\mathsf{P}_{110}\nonumber \\
= & \left\langle \mathsf{P}_{110}\right\rangle
\end{align}
and find that
\begin{equation}
\mathsf{P}_{100}\otimes\mathsf{P}_{010}=\mathsf{P}_{110}.
\end{equation}
This result indicates that by fusing two particles carrying $\mathbb{Z}_{N_{1}}$
and $\mathbb{Z}_{N_{2}}$ gauge charges respectively we obtain a single
particle that carries both $\mathbb{Z}_{N_{1}}$ and $\mathbb{Z}_{N_{2}}$
gauge charges.

\subsection{Two $\mathbb{Z}_{N_{1}}$-particles}

The second example is the fusion of two $\mathbb{Z}_{N_{1}}$-particles:
\begin{align}
\left\langle \mathsf{P}_{100}\otimes\mathsf{P}_{100}\right\rangle = & \frac{1}{\mathcal{Z}}\int\mathcal{D}\left[A^{i}\right]\mathcal{D}\left[B^{i}\right]\exp\left({\rm i}S\right)\exp\left({\rm i}\int_{\gamma}A^{1}\right)\times\exp\left({\rm i}\int_{\gamma}A^{1}\right)\nonumber \\
= & \frac{1}{\mathcal{Z}}\int\mathcal{D}\left[A^{i}\right]\mathcal{D}\left[B^{i}\right]\exp\left({\rm i}S\right)\exp\left({\rm i}2\int_{\gamma}A^{1}\right).
\end{align}
Integrating out $B^{1}$, $B^{2}$, and $A^{3}$, we obtain flat connection
conditions for $A^{1}$, $A^{2}$, and $B^{3}$ respectively:
\begin{equation}
\frac{N_{1}}{2\pi}dA^{i}=0\Rightarrow\oint A^{1}=\frac{2\pi m_{1}}{N_{1}},
\end{equation}
\begin{equation}
\frac{N_{2}}{2\pi}dA^{2}=0\Rightarrow\oint A^{2}=\frac{2\pi m_{2}}{N_{2}},
\end{equation}
\begin{equation}
\frac{N_{3}}{2\pi}dB^{3}=0\Rightarrow\oint B^{3}=\frac{2\pi m_{3}}{N_{3}},
\end{equation}
with $m_{1,2,3}\in\mathbb{Z}$. Now $\left\langle \mathsf{P}_{100}\otimes\mathsf{P}_{100}\right\rangle $
becomes
\begin{align}
\left\langle \mathsf{P}_{100}\otimes\mathsf{P}_{100}\right\rangle = & \frac{1}{\exp\left({\rm i}\int\frac{pN_{1}N_{2}N_{3}}{\left(2\pi\right)^{2}N_{123}}\widetilde{A}^{1}\widetilde{A}^{2}\widetilde{B}^{3}\right)}\cdot\exp\left({\rm i}\int\frac{pN_{1}N_{2}N_{3}}{\left(2\pi\right)^{2}N_{123}}\widetilde{A}^{1}\widetilde{A}^{2}\widetilde{B}^{3}\right)\exp\left({\rm i}2\int_{\gamma}\widetilde{A}^{1}\right)\nonumber \\
= & \exp\left({\rm i}2\int_{\gamma}\widetilde{A}^{1}\right)\nonumber \\
= & \exp\left(\frac{{\rm i}2\cdot2\pi m_{1}}{N_{1}}\right)
\end{align}
where $\widetilde{A}^{1}$, $\widetilde{A}^{2}$, and $\widetilde{B}^{3}$
are gauge field configurations satisfying the above flat connection
conditions. Since in this case the gauge group is $G=\prod_{i=1}^{3}\mathbb{Z}_{N_{i}}=\left(\mathbb{Z}_{2}\right)^{3}$,
we have
\begin{align}
\left\langle \mathsf{P}_{100}\otimes\mathsf{P}_{100}\right\rangle = & \exp\left(\frac{{\rm i}2\cdot2\pi m}{2}\right)\nonumber \\
= & 1\nonumber \\
= & \frac{1}{\mathcal{Z}}\int\mathcal{D}\left[A^{i}\right]\mathcal{D}\left[B^{i}\right]\exp\left({\rm i}S\right)\times1\nonumber \\
= & \left\langle \mathsf{1}\right\rangle ,
\end{align}
i.e.,
\begin{equation}
\mathsf{P}_{100}\otimes\mathsf{P}_{100}=\mathsf{1}.
\end{equation}
This result tells us that $\mathsf{P}_{100}$ is the anti-particle
of itself which is expected since $\mathsf{P}_{100}$ carries one
unit of $\mathbb{Z}_{N_{1}}=\mathbb{Z}_{2}$ gauge charge.

\subsection{Two $\mathbb{Z}_{N_{1}}$-loops}

In this third example, we give a more complicated example of fusion
of two $\mathbb{Z}_{N_{1}}$-loops.
\begin{align}
\left\langle \mathsf{L}_{100}\otimes\mathsf{L}_{100}\right\rangle = & \frac{1}{\mathcal{Z}}\int\mathcal{D}\left[A^{i}\right]\mathcal{D}\left[B^{i}\right]\exp\left({\rm i}S\right)\nonumber \\
 & \times2\exp\left[{\rm i}\int_{\sigma}B^{1}+\frac{1}{2}\frac{2\pi q}{N_{1}}\left(d^{-1}A^{2}B^{3}+d^{-1}B^{3}A^{2}\right)\right]\delta\left(\int_{\gamma}A^{2}\right)\delta\left(\int_{\sigma}B^{3}\right)\nonumber \\
 & \times2\exp\left[{\rm i}\int_{\sigma}B^{1}+\frac{1}{2}\frac{2\pi q}{N_{1}}\left(d^{-1}A^{2}B^{3}+d^{-1}B^{3}A^{2}\right)\right]\delta\left(\int_{\gamma}A^{2}\right)\delta\left(\int_{\sigma}B^{3}\right)
\end{align}
By integrating out $B^{1}$, $B^{2}$, and $A^{3}$, we can write
down
\begin{align}
\left\langle \mathsf{L}_{100}\otimes\mathsf{L}_{100}\right\rangle = & \frac{1}{\exp\left({\rm i}\int\frac{pN_{1}N_{2}N_{3}}{\left(2\pi\right)^{2}N_{123}}\widetilde{A}^{1}\widetilde{A}^{2}\widetilde{B}^{3}\right)}\cdot\exp\left({\rm i}\int\frac{pN_{1}N_{2}N_{3}}{\left(2\pi\right)^{2}N_{123}}\widetilde{A}^{1}\widetilde{A}^{2}\widetilde{B}^{3}\right)\nonumber \\
 & \times4\exp\left[{\rm i}2\int_{\sigma}\frac{1}{2}\frac{2\pi}{N_{1}}\frac{pN_{1}N_{2}N_{3}}{\left(2\pi\right)^{2}N_{123}}\left(d^{-1}\widetilde{A}^{2}\widetilde{B}^{3}+d^{-1}\widetilde{B}^{3}\widetilde{A}^{2}\right)\right]\nonumber \\
 & \times\delta\left(\int_{\gamma}\widetilde{A}^{2}\right)\delta\left(\int_{\sigma}\widetilde{B}^{3}\right)\delta\left(\int_{\gamma}\widetilde{A}^{2}\right)\delta\left(\int_{\sigma}\widetilde{B}^{3}\right).
\end{align}

We first calculate $\int_{\sigma}d^{-1}\widetilde{A}^{2}\widetilde{B}^{3}$
and $\int_{\sigma}d^{-1}\widetilde{B}^{3}\widetilde{A}^{2}$. We notice
that $\sigma$ can be written as $\sigma=\gamma\times S^{1}$. By
definition, $d^{-1}\widetilde{A}^{2}=\int_{\left[a,b\right]\in\gamma}\widetilde{A}^{2}$
which is a $0$-form with $\left[a,b\right]$ being an open interval
on $\gamma$. Since $\int_{\gamma}\widetilde{A}^{2}=\frac{2\pi m_{2}}{N_{2}}$,
$\int_{\left[a,b\right]\in\gamma}\widetilde{A}^{2}=\frac{2\pi k_{2}}{N_{2}}$
with $k_{2}$ is an integer and there exists $k_{2}^{\prime}$ such
that $k_{2}+k_{2}^{\prime}=m_{2}$. We conclude that $\int_{\sigma}d^{-1}\widetilde{A}^{2}\widetilde{B}^{3}=d^{-1}\widetilde{A}^{2}\int_{\sigma}B^{3}=\frac{2\pi k_{2}}{N_{2}}\frac{2\pi m_{3}}{N_{3}}$.
On the other hand, $d^{-1}\widetilde{B}^{3}=\int_{\mathcal{A}\in\sigma}\widetilde{B}^{3}$
as a $1$-form, where $\mathcal{A}$ is an open area on $\sigma$.
Similarly, we have $\int_{\sigma}d^{-1}\widetilde{B}^{3}\widetilde{A}^{2}=\int_{S^{1}}d^{-1}\widetilde{B}^{3}\int_{\gamma}\widetilde{A}^{2}=\frac{2\pi k_{3}}{N_{3}}\frac{2\pi m_{2}}{N_{2}}$
with $k_{3}\in\mathbb{Z}$ and there exists $k_{3}^{\prime}$ such
that $k_{3}+k_{3}^{\prime}=m_{3}$.

For the Kronecker delta functions, we have
\begin{equation}
\delta\left(\int_{\gamma}\widetilde{A}^{2}\right)=\delta\left(\frac{2\pi m_{2}}{N_{2}}\right)=\frac{1}{N_{2}}\left[1+\exp\left({\rm i}\frac{2\pi m_{2}\cdot1}{N_{2}}\right)+\exp\left({\rm i}\frac{2\pi m_{2}\cdot2}{N_{2}}\right)+\cdots+\exp\left({\rm i}\frac{2\pi m_{2}\cdot\left(N_{2}-1\right)}{N_{2}}\right)\right],
\end{equation}
\begin{equation}
\delta\left(\int_{\gamma}\widetilde{B}^{3}\right)=\delta\left(\frac{2\pi m_{3}}{N_{3}}\right)=\frac{1}{N_{3}}\left[1+\exp\left({\rm i}\frac{2\pi m_{3}\cdot1}{N_{3}}\right)+\exp\left({\rm i}\frac{2\pi m_{3}\cdot2}{N_{3}}\right)+\cdots+\exp\left({\rm i}\frac{2\pi m_{3}\cdot\left(N_{2}-1\right)}{N_{3}}\right)\right].
\end{equation}
Remind that $N_{2}=N_{3}=2$, so
\begin{equation}
\delta\left(\int_{\gamma}\widetilde{A}^{2}\right)=\delta\left(\frac{2\pi m_{2}}{N_{2}}\right)=\frac{1}{2}\left[1+\exp\left({\rm i}\frac{2\pi m_{2}}{2}\right)\right]=\begin{cases}
1 & ,m_{2}=0\mod2\\
0 & ,m_{2}=1\mod2
\end{cases}.
\end{equation}
\begin{equation}
\delta\left(\int_{\gamma}\widetilde{A}^{3}\right)=\delta\left(\frac{2\pi m_{3}}{N_{3}}\right)=\frac{1}{2}\left[1+\exp\left({\rm i}\frac{2\pi m_{3}}{2}\right)\right]=\begin{cases}
1 & ,m_{3}=0\mod2\\
0 & ,m_{3}=1\mod2
\end{cases}.
\end{equation}
It is easy to verify that $\delta\left(\int_{\gamma}\widetilde{A}^{2}\right)\delta\left(\int_{\sigma}\widetilde{B}^{3}\right)\delta\left(\int_{\gamma}\widetilde{A}^{2}\right)\delta\left(\int_{\sigma}\widetilde{B}^{3}\right)=\delta\left(\int_{\gamma}\widetilde{A}^{2}\right)\delta\left(\int_{\sigma}\widetilde{B}^{3}\right)$.

With the above results, we have
\begin{align}
\left\langle \mathsf{L}_{100}\otimes\mathsf{L}_{100}\right\rangle = & 4\exp\left[{\rm i}2\cdot\frac{1}{2}\frac{2\pi}{N_{1}}\frac{pN_{1}N_{2}N_{3}}{\left(2\pi\right)^{2}N_{123}}\left(\frac{2\pi k_{2}}{N_{2}}\frac{2\pi m_{3}}{N_{3}}+\frac{2\pi k_{3}}{N_{3}}\frac{2\pi m_{2}}{N_{2}}\right)\right]\nonumber \\
 & \times\frac{1}{2}\left[1+\exp\left({\rm i}\frac{2\pi m_{2}}{2}\right)\right]\times\frac{1}{2}\left[1+\exp\left({\rm i}\frac{2\pi m_{3}}{2}\right)\right]\nonumber \\
= & 1+\exp\left({\rm i}\frac{2\pi m_{2}}{2}\right)+\exp\left({\rm i}\frac{2\pi m_{3}}{2}\right)+\exp\left({\rm i}\frac{2\pi m_{2}}{2}+{\rm i}\frac{2\pi m_{3}}{2}\right).
\end{align}
We can immediately find that
\begin{align}
\left\langle \mathsf{L}_{100}\otimes\mathsf{L}_{100}\right\rangle = & \frac{1}{\mathcal{Z}}\int\mathcal{D}\left[A^{i}\right]\mathcal{D}\left[B^{i}\right]\exp\left({\rm i}S\right)\nonumber \\
 & \times\left[1+\exp\left({\rm i}\int_{\gamma}A^{2}\right)+\exp\left({\rm i}\int_{\sigma}B^{3}\right)+\exp\left({\rm i}\int_{\gamma}A^{2}+{\rm i}\int_{\sigma}B^{3}\right)\right]\nonumber \\
= & \left\langle \mathsf{1}\oplus\mathsf{P}_{010}\oplus\mathsf{L}_{001}\oplus\mathsf{L}_{001}^{010}\right\rangle .
\end{align}
Therefore, we can conclude that
\begin{equation}
\mathsf{L}_{100}\otimes\mathsf{L}_{100}=\mathsf{1}\oplus\mathsf{P}_{010}\oplus\mathsf{L}_{001}\oplus\mathsf{L}_{001}^{010}.
\end{equation}
This is a non-Abelian fusion rule which tells us that if we fuse two
$\mathbb{Z}_{N_{1}}$-loops we would obtain the superposition of a
vacuum, a $\mathbb{Z}_{N_{2}}$-particle, a $\mathbb{Z}_{N_{3}}$-loop,
and a $\mathbb{Z}_{N_{3}}$-loop decorated by a $\mathbb{Z}_{N_{2}}$-particle.

\subsection{$\mathbb{Z}_{N_{1}}$-loop and $\mathbb{Z}_{N_{2}}$-loop}

In the fourth example, we continue to consider $\mathsf{L}_{100}\otimes\mathsf{L}_{010}$:
\begin{align}
\left\langle \mathsf{L}_{100}\otimes\mathsf{L}_{010}\right\rangle = & \frac{1}{\mathcal{Z}}\int\mathcal{D}\left[A^{i}\right]\mathcal{D}\left[B^{i}\right]\exp\left({\rm i}S\right)\nonumber \\
 & \times2\exp\left[{\rm i}\int_{\sigma}B^{1}+\frac{1}{2}\frac{2\pi q}{N_{1}}\left(d^{-1}A^{2}B^{3}+d^{-1}B^{3}A^{2}\right)\right]\delta\left(\int_{\gamma}A^{2}\right)\delta\left(\int_{\sigma}B^{3}\right)\nonumber \\
 & \times2\exp\left[{\rm i}\int_{\sigma}B^{2}-\frac{1}{2}\frac{2\pi q}{N_{2}}\left(d^{-1}B^{3}A^{1}+d^{-1}A^{1}B^{3}\right)\right]\delta\left(\int_{\sigma}B^{3}\right)\delta\left(\int_{\gamma}A^{1}\right).
\end{align}
After integrating out $B^{1}$, $B^{2}$, and $A^{3}$ and plugging
these constraints of discretized gauge fields back to the path integral
and recalling $N_{1}=N_{2}=N_{3}=2$, we get
\begin{align}
\left\langle \mathsf{L}_{100}\otimes\mathsf{L}_{010}\right\rangle = & \left\langle \exp\left[{\rm i}\int_{\sigma}B^{1}+B^{2}+\frac{1}{2}\frac{2\pi q}{N_{1}}\left(d^{-1}A^{2}B^{3}+d^{-1}B^{3}A^{2}\right)-\frac{1}{2}\frac{2\pi q}{N_{2}}\left(d^{-1}B^{3}A^{1}+d^{-1}A^{1}B^{3}\right)\right]\right\rangle \nonumber \\
 & \times\frac{4}{8}\times\left[1+\exp\left({\rm i}\frac{2\pi m_{1}}{N_{1}}\right)+\exp\left({\rm i}\frac{2\pi m_{2}}{N_{2}}\right)+\exp\left({\rm i}\frac{2\pi m_{1}}{N_{1}}+{\rm i}\frac{2\pi m_{2}}{N_{2}}\right)\right]\times\left[1+\exp\left({\rm i}\frac{2\pi m_{3}}{N_{3}}\right)\right]\nonumber \\
= & \left\langle \mathsf{L}_{110}\oplus\mathsf{L}_{110}^{100}\right\rangle .
\end{align}
To see this, we can write down
\begin{align}
\left\langle \mathsf{L}_{110}\oplus\mathsf{L}_{110}^{100}\right\rangle = & \frac{1}{\mathcal{Z}}\int\mathcal{D}\left[A^{i}\right]\mathcal{D}\left[B^{i}\right]\exp\left({\rm i}S\right)\nonumber \\
 & \times\left\{ 2\exp\left[{\rm i}\int_{\sigma}B^{1}+B^{2}+\frac{1}{2}\frac{2\pi q}{N_{1}}\left(d^{-1}A^{2}B^{3}+d^{-1}B^{3}A^{2}\right)-\frac{1}{2}\frac{2\pi q}{N_{2}}\left(d^{-1}B^{3}A^{1}+d^{-1}A^{1}B^{3}\right)\right]\right.\nonumber \\
 & \left.\,+2\exp\left[{\rm i}\int_{\gamma}A^{1}+{\rm i}\int_{\sigma}B^{1}+B^{2}+\frac{1}{2}\frac{2\pi q}{N_{1}}\left(d^{-1}A^{2}B^{3}+d^{-1}B^{3}A^{2}\right)-\frac{1}{2}\frac{2\pi q}{N_{2}}\left(d^{-1}B^{3}A^{1}+d^{-1}A^{1}B^{3}\right)\right]\right\} \nonumber \\
 & \times\delta\left(\int_{\gamma}A^{2}-A^{1}\right)\delta\left(\int_{\sigma}B^{3}\right)\nonumber \\
= & \frac{1}{\mathcal{Z}}\int\mathcal{D}\left[A^{i}\right]\mathcal{D}\left[B^{i}\right]\exp\left({\rm i}S\right)\nonumber \\
 & \times4\exp\left[{\rm i}\int_{\sigma}B^{1}+B^{2}+\frac{1}{2}\frac{2\pi q}{N_{1}}\left(d^{-1}A^{2}B^{3}+d^{-1}B^{3}A^{2}\right)-\frac{1}{2}\frac{2\pi q}{N_{2}}\left(d^{-1}B^{3}A^{1}+d^{-1}A^{1}B^{3}\right)\right]\nonumber \\
 & \times\frac{1}{2}\left[1+\exp\left({\rm i}\int_{\gamma}A^{1}\right)\right]\times\delta\left(\int_{\gamma}A^{2}-A^{1}\right)\delta\left(\int_{\sigma}B^{3}\right).
\end{align}
Integrate out the Lagrange multipliers so we have
\begin{align}
\left\langle \mathsf{L}_{110}\oplus\mathsf{L}_{110}^{100}\right\rangle = & \left\langle \exp\left[{\rm i}\int_{\sigma}B^{1}+B^{2}+\frac{1}{2}\frac{2\pi q}{N_{1}}\left(d^{-1}A^{2}B^{3}+d^{-1}B^{3}A^{2}\right)-\frac{1}{2}\frac{2\pi q}{N_{2}}\left(d^{-1}B^{3}A^{1}+d^{-1}A^{1}B^{3}\right)\right]\right\rangle \nonumber \\
 & \times\frac{4}{8}\times\left[1+\exp\left({\rm i}\frac{2\pi m_{1}}{N_{1}}\right)+\exp\left({\rm i}\frac{2\pi m_{2}}{N_{2}}-{\rm i}\frac{2\pi m_{1}}{N_{1}}\right)+\exp\left({\rm i}\frac{2\pi m_{2}}{N_{2}}\right)\right]\times\left[1+\exp\left({\rm i}\frac{2\pi m_{3}}{N_{3}}\right)\right]
\end{align}
which is exactly $\left\langle \mathsf{L}_{100}\otimes\mathsf{L}_{010}\right\rangle $.
Therefore, we can conclude that
\begin{equation}
\mathsf{L}_{100}\otimes\mathsf{L}_{010}=\mathsf{L}_{110}\oplus\mathsf{L}_{110}^{100}.
\end{equation}
This is another non-Abelian fusion rule. The output of fusion of a
$\mathbb{Z}_{N_{1}}$-loop and a $\mathbb{Z}_{N_{2}}$-loop is the
superposition of a \emph{pure }$\left(\mathbb{Z}_{N_{1}},\mathbb{Z}_{N_{2}}\right)$-loop, $\mathsf{L}_{110}$, and a $\left(\mathbb{Z}_{N_{1}},\mathbb{Z}_{N_{2}}\right)$-loop decorated by a $\mathbb{Z}_{N_{1}}$-particle, $\mathsf{L}_{110}^{100}$. We should notice the following equivalence relation as indicated in Table~\ref{tab:AAB_decorated_loop}: $\mathsf{L}_{110}=\mathsf{L}_{110}^{110}$ and $\mathsf{L}_{110}^{100}=\mathsf{L}_{110}^{010}$.

\subsection{Two $\mathbb{Z}_{N_{1}}$-loops decorated by $\mathbb{Z}_{N_{3}}$-particle\label{appendix:two_vacuum_example}}

In this example, we consider $\mathsf{L}_{100}^{001}\otimes\mathsf{L}_{100}^{001}$.
In path integral, this fusion process is written as
\begin{align}
\left\langle \mathsf{L}_{100}^{001}\otimes\mathsf{L}_{100}^{001}\right\rangle = & \frac{1}{\mathcal{Z}}\int\mathcal{D}\left[A^{i}\right]\mathcal{D}\left[B^{i}\right]\exp\left({\rm i}S\right)\nonumber \\
 & \times4^{2}\times\exp\left[{\rm i}2\int_{\sigma}B^{1}+{\rm i}2\int_{\sigma}\frac{1}{2}\frac{2\pi q}{N_{1}}\left(d^{-1}A^{2}B^{3}+d^{-1}B^{3}A^{2}\right)\right.\nonumber \\
 & \left.+{\rm i}2\int_{\gamma}A^{3}+{\rm i}2\int_{\gamma}\frac{1}{2}\frac{2\pi q}{N_{3}}\left(d^{-1}A^{1}A^{2}-d^{-1}A^{2}A^{1}\right)\right]\label{eq:two_zn1_loops_zn3_particle_path_integral}\\
 & \times\delta\left(\int_{\gamma}A^{2}\right)\delta\left(\int_{\sigma}B^{3}\right)\delta\left(\int_{\gamma}A^{1}\right).
\end{align}
We integrate out the Lagrange multipliers and denote the remaining
gauge fields as $\widetilde{A}^{1}$, $\widetilde{A}^{2}$, and $\widetilde{B}^{3}$.
Since $\widetilde{A}^{1}$, $\widetilde{A}^{2}$, and $\widetilde{B}^{3}$
are forced to be $\mathbb{Z}_{2}$-valued, we have
\begin{equation}
\exp\left[{\rm i}2\int_{\sigma}\frac{1}{2}\frac{2\pi q}{N_{1}}\left(d^{-1}\widetilde{A}^{2}\widetilde{B}^{3}+d^{-1}\widetilde{B}^{3}\widetilde{A}^{2}\right)\right]=1,
\end{equation}
\begin{equation}
\exp\left[{\rm i}2\int_{\gamma}\frac{1}{2}\frac{2\pi q}{N_{3}}\left(d^{-1}\widetilde{A}^{1}\widetilde{A}^{2}-d^{-1}\widetilde{A}^{2}\widetilde{A}^{1}\right)\right]=1.
\end{equation}
Therefore,
\begin{align}
\left\langle \mathsf{L}_{100}^{001}\otimes\mathsf{L}_{100}^{001}\right\rangle = & 4^{2}\times\frac{1}{2}\left[1+\exp\left({\rm i}\int_{\gamma}\widetilde{A}^{1}\right)\right]\times\frac{1}{2}\left[1+\exp\left({\rm i}\int_{\gamma}\widetilde{A}^{2}\right)\right]\times\frac{1}{2}\left[1+\exp\left({\rm i}\int_{\gamma}\widetilde{B}^{3}\right)\right]\nonumber \\
= & 2\cdot\left\langle \mathsf{1}\oplus\mathsf{P}_{100}\oplus\mathsf{P}_{010}\oplus\mathsf{L}_{001}\oplus\mathsf{P}_{110}\oplus\mathsf{L}_{001}^{100}\oplus\mathsf{L}_{001}^{010}\oplus\mathsf{L}_{001}^{110}\right\rangle
\end{align}
and we conclude that this fusion rule is
\begin{align}
\mathsf{L}_{100}^{001}\otimes\mathsf{L}_{100}^{001}= & 2\cdot\left(\mathsf{1}\oplus\mathsf{P}_{100}\oplus\mathsf{P}_{010}\oplus\mathsf{L}_{001}\oplus\mathsf{P}_{110}\oplus\mathsf{L}_{001}^{100}\oplus\mathsf{L}_{001}^{010}\oplus\mathsf{L}_{001}^{110}\right).\label{eq:two_zn1_loops_zn3_particle_two_vacuum}
\end{align}
We notice that the fusion of two $\mathsf{L}_{100}^{001}$'s produces
two vacuum. In fact, for $\mathsf{L}_{010}^{001}$ and $\mathsf{L}_{110}^{001}$,
the fusion of their two copies also leads to the same output as $\mathsf{L}_{100}^{001}\otimes\mathsf{L}_{100}^{001}$,
as shown in Table.~\ref{tab:fusion_AAB_Z2Z2Z2_full}. In Eq.~(\ref{eq:two_zn1_loops_zn3_particle_two_vacuum}),
the fusion output is two copies of the direct sum of all Abelian excitations.
For simplicity, we denote $\mathbf{Ab}\equiv\mathsf{1}\oplus\mathsf{P}_{100}\oplus\mathsf{P}_{010}\oplus\mathsf{L}_{001}\oplus\mathsf{P}_{110}\oplus\mathsf{L}_{001}^{100}\oplus\mathsf{L}_{001}^{010}\oplus\mathsf{L}_{001}^{110}$.
Loosely speaking, $\mathbf{Ab}$ in the fusion output is be resulted
from the three Kronecker delta functions in Eq.~(\ref{eq:two_zn1_loops_zn3_particle_path_integral}).
The reason for the \emph{two} copies of $\mathbf{Ab}$ is that the
factor of $\mathsf{L}_{100}^{001}$ is $4$:
\begin{equation}
4\times4\times\left(\frac{1}{2}\right)^{3}=2.
\end{equation}
As for the factor in the front of $\mathsf{L}_{100}^{001}$, it come from the fact
that
\begin{equation}
\mathsf{L}_{100}\otimes\mathsf{P}_{001}=\mathsf{L}_{100}^{001}
\end{equation}
in which
\begin{align}
\mathsf{L}_{100}= & 2\exp\left[{\rm i}\int_{\sigma}B^{1}+\frac{1}{2}\frac{2\pi q}{N_{1}}\left(d^{-1}A^{2}B^{3}+d^{-1}B^{3}A^{2}\right)\right]\nonumber \\
 & \times\delta\left(\int_{\gamma}A^{2}\right)\delta\left(\int_{\sigma}B^{3}\right)
\end{align}
and
\begin{align}
\mathsf{P}_{001}= & 2\exp\left[{\rm i}\int_{\gamma}A^{3}+\frac{1}{2}\frac{2\pi q}{N_{3}}\left(d^{-1}A^{1}A^{2}-d^{-1}A^{2}A^{1}\right)\right]\delta\left(\int_{\gamma}A^{1}\right)\delta\left(\int_{\gamma}A^{2}\right).
\end{align}
One may wonder if we could assume there is only one vacuum after $\mathsf{L}_{100}^{001}\otimes\mathsf{L}_{100}^{001}$
then determined the factor of $\mathsf{L}_{100}^{001}$. Unfortunately,
such factor would violate the requirement that fusion coefficients
are integers. In conclusion, the two vacuum output of $\mathsf{L}_{100}^{001}\otimes\mathsf{L}_{100}^{001}$
is a result from field-theoretical aspect. We hope future work could provide a deeper understanding for this result.

\twocolumngrid

%\bibliography{braid}


\end{document}